\documentclass[aps,prc,floatfix,groupedaddress,showpacs,amsfonts]{revtex4}
\usepackage{bm}
\usepackage{epsfig}
\usepackage{amsmath}

\newcommand*{\cm}{center of mass }

\newcommand*{\bK}{{\mathbf K}}

\newcommand*{\Gcu}{\underline{{\mathcal G}}}

\newcommand*{\Ucu}{\underline{{\mathcal U}}}

\newcommand*{\gu}{\underline{g}}
\newcommand*{\tu}{\underline{t}}

\newcommand*{\Gc}{\mathcal{G}}

\newcommand*{\Uc}{\mathcal{U}}

\newcommand*{\bp}{{\mathbf p}}
\newcommand*{\bq}{{\mathbf q}}
\newcommand*{\bQ}{{\mathbf Q}}
\newcommand*{\Lc}{\mathcal{L}}
\newcommand*{\Mc}{\mathcal{M}}

\newcommand*{\be}{\begin{equation}}
\newcommand*{\ee}{\end{equation}}
\newcommand*{\bay}{\begin{eqnarray}}
\newcommand*{\eay}{\end{eqnarray}}

\begin{document}
\title{\bf Generalized Faddeev equations in the AGS form for deuteron stripping with explicit inclusion of target excitations and Coulomb interaction}

\author{A. M. Mukhamedzhanov $^{1}$}
\author{V. Eremenko$^{1,2}$}
\author{A. I. Sattarov$^{3}$}
%\email{akram@comp.tamu.edu} 
\affiliation{$^{1}$Cyclotron Institute, Texas
A\&M University, College Station, TX 77843, USA }
\affiliation{$^{2}$D.V. Skobeltsyn Institute of Nuclear Physics, Lomonosov Moscow State University, Moscow,
Russia  }
\affiliation{$^{3}$Department of Physics,
Texas A\&M University, College Station, TX 77843, USA }

\date{\today}

\begin{abstract}
Theoretical description of reactions in general, and the theory for $(d,p)$ reactions, in particular, needs to advance into the new century. Here deuteron stripping processes off a target nucleus consisting of ${A}$ nucleons are treated within the framework of the few-body integral equations theory. The generalized Faddeev equations in the AGS form, which take into account the target excitations, with realistic optical potentials provide the most advanced and complete description of the deuteron stripping. The main problem in practical application of such equations is the screening of the Coulomb potential, which works only for light nuclei. In this paper we present a new formulation of the Faddeev equations in the AGS form taking into account the target excitations with explicit inclusion of the Coulomb interaction. By projecting the $(A+2)$-body operators onto target states, matrix three-body integral equations are derived which allow for the incorporation of
the excited states of the target nucleons. Using the explicit equations for the partial Coulomb scattering wave functions in the momentum space we present the AGS equations in the Coulomb distorted wave representation without screening procedure. We also use the explicit expression for the off-shell two-body Coulomb scattering $T$-matrix which is needed to calculate the effective potentials in the AGS equations. The integrals containing the off-shell Coulomb T-matrix are regularized to make the obtained equations suitable for calculations. For $NN$ and nucleon-target nuclear interactions we assume the separable potentials what significantly simplifies solution of the AGS equations.   
\end{abstract}

\maketitle

%%%%%%%%%%%%%%%%%%%%%%%%%%%%%%%%%%%%%%%%%%%%%%%%%%%%%%%%%%%%%%%%%%%%
\newpage
\section{INTRODUCTION}
Scattering of composite projectiles such as the deuteron off a
composite target nucleus can in principle be described by the N-body
scattering theories (see, e.g., Ref.\ \cite{Sandhas78}), which allow
the simultaneous treatment of all reaction channels (including
partial and complete breakup) in an exact, unique,
N-particle-unitarity-preserving manner. These N-body equations can
be reduced to two-cluster equations, as proposed for the first time
in Ref. \cite{GrassSand67}, the only input necessary being the
elementary two-body transition amplitudes. However, due to the
inherent complexity of the resulting equations it is always desirable 
to take resort to some kind of simplifying approach. For
reactions in which, besides two-body, also three-body channels are
known to be important a promising candidate is the three-body Faddeev
integral equations theory written in the Alt-Grassberger-Sandhas (AGS) form \cite{AGS67,alt1978}.
But even the three-body AGS equations are so complicated that a more simplified CDCC approach became quite popular in the last 30 years (see \cite{rawitscher,austern87,Austern96} and references therein). Of course, this approach does no longer constitute a rigorous theory, the approximative character becoming the more apparent the more important the internal
degrees of freedom are which can be excited in the energy domain
under consideration.

If the internal structure of the target can no longer be 
neglected, in order to still be able to work with the manageable three-body
theory, the possibility of excitation must - at least approximately
- be taken into account. For the first time it has been done in \cite{alt2007}.
Deuteron stripping processes off a target nucleus consisting of ${A}$ 
nucleons were treated within the framework of the generalized few-body Faddeev 
integral equations theory written in the AGS form. Generalization of the AGS equations is 
achieved by taking into account the excitation of the target. To reduce the $(A+2)$-particle 
problem to the much simpler three-body problem all the operators acting in the $(A+2)$-particle 
space were projected onto the three-particle space. Obtained generalized Faddeev equations couple all rearrangement (reaction), inelastic and elastic amplitudes. The transition amplitudes for all interesting three-body processes were obtained, whether the target nucleus is in its ground
or in some excited state before and/or after the collision. 
The practical application was done for the deuteron stripping reaction 
${}^{12}{\rm C}(d,p){}^{13}{\rm C}$ at the deuteron bombarding energies of $4.66,\,15$ and $56$ MeV. 
The two-body-type T-operators for the nucleon-nucleus subsystem were 
calculated from multichannel equations to account for the
excitation and de-excitation of the nucleus in nucleon-nucleus
scattering. When inserted in the three-body-type integral equations
this feature is automatically introduced also into the three-body
dynamics. The complexity of the AGS equations is significantly reduced by 
using separable potential approach for $NN$ and $NA$ interactions. 
However, the Coulomb $p-A$ interaction was neglected when solving the AGS equations
and was taken into account only approximately by multiplying each calculated purely nuclear, 
partial wave reaction amplitudes by the initial- and final-state Coulomb distortion factors
before summing up the partial wave series. Definitely, in such an approach we neglect the Coulomb-modified vertex form factors describing the subsystem $p-A$ and effective potentials in the AGS approach described by the triangular diagrams containing the Coulomb $p-A$ scattering amplitude as the four-ray vertex. 

After our work \cite{alt2007} a significant advance in the application of the Faddeev three-body
approach was achieved in works \cite{deltuva2009,deltuva2009a,deltuva2009b}, where AGS equations for stripping reactions on different targets were solved with realistic potentials. In works \cite{nunes2011,filomenadeltuva} the AGS calculations were compared with the adiabatic distorted wave approach (ADWA) for deuteron stripping reactions to estimate the accuracy of the conventional ADWA, which provides a simplified but practical version of the CDCC \cite{johnson}. The advantage of these works compared to our work \cite{alt2007} was usage of the realistic potentials. Also the Coulomb interaction was included using the screening procedure, which has been applied earlier in our work for $p+ d$ scattering \cite{alt2002}. Inclusion of the screening procedure requires higher screening radii when charge of the target increases to get the convergence. That is why the application of the Coulomb potential screening procedure was successful only for targets with charge $Z \leq 20$.
Besides, the screening procedure cannot be always a reliable remedy to solve problems with charged particles, because limiting the range of the potential may lead to the loss of information about the very nature of the field creating the Coulomb potential.
Neglect of the internal structure of the target is another setback in the AGS calculations in \cite{deltuva2009,deltuva2009a,deltuva2009b,nunes2011,filomenadeltuva}. 

In this paper we present a new formulation of the generalized Faddeev equations in the AGS form for the deuteron stripping, which includes explicitly the Coulomb interactions and target excitations. The Coulomb interaction in the AGS approach appears in the three-ray vertex form factors in the effective potentials and in the four-ray vertex in the triangular diagrams. Applying the two-potential equation allows us to remove the non-compact singularity in the triangular diagram describing the elastic scattering and containing the $p-A$ Coulomb scattering amplitude. Besides the AGS equations can be rewritten in the Coulomb distorted wave representation
\cite{alt1980,muk2000,muk2001}, in which the reaction amplitudes and the effective potentials are sandwiched by the Coulomb distorted waves in the initial and final states. Applying the regularization procedure we obtain the expression for the effective potentials in the AGS equations, which are free of the singularities caused by the Coulomb distortions in the initial and final states. We also investigate the off-shell Coulomb scattering amplitude 
and show that the Coulomb-modified form factors in the transfer amplitudes and in the triangular diagrams don't contain non-integrable singularities. The target excitation is taken into account following the formalism developed in \cite{alt2007}. The final generalized matrix AGS equations are written in the form which includes explicitly Coulomb interactions, target excitations and spins of the particles. Because the solution of the AGS equations is greatly simplified for separable potentials, we use separable potential approach assuming that the adopted separable potentials will approximate realistic potentials $NN$ and $NA$ potentials. The calculations of the obtained equations and comparison with the experimental data will be presented in the following up papers.   

It is important to underscore that AGS equations in the Coulomb distorted wave representation 
have been derived for the first time. Previously possibility of derivation of these equations 
in the Coulomb distorted wave representation was considered in \cite{alt1980,muk2001}, but obtained
equations were not genuine integral equations because the amplitudes under the integral and on the left-hand side were different functions. Here, using a different off-shell continuation we derive 
genuine AGS equations in the Coulomb distorted wave representation. Another important topic of our research is investigation of the singularities of the integrals containing the off-shell Coulomb scattering amplitude. The compactness of the AGS equations for charged particles with repulsive interactions was proved in \cite{muk2000,muk2001}. However, the practical application of this result requires regularization of the integrals containing 
the off-shell Coulomb scattering amplitudes. Note that in the case under consideration, when only two particles are charged, only one off-shell Coulomb scattering amplitude of the proton-nucleus scattering is needed. Regularization of the integrals containing the off-shell Coulomb scattering amplitude is also important because in a standard procedure involving the screening of the Coulomb potential is tacitly assumed that, in the limit of the screening radius $R \to \infty$, all the integrals containing Coulomb scattering amplitude have well-defined limits and that the Coulomb screening affects only the Coulomb distorted waves in the initial and final states. We show how to deal with all the integrals containing the Coulomb off-shell T-matrix. The obtained equations are suitable for calculations and the results will be presented in the following up papers.    

The advantage of the developed approach is that, owe to the explicit including of the Coulomb interaction, it can be applied for analysis of the deuteron stripping on heavier nuclei. Such reactions provide a unique tool to study $(n,\,\gamma)$ processes on exotic nuclei, which are important for nuclear astrophysics and applied physics.

The plan of this paper is as follows. In Sect. \ref{reduction} we first derive
the matrix three-body equations for the (A+2)-body operators
projected onto target states. 
In Sect. \ref{sepmultchpottwobodyequations} 
by choosing (quasi-)separable ansaetze
for the multichannel potentials these three-body equations are
converted into effective-two body equations in the usual manner. 
In Sect. \ref{Couldistwaverepr} the AGS equations in the Coulomb distorted wave 
representation are derived. In Sect. \ref{angmomentdecomp} we present the final expressions for the modified AGS equations after angular momentum decomposition. In Appendices A-D the partial Coulomb scattering wave functions, regularized matrix elements sandwiched by the Coulomb distorted waves, the Coulomb off-shell scattering amplitude, the Coulomb modified form factors and the pole singularity of the exchange Coulomb triangular diagram are considered. Throughout the paper the consideration is done in the \cm of the three-body system $\alpha + \beta + \gamma$, that is the sum of momenta of all three particles is always zero. We use the system of units in which $\hbar=c=1$.

\section{Reduction of the (A+2)-particle to a three-particle problem}
\label{reduction}

In this section we consider the reduction of the system consisting of two nucleons (denoted 
as particles 1 (proton) and 3 (neutron)) and a nucleus consisting of A nucleons (particle 2) to the three-body system. Presentation here extends the formalism presented in \cite{alt2007} by including the proton-target Coulomb interaction and 
removing some inconsistencies in \cite{alt2007}.     

The Hamiltonian is given as
\begin{equation}
{\it H}={\it H}_{int}+{\it H}_0+{\it V},
\end{equation}          
\begin{equation}
{\it H}_0={\it{\bK}}^2_\alpha /2\mu_\alpha + {\it{\bQ}}^2_\alpha /2 M_\alpha,
\end{equation}          
\begin{equation}
{\it V}={\it V}_1+{\it V}_2+{\it V}_3,
\end{equation}          
where ${\sl H}_{int} = T_{int} + V_{int}\,$ is the internal Hamiltonian of nucleus $2$; $\,T_{int}$ and $V_{int}$ are the internal kinetic energy operator and internal potential of nucleus $2$. ${\it 
H}_0$ is the Hamiltonian of the relative motion of the non-interacting 
particles 1, 3 and the center of mass of particle 2. That is, ${\it 
{\bK}}_\alpha$ is the momentum operator for the relative motion of particles 
$\beta$ and $\gamma$ and $\mu_\alpha=m_\beta 
m_\gamma/m_{\beta \gamma}$ the corresponding reduced mass, $m_{\beta\gamma}= m_{\beta} + m_{\gamma}$; $\,\,{\it 
{\bQ}}_\alpha$ is the relative momentum operator for the motion of particle 
$\alpha$ and the center of mass of $(\beta,\gamma)$ with 
$M_\alpha=m_\alpha\,m_{\beta\gamma}/(m_\alpha+m_\beta+m_\gamma)$; 
$m_\nu$ denotes the mass of particle $\nu$. The potentials ${\it V}_1$ 
and ${\it V}_3$ describe the interaction of the nucleons 3 and 1, 
respectively, with each of 
the constituents of nucleus 2, and ${\it V}_2$ is the internucleon 
potential.  Potential ${\it V}_3 = {\it V}_3^{S}  + {\it V}_3^{C}$, where 
${\it V}_3^{S}$  and ${\it V}_3^{C}$  are the short-range and the Coulomb part
of the proton-target interaction, respectively. The Coulomb potential depends only on the distance
between the proton (particle $1$) and the \cm of nucleus $2$.
For simplicity, in this section we disregard the Coulomb interaction, which will be 
explicitly included in the next sections.  

Consider the case that nucleus 2 can exist in several internal states 
$\rho$ ($\rho=1,2\dots ,N$), assumed to be orthogonal, with wave 
functions $|\varphi_{2}^\rho>$ and energies $\epsilon^\rho \geq 0$, that is,
\begin{equation}
{\it H}_{int}|\varphi_{2}^\rho>=\epsilon^\rho |\varphi_{2}^\rho>.
\end{equation}

The notation is such that $\rho =1$ corresponds to the ground state 
with $\epsilon^1=0$. The index $\rho$ is supposed to contain the 
complete specification of the internal state, in particular also its 
spin, isospin etc. In concrete calculations it is not necessary to limit 
oneself to genuine bound states; also so-called quasi-states simulating 
the contribution from the continuous spectrum of the internal 
Hamiltonian of the nucleus might be included among the $\{|\varphi_{2}^\rho>\}$.

To reduce the (A+2)-particle problem to the much simpler three-body 
problem we project all operators acting in the (A+2)-particle space onto 
the three-particle space. In this way they become $N \times N$ matrix operators:
\begin{equation}
{\b H} =[H^{\rho\sigma}]=[<\varphi_{2}^\rho |{\it H}| \varphi_{2}^\sigma >],
\end{equation}
\begin{equation}
{\b H}_0=[H_0^{\rho\sigma}]=[<\varphi_{2}^\rho |{\it H}_0| 
\varphi_{2}^\sigma >]=
[\delta_{\rho\sigma}({\it {\bK}}^2_\alpha /2\mu_\alpha +{\it {\bQ}}^2_\alpha /2 
M_\alpha)],
\end{equation}
\begin{equation}
{\b V}_\alpha=[V_\alpha^{\rho\sigma}]=[<\varphi_{2}^\rho |{\it V}_\alpha| 
\varphi_{2}^\sigma >],
\end{equation}
 $[\delta_{\rho\sigma}\,{\bQ}_\alpha]=[<\varphi_{2}^\rho |{\it {\bQ}}_\alpha| \varphi_{2}^\sigma >]$ and 
$[\delta_{\rho\sigma}\,{\bK}_\alpha]=[<\varphi_{2}^\rho |{\it {\bK}}_\alpha| \varphi_{2}^\sigma >]$. The resolvent matrices corresponding to the restricted full and free Hamiltonian 
matrices are
\begin{equation} 
{\Gcu}(z)=(z-{\b H})^{-1},
\end{equation}
\begin{equation}
{\Gcu}_0(z)=(z-{\b H}_0)^{-1}.
\end{equation}
All the operators acting in the $A+2$ space and projected onto the three-body space are underlined,
while their matrix elements, which have upper indices characterizing the excited states of nucleus $A$, are not. 
Note that since the interaction between the nucleons 1 and 3 does not 
depend on the internal state of nucleus 2, the potential matrix $V_2$ is 
diagonal: $[V_2^{\rho\sigma}]=[\delta_{\rho\sigma}\,V_2]$, $V_2$ being a 
scalar function.

Next we introduce the channel Hamiltonian matrix for channel $\alpha $:
\begin{equation}
\hat{\b H}_\alpha=[\hat{H}_\alpha^{\rho\sigma}]=[\big({\bK}^2_\alpha 
/(2\,\mu_\alpha)+V_\alpha \big)^{\rho\sigma}].
\label{Halpha2}
\end{equation}

The plane wave $|{\bf q}_\alpha>$ is eigenfunction of the operator 
${{\bQ}}_\alpha$ to the eigenvalue ${\bf q}_\alpha$ and 
\begin{equation}
{\bQ}^2_\alpha 
/(2\,M_\alpha)|{\bf 
q}_\alpha>=q^2_\alpha/(2\,M_{\alpha})|{\bf q}_\alpha>.
\label{qplanewave1}
\end{equation}

Let us introduce the bound state of particles $\beta=2$ and $\gamma$ with the quantum numbers 
collectively denoted by $n_{\alpha}$ with the wave function $\varphi_{\alpha n_{\alpha}}$, $\,\alpha \not= 2$. It satisfies
\begin{align} 
\big({\hat E}_{\alpha n_{\alpha}} -  {\hat H}_\alpha - H_{int} \big)\,\varphi_{\alpha n_{\alpha} }=0,
\label{eqvarphiaq1}
\end{align}
${\hat H}_{\alpha}= {\bK}^2_\alpha 
/(2\,\mu_{\alpha}) +V_\alpha$  and ${\hat E}_{\alpha n_{\alpha}} <0$ is the binding energy of the bound state $\{\alpha,\,n_{\alpha}\}$.
Multiplying this equation from the left by the bound state wave function $\varphi_{2}^{\rho}$ of nucleus $2$ in the excited state $\rho$ and inserting $\sum\limits_{\sigma}|\varphi_{2}^{\sigma}><\varphi_{2}^{\sigma}|$ we obtain the coupled equations
\begin{equation}
\sum_{\sigma}\hat{H}_\alpha^{\rho\sigma}| \phi^\sigma_{\alpha n_{\alpha}}>=\hat{E}_{\alpha n_{\alpha}}^{\rho}\,| \phi^\rho_{\alpha n_{\alpha}}>
\label{coupledeqn1}
\end{equation}
for the overlap functions of the bound state wave function $\varphi_{\alpha n_{\alpha}}$ and the bound state wave function $\varphi_{2}^{\rho}$ of nucleus $2$
\begin{align}
|\phi^\rho_{\alpha n_{\alpha} }> = <\varphi_{2}^{\rho}|\varphi_{\alpha n_{\alpha} }>,  \qquad \rho = 1,\dots , N.
\label{overlap1}
\end{align}          

A priori, we have infinite set of the coupled overlap functions, but here we restrict the number of the excited  states of nucleus $2$, constraining, correspondingly, the number of the coupled overlap functions by $N$. Also
\begin{align}
&E = q_{\alpha}^{2}/(2\,M_{\alpha}) + {\hat E}_{\alpha n_{\alpha}}^{\rho} + \epsilon^{\rho}
= q_{\alpha}^{2}/(2\,M_{\alpha}) + {\hat E}_{\alpha n_{\alpha}},  \qquad\qquad \alpha \not=2, \nonumber\\
&E = q_{\alpha}^{2}/(2\,M_{\alpha}) + {\hat E}_{\alpha n_{\alpha}} + \epsilon^{\rho}, \qquad \alpha=2.
\label{E1}
\end{align}
is the total energy of the three-body system $\alpha + \beta + \gamma$ with the pair $\alpha$ in the bound state $n_{\alpha}$ and 
\begin{align}
\hat{E}_{\alpha n_{\alpha}}^{\rho} = \hat{E}_{\alpha n_{\alpha}} - \epsilon^{\rho} <0.
\label{Erhoalha1}
\end{align}
Assume that $\beta=2$ and $\gamma$ is a nucleon. Then $\hat{E}_{\alpha n_{\alpha}}^{\rho}$ is the binding energy for the decay of the bound state $(\beta \gamma)_{n_{\alpha}} \to \beta^{\rho} + \gamma$, where $\beta^{\rho}=2^{\rho}$ is the nucleus $2$ being in the excited state $\rho$.
Here we denote $\hat{E}_{\alpha n_{\alpha}}= \hat{E}_{\alpha n_{\alpha}}^{1}$, $\,\,\alpha \not=2$. For $\alpha=2$ $\,\,\hat{E}_{\alpha n_{\alpha}}$ is the deuteron binding energy. 
   
Note that from normalization $<\varphi_{\alpha n'_{\alpha}}|\varphi_{\alpha n_{\alpha}}>= \delta_{n'_{\alpha}n_{\alpha}}$ we get 
\begin{equation}
<\varphi_{\alpha n'_{\alpha}}|\varphi_{\alpha n_{\alpha}}>=
\sum^N_{\sigma=1}  <\phi^\sigma_{\alpha n'_{\alpha}} 
|\phi^\sigma_{\alpha n_{\alpha}}> =
\delta_{n'_{\alpha}n_{\alpha}}.
\label{normal1}
\end{equation}
For the two-nucleon subsystem $\alpha =2$, Eq. (\ref{coupledeqn1}) reduces to the familiar eigenvalue equation since both nucleons are structureless.

We define a matrix of two-body T-operators ${\b t}_\alpha(z)$ in the three-body space  for subsystem 
$\alpha \not= 3$ via the Lippmann-Schwinger equation
\begin{equation}
{\b t}_\alpha(z)={\b V}_\alpha+{\b V}_\alpha {\Gcu}_0(z){\b t}_\alpha(z).
\label{LSM}
\end{equation}
This amplitude is obtained from the standard nucleon-nucleus scattering amplitude
\begin{equation}
t_\alpha(z)=V_\alpha+ V_\alpha\,\Gc_0(z)\,t_\alpha(z)
\label{LSM}
\end{equation}
by projecting it onto the target $A$ bound states,
where
\begin{align}
\Gc_{0}(z)= \frac{1}{z- H_{0} - H_{int}}
\label{g01}
\end{align} 
is the free Green function containing the internal Hamiltonian $H_{int}$ of nucleus $2$. 

The elements $t^{\rho\sigma}_\alpha(z)$ are related to the corresponding 
ones of the matrix of two-particle
T-operators ${\hat{\b t}}_\alpha$ read in the two-particle space as
\begin{equation}
t^{\rho\sigma}_\alpha(z)=\int \frac{{\rm d}\, {\bp}_\alpha}{(2\,\pi)^{3}}|{\bf p}_\alpha > 
{\hat t}^{\rho\sigma}_\alpha(z-p^2_\alpha/2M_\alpha)\,<{\bf p}_\alpha |.
\end{equation}
Equation (\ref{LSM}) is equivalent to the following coupled system of 
Lippmann-Schwinger equations for the operators 
$\hat{t}^{\rho\sigma}_\alpha(z)$:
\begin{equation}
\hat{t}^{\rho\sigma}_\alpha(\hat{z}_{\alpha})=V^{\rho\sigma}_\alpha+
\sum^{N}_{\tau=1}V^{\rho\tau}_\alpha\frac{1}{\hat{z}_{\alpha}-\epsilon^\tau-
{\rm {\bf K}}^2_\alpha/(2\mu_\alpha)}\,\hat{t}^{\tau\sigma}_\alpha(\hat{z}_{\alpha}),
\label{LSM1}
\end{equation}
where the energy shift accounts for the different reaction thresholds 
due to the excitation of the nucleus. On the energy shell, $p_{\alpha}= q_{\alpha}$,
$\,\,z =E$ and 
\begin{align}
{\hat z}_{\alpha} = z - q_{\alpha}^{2}/(2\,M_{\alpha}).
\label{zalpha1}
\end{align}
Also 
\begin{equation}
\hat
{{\tu}}_2(\hat{z}_{2})=[\delta_{\rho\sigma}{\hat t}_2(\hat{z}_{2})]
\end{equation}
with
\begin{equation}
\hat{t}_2(\hat{z}_{2})=V_2+V_2\frac{1}{\hat{z}_{2}-{\rm {\bf K}}^2_2/(2\mu_2)}\,\hat{t}_2(\hat{
z}_{2})
\end{equation}
being the purely elastic nucleon-nucleon T-operator.

After we have defined all the necessary ingredients we can introduce the transition operators
satisfying the AGS equations. In this paper we use the modified transition operators. To explain this modifications we consider the standard transition operators \cite{GrassSand67}:
\begin{align}
U_{\beta \alpha}(z)= -\,{\overline \delta}_{\beta \alpha}\,\big(H_{0} + T_{int} - z \big) + V' - V_{\alpha} - V_{int} - {\overline \delta}_{\beta \alpha}\,(V_{\beta} + V_{int}) + {\overline V}_{\beta}\,{\Gc}\,{\overline V}_{\alpha}.
\label{standtransoper1}
\end{align}  
Here, ${\mathaccent 22 \delta}_{\beta \alpha}=1-\delta_{\beta \alpha}$ 
is the anti-Kronecker symbol.
\begin{align}
V'= V_{\alpha} + V_{\beta} + V_{\gamma} + V_{int},
\label{Vmodpotent1}
\end{align}
where $V_{int}$ is the interaction potential of nucleons of target $A$, $\,{\overline V}_{\alpha}= V'- V_{\alpha} - V_{int}$. Let us first consider the nondiagonal transition operators ($\beta \not= \alpha$):
\begin{align}
U_{\beta \alpha}(z)= -\,\big(H_{0} + T_{int} - z \big) + V' - V_{\alpha} - V_{int} - V_{\beta} - V_{int} + {\overline V}_{\beta}\,{\Gc}\,{\overline V}_{\alpha} = -\big(H_{0} + H_{int} - z  \big)  + V - V_{\alpha} -  V_{\beta} + {\overline V}_{\beta}\,{\Gc}\,{\overline V}_{\alpha},
\label{standtransoper2}
\end{align} 
where the full resolvent $\,{\Gc}(z)= (z - H)^{-1}.\,$
Similarly for the diagonal transition ($\beta =\alpha$) 
\begin{align}
U_{\alpha \alpha} = {\overline V}_{\alpha} + {\overline V}_{\alpha}\,{\Gc}\,{\overline V}_{\alpha}.
\label{diagtransoperator1}
\end{align}
Note that the on-shell reaction amplitude is given by the matrix element $<\Phi_{\beta\,n_{\beta}}|U_{\beta \alpha}(z)|\Phi_{\alpha\,n_{\alpha}}>$, where the channel wave function $|\Phi_{\alpha\,n_{\alpha}}>= \varphi_{\alpha\,n_{\alpha}}\,|{\bq}_{\alpha}>$, $\,\,|{\bq}_{\alpha}>\,$ is the plane wave in the initial channel $\alpha$. Because $\big(H_{0} + H_{int} + V_{\alpha}  - z \big)\,\Phi_{\alpha\,n_{\alpha}}=0$, the matrix element $<\Phi_{\beta\,n_{\beta}}|U_{\beta \alpha}(z)|\Phi_{\alpha\,n_{\alpha}}>=  <\Phi_{\beta\,n_{\beta}}|U_{\beta \alpha}^{(-)}(z)|\Phi_{\alpha\,n_{\alpha}}>$, where  $U_{\beta \alpha}^{(+)}(z)$ is the standard transition operator $U_{\beta \alpha}^{(-)}(z)= {\overline V}_{\beta} + {\overline V}_{\beta}\,G\,{\overline V}_{\alpha}$.
 
Thus we can rewrite the standard transition operator in the form, in which $H_{int}$ can be combined with $H_{0}$ as a ``modified free motion'' Hamiltonian $H_{0} + H_{int}$. After that we project the transition operator $U_{\beta \alpha}$ onto the eigenfunctions of $H_{int}$, what allows us to eliminate from the consideration $H_{int}$ replacing it by the corresponding excitation energy of the target $\epsilon^{\rho}$. The modified transition operators $\Ucu_{\beta\alpha}(z)$ also satisfy the AGS equations \cite{AGS67} (the Coulomb interaction
between particles $1$ and $2$ is, for the moment, disregarded) : 
\begin{equation}
\Ucu_{\beta\alpha}(z)={\bar {\delta}}_{\beta\alpha}\Gcu^{-1}_0(z)
+\sum_\gamma \bar {\delta}_{\gamma\alpha}
\Ucu_{\beta\gamma}(z)\Gcu_0(z)\tu_\gamma(z). 
\label{AGSmod1}
\end{equation}  
It is apparent that the general form of 
(\ref{AGSmod1}) coincides with that for three point particles, the only 
difference being that in the present case all operators are now N x N 
matrix operators. The physical amplitude for the transition from an 
incoming state in channel $\alpha$ with relative momentum ${\bf 
q}_\alpha$, with the bound state of particles $\beta$ and $\gamma$ being 
characterized by quantum numbers $n_{\alpha}$, to an outgoing state in channel 
$\beta$ characterized by relative momentum ${\bf q}_\beta$ and 
bound-state quantum numbers $n_{\beta}$, irrespective of the internal excitation 
state of nucleus 2 and taking into account all intermediate-state 
excitations and de-excitations of the nucleus due to
scattering with each of the nucleons, is defined as matrix element of 
$\Ucu_{\alpha\beta}$ between the channel states $\Phi_{\alpha n_{\alpha},{\bf q}_\alpha}$:
\begin{equation}
X_{\beta n_{\beta},\alpha n_{\alpha}}({\bf q}'_\beta, {\bf q}_\alpha; 
E+i0)=<\Phi_{\beta n_{\beta},{\bf q}'_\beta}|\Uc_{\beta\alpha}(E+i0)|\Phi_{\alpha n_{\alpha},{\bf q}_\alpha}>
\label{XAGS1}
\end{equation}
with $E$ being the total energy of the system. The introduced channel wave 
functions $|\Phi_{\alpha n_{\alpha};\bq_\alpha }>$ have the
form of a column matrix,
\begin{equation}
|\Phi_{\alpha n_{\alpha};\bq_\alpha }>=[|\varphi^\rho_{\alpha n_{\alpha}}>
|{\bq}_\alpha>]= \left\{
\begin{array}{c}
|\varphi^1_{\alpha n_{\alpha}}> |{\bq}_\alpha> \\
|\varphi^2_{\alpha n_{\alpha}}> |{\bq}_\alpha> \\
\vdots \\
|\varphi^N_{\alpha n_{\alpha}}> |{\bq}_\alpha>
\end{array}
\right\}. \label{wfunc1}
\end{equation}
Equation (\ref{XAGS1}) represents the sum of 
all contributions
from all excitation states of the nucleus in the incoming and the 
outgoing state which are
allowed by the on-shell condition
\begin{align}
&E=q^2_\alpha/2M_\alpha+ {\hat E}^\rho_{\alpha n_{\alpha}} + \epsilon^\rho = q^2_\alpha/2M_\alpha+ \hat{E}_{\alpha 
n_{\alpha}}                               \nonumber\\
&= q'^2_\beta/2M_\beta+{\hat E}^\sigma_{\beta n_{\beta}} + \epsilon^\sigma=
q'^2_\beta/2M_\beta+\hat{E}_{\beta n_{\beta}}.
\label{E2}
\end{align}
The component
\begin{equation}
X^{\sigma\rho}_{\beta n_{\beta},\alpha n_{\alpha}}({\bf q}'_\beta,{\bf q}_\alpha;E+i0)=
<{\bf q}'_\beta|<\phi^\sigma_{\beta 
n_{\beta}}|U^{\sigma\rho}_{\beta\alpha}(E+i0)|\phi^\rho_{\alpha n_{\alpha}}>|{\bf 
q}_\alpha>,
\label{physamp1}
\end{equation}
where $[U^{\sigma\rho}_{\beta\alpha}]= [<\varphi_{2}^{\sigma}|\Uc_{\beta\alpha}|\varphi_{2}^{\rho}>]$,
describes the transition from an incoming $\alpha$-channel configuration 
$\{{\bf q}_\alpha, n_{\alpha}\}$, with the nucleus being in excitation state 
$\rho$ and satisfying the on-shell constraint, to an outgoing channel 
$\beta$-channel configuration characterized by $\{{\bf q}'_\beta, n_{\beta}\}$ 
and internal excitation state $\sigma$, taking into account all 
intermediate-state excitations and de-excitations of the nucleus due to 
scattering with each of the nucleons. ${\bf q}_\alpha$ and ${\bf q}'_\beta$ are the on-shell relative momenta of particles in the initial channel $\alpha$ and final channel $\beta$.

\section{Separable multichannel potentials and effective-two body
equations}
\label{sepmultchpottwobodyequations}

As is well known, the solution of the AGS equations (\ref{AGSmod1})
is greatly simplified if the subsystem transition operators
$\tu_\alpha$ are represented in separable form. For the
nucleon-nucleus subsystem $\alpha(\not= 2)$ we assume the interaction
$V^{\rho\sigma}_\alpha$ leading from a two-body state with
internal excitation state $\rho$ of the nucleus, to a two-body state
with internal state $\sigma$ of the latter, to be described by a
(quasi-separable) multichannel potential of the form
\begin{equation}
V^{\rho\sigma}_{n'_{\alpha}n_{\alpha}} = \sum_{t'_{\alpha}t_{\alpha}}^{A_{\alpha}}\,|\chi^\rho_{\alpha n'_{\alpha} t'_{\alpha}}
>\lambda^{\rho\sigma}_{\alpha; n'_{\alpha} t'_{\alpha} n_{\alpha} t_{\alpha}}<\chi^\sigma_{\alpha n_{\alpha} t_{\alpha}}|
\label{Matpot}
\end{equation}
for $\rho,\sigma=1,2,\dots N$. Here, $n_{\alpha}$ ($n'_{\alpha}$) denote the
quantum numbers of the two-body state in the pair $\alpha$ before (after) the interaction, $t_{\alpha}\,$ ($t'_{\alpha}$) is the number of the separable expansion term before (after) interaction with the total number of the expansion terms $A_{\alpha}$. A priori, $A_{\alpha} \geq N_{\alpha}$, where $N_{\alpha}$ is the number of the bound states in the pair $\alpha$. Terms with $t_{\alpha} > N_{\alpha}$ represent auxiliary terms, which don't correspond to bound states but are needed to improve the accuracy. 

The form factor for the state with quantum
numbers $n_{\alpha}$, term number of separable expansion $t_{\alpha}$ and internal excitation $\sigma$ is denoted by
$|\chi^\sigma_{\alpha n_{\alpha} t_{\alpha}}>$. The coupling matrix
$[\lambda^{\rho\sigma}_{\alpha;n'_{\alpha} t'_{\alpha} n_{\alpha}  t_{\alpha}}]$ is chosen to be symmetric, that is,
$\lambda^{\rho\sigma}_{\alpha;n'_{\alpha} t'_{\alpha} n_{\alpha} t_{\alpha}}=\lambda^{\sigma\rho}_{\alpha;n_{\alpha}t_{\alpha} n'_{\alpha} t'_{\alpha}}$
in order to ensure the hermiticity of the potential. Since, as
mentioned above, the nucleon-nucleon interaction does not depend on
the internal state of the spectator nucleus and is therefore
diagonal over the upper-scripts, we have for the matrix elements of the coupling matrix
\begin{equation}
\lambda^{\rho\sigma}_{2;n'_{\alpha} t'_{\alpha} n_{\alpha}  t_{\alpha}}=\delta_{\rho\sigma}\lambda_{2;n'_{\alpha}t'_{\alpha} n_{\alpha}  t_{\alpha}}.
\label{lambnn}
\end{equation}
The solution of the Lippmann-Schwinger equation (\ref{LSM1}) for the
above potential matrix
(\ref{Matpot}) has the form
\begin{equation}
\hat{t}^{\rho\sigma}_\alpha(\hat{z}_{\alpha})=\sum_{t'_{\alpha}t_{\alpha}}^{A_{\alpha}}\,|\chi^\rho_{\alpha
n'_{\alpha} t'_{\alpha}}>\Delta^{\rho\sigma}_{\alpha;n'_{\alpha}t'_{\alpha} n_{\alpha}  t_{\alpha}}(\hat{z}_{\alpha})<\chi^\sigma_{\alpha
n_{\alpha} t_{\alpha}}|,
\end{equation}
where the $\Delta^{\rho\sigma}_{\alpha;n'_{\alpha}t'_{\alpha} n_{\alpha} t_{\alpha}}(\hat{z}_{\alpha})$ satisfy the
coupled equations
\begin{equation}
\Delta^{\rho\sigma}_{\alpha;n'_{\alpha}t'_{\alpha} n_{\alpha}  t_{\alpha}}({\hat z}_{\alpha})=
\lambda^{\rho\sigma}_{\alpha;n'_{\alpha}t'_{\alpha} n_{\alpha}  t_{\alpha}} +
\sum^N_{\tau=1}\,\sum\limits_{n''_{\alpha}}\,\sum_{t''_{\alpha}}^{A_{\alpha}}\,\lambda^{\rho\tau}_{\alpha;n'_{\alpha}t'_{\alpha}n''_{\alpha} t''_{\alpha}}<\chi^\tau_{\alpha
n''_{\alpha} t''_{\alpha}}|\frac{1}{\hat{z}_{\alpha} - \epsilon^\tau-K^2_\alpha/2\mu_\alpha}|
\chi^\tau_{\alpha n''_{\alpha} t''_{\alpha}}> \Delta^{\tau\sigma}_{\alpha;n''_{\alpha}t''_{\alpha} n_{\alpha}  t_{\alpha}}(\hat{z}_{\alpha})
\label{taueq}
\end{equation}
for $\alpha \not=2$, with a similar equation for $\alpha=2$ which is,
however, uncoupled with respect to the indices characterizing the
internal state of the nucleus.  

Now we will derive the expression for the overlap function $|\phi^\rho_{\alpha n_{\alpha}} >$
in the separable representation. To this end we can rewrite Eq. (\ref{coupledeqn1}) in the more convenient form:
\begin{equation}
\Big(\hat{E}_{\alpha n_{\alpha}}^{\rho}\, - \frac{{\rm {\bf K}}_{\alpha}^{2}}{2\,\mu_{\alpha}} \Big)\,| \phi^\rho_{\alpha n_{\alpha}}>=
\sum_{\sigma}\,{\it V}_\alpha^{\rho\sigma}| \phi^\sigma_{\alpha n_{\alpha}}>.
\label{coupledeqn2}
\end{equation}
From this equation taking into account Eq. (\ref{Matpot}) for the multichannel potential we get
for the overlap function with quantum numbers $n_{\alpha}$ of the
pair $\alpha\not=2$ and with the nucleus being in excited state
$\rho$  
\begin{align}
| \phi^\rho_{\alpha n_{\alpha}}>\,=\, \frac{1}{\hat{E}_{\alpha n_{\alpha}}^{\rho}\, - \frac{{\rm {\bf K}}_{\alpha}^{2}}{2\,\mu_{\alpha}}}\,\sum\limits_{t_{\alpha}=1}^{A_{\alpha}}\,c_{{n_\alpha } t_{\alpha}}^\rho\,|\chi^\rho_{\alpha n_{\alpha} t_{\alpha}}>,
\label{coupledeqn2}
\end{align}
where
\begin{align}
c_{{n_\alpha } t_{\alpha}}^\rho  = \sum\limits_{\sigma  = 1}^N\,\sum\limits_{n'_{\alpha}}\,{\sum\limits_{t'_{\alpha}=1}^{A_{\alpha}}\,{\lambda _{\alpha;\, n_{\alpha} t_{\alpha} n'_{\alpha} t'_{\alpha}}^{\rho\sigma} } }\,< \chi _{\alpha n'_\alpha t'_{\alpha}}^\sigma |\phi _{\alpha n'_\alpha}^\sigma>.
\label{csigmana1}
\end{align}

From normalization condition (\ref{normal1}) we get  
\begin{align}
\sum\limits_{\rho=1}^{N}\,\sum\limits_{t'_{\alpha},t_{\alpha}=1}^{A_{\alpha}}\,c_{n_{\alpha} t'_{\alpha}}^{\rho\,*}\,c_{n_{\alpha} t_{\alpha}}^{\rho}\,<\chi^\rho_{\alpha n_{\alpha} t'_{\alpha}} |\Big[\frac{1}{\hat{E}_{\alpha n_{\alpha}}^{\rho}\, - \frac{{\rm {\bf K}}_{\alpha}^{2}}{2\,\mu_{\alpha}}}\Big]^{2}\,|\chi^\rho_{\alpha n_{\alpha} t_{\alpha}}> =1.
\label{normalizcond1}
\end{align}

The denominator in this equation is nonsingular because ${\hat E}_{\alpha n_{\alpha}}^{\rho} <0$.
Eq. (\ref{coupledeqn2}) is extremely important for solution of the Faddeev equations for deuteron stripping in the separable representation. The overlap function is not an eigenfunction $|\phi^\rho_{\alpha n_{\alpha}}>$ of any Hermitian Hamiltonian and, hence, not normalized to unity. Its square of the norm is the spectroscopic factor of the configuration $\beta^{\rho} + \gamma$ in the bound state $\alpha_{n_{\alpha}}$:
\begin{align}
<\varphi^\rho_{\alpha n_{\alpha}}|\varphi^\rho_{\alpha n_{\alpha}}> =S_{({\beta^{\rho} \gamma})_{n_{\alpha}}}.
\label{spectrfactor1}
\end{align}

In the three-body model, in which the antisymmetrization between the nucleon $\gamma$ and nucleons of nucleus $\beta=A$ is neglected,  the sum rule is $\sum\limits_{\rho=1}^{N}\,S_{({\beta^{\rho} \gamma})_{n_{\alpha}}}= 1$.
If we take this antisymmetrization into account, for what we need to go beyond of the three-body approach, in the isospin formalism the sum rule is $\sum\limits_{\rho=1}^{N}\,S_{({\beta^{\rho} \gamma})_{n_{\alpha}}}=A+1$. For antisymmetrization separately with respect to protons and neutrons the sum rules is $\sum\limits_{\rho=1}^{N}\,S_{({\beta^{\rho} \gamma})_{n_{\alpha}}}= N_{A}+1$ for $\gamma=3$ and $\sum\limits_{\rho=1}^{N}\,S_{({\beta^{\rho} \gamma})_{n_{\alpha}}}= Z_{A}+1$ for $\gamma=1$, where $N_{A}\,(Z_{A})$ is the number of neutrons (protons) in nucleus $A$. To take into account the antisymmetrization we can include the antisymmetrization factor into the coefficients $c_{n_{\alpha} t_{\alpha}}^{\rho}$. 

Because we use the three-body approach we suggest the following procedure. 
For $\alpha \not=3$ the overlap function $\varphi^\rho_{\alpha n_{\alpha}}$ has a pole at 
$k_{\alpha}^{2}=-\,2\,\mu_{\alpha}\,{\hat E}_{\alpha n_{\alpha}}^{\rho}$.
Its residue in the pole can be expressed in terms of the asymptotic normalization coefficient (ANC) for the virtual decay $\alpha_{n_{\alpha}} \to \beta^{\rho} + \gamma$. The neutron ANCs can be determined from the analysis of the experimental data. The overall normalization due to the antisymmetrization can be included into the coefficients $c_{{n_\alpha }}^\rho$. For $\alpha=3$ the singularity at $\,k_{\alpha}^{2}=-\,2\,\mu_{\alpha}\,{\hat E}_{\alpha n_{\alpha}}^{\rho}\,$ is the branching point but it is a pole for the two-body $p+A$ scattering T-matrix and its residue is expressed in terms of the proton ANC. For $\alpha=2$ the overlap function is the deuteron bound state wave function, which is decoupled from the indices characterizing the excitation of nucleus $A$. 

Now we proceed to the generalized AGS equations. Let $\zeta_{\alpha}= \{n_{\alpha},\,t_{\alpha}\}$
collectively denotes the quantum numbers of the state of the pair $\alpha$ and the number of the separable expansion term.   Allowance for the Coulomb proton-nucleus interaction leads for the physical transition amplitudes, which are the matrix elements of the operators $\Uc_{\alpha\beta}^{\sigma\rho}$: 
\begin{equation}
X^{\sigma\rho}_{\beta\,\zeta_{\beta},\alpha\,
\zeta_{\alpha}}({\bq}'_\beta,{\bq}_\alpha;z)=c^{\sigma *}_{\zeta_{\beta}}\, c^{\rho}_{\zeta_{\alpha}}
<{\bq}'_\beta|<g^\sigma_{\beta\,\zeta_{\beta}}|\Gc^{\sigma\sigma}_0(z)
\Uc^{\sigma\rho}_{\beta\,\alpha}(z)\,\Gc^{\rho\rho}_0(z)|g^\rho_{\alpha\,
\zeta_{\alpha}}>| {\bq}_\alpha>,     \qquad t_{\beta} \leq N_{\beta},\,\,\,t_{\alpha} \leq N_{\alpha},
\label{XCoul1}
\end{equation}
where for $\,\alpha=3\,\,$  $g^\rho_{\alpha\,\zeta_{\alpha}}\,$  is the Coulomb-modified form factor, see Eq. (\ref{gCoulmodif1}) below. For $\alpha \not= 3$ $\,\,g^\rho_{\alpha\,\zeta_{\alpha}}= \chi^\rho_{\alpha\,\zeta_{\alpha}}$. 
The total physical amplitude for transition from channel $\alpha$
with bound state quantum numbers $n_{\alpha}$ and internal state $\rho$ of
the nucleus $A$ to channel $\beta$ with a configuration characterized by
$n_{\beta}$ and $\sigma$ is determined by 
\begin{align}
X^{\sigma\rho}_{\beta\,n_{\beta},\alpha\,n_{\alpha}}({\bq}'_\beta,{\bq}_\alpha;z)= \sum\limits_{t_{\beta}=1}^{N_{\beta}}\,\sum\limits_{t_{\alpha}=1}^{N_{\alpha}}\,X^{\sigma\rho}_{\beta\, \zeta_{\beta},\alpha\,\zeta_{\alpha}}({\bq}'_\beta,{\bq}_\alpha;z).
\label{XPhysamplnoCoul1}
\end{align}

Note that 
\begin{align}
\Gc^{\sigma\sigma}_{0;\beta\,\zeta_{\beta},\alpha\,\zeta_{\alpha}}(z)= \delta_{\beta \alpha}\,\Delta_{\alpha;\,\zeta'_{\alpha} \zeta_{\alpha}}^{\sigma\sigma}(z),
\label{Gc1}
\end{align}
where $\zeta_{\beta} = \zeta'_{\alpha}$.
The amplitudes $X^{\sigma\rho}_{\beta\, \zeta_{\beta},\alpha\,
\zeta_{\alpha}}$ satisfy the off-shell effective-two body equations
\begin{align}
&X^{\sigma\rho}_{\beta\,\zeta_{\beta},\alpha\,\zeta_{\alpha}}({\bp}'_\beta,{\bp}_\alpha;z)=
\delta_{\sigma\rho}Z^{\rho\rho}_{\beta\,\zeta_{\beta},\alpha\,\zeta_{\alpha}}({\bp}'_\beta,{\bp}_\alpha;z)\,+ \,\sum_{\gamma}\,\sum_{ n'_{\gamma} n_{\gamma}}\,\sum\limits_{t'_{\gamma},t_{\gamma}=1}^{A_{\gamma}}\,\sum^{N}_{\tau=1}\int \frac{{\rm d}{\bp}_\gamma}{(2\,\pi)^{3}}\,Z^{\sigma\sigma}_{\beta\,\zeta_{\beta},\gamma\,\zeta'_{\gamma}}({\bp}'_\beta,{\bp}_\gamma;z)
\nonumber\\
& \times \Delta^{\sigma\tau}_{\gamma;\,\zeta'_{\gamma}\,\zeta_{\gamma}}(z-p^2_\gamma/2M_\gamma)\, X^{\tau\rho}_{\gamma\,\zeta_{\gamma},\alpha\,\zeta_{\alpha}}({\bp}_\gamma,{\bp}_\alpha;z).
\label{AGSMI} 
\end{align}
Here ${\bp}'_{\beta}$ and ${\bp}_{\alpha}$ are the off-shell momenta. The off-shell amplitudes are needed later on when deriving the AGS equations in the Coulomb distorted wave representation. 

For only two charged particles (with the charges of the same sign), particles $1$ and $2$ in the case under consideration, the effective potentials are given by \cite{muk2000}
\begin{equation}
Z^{\sigma\sigma}_{\beta\,\zeta_{\beta},\alpha\,
\zeta_{\alpha}}({\bp}'_\beta,{\bp}_\alpha;z)=\bar{\delta}_{\alpha\beta}
c^{\sigma*}_{\zeta_{\beta}}\,c^\sigma_{\zeta_{\alpha}} <{\bp}'_\beta |<\chi^\sigma_{\beta\,
\zeta_{\beta}}|(1 - \delta_{\beta\,\alpha}\,\delta_{3\,\alpha})\,\Gc_{3}^{C\,{\sigma\sigma}}(z)- \delta_{\beta\,\alpha}\,{\overline \delta}_{3\,\alpha}\,\Gc^{\sigma\,\sigma}_{0}|\chi^\sigma_{\alpha\,\zeta_{\alpha}}>|{\bp}_\alpha>.
\label{Zeffpotent1}
\end{equation} 
Here the Coulomb Green function operator is 
\begin{equation} 
{\Gcu}^{C}_{3}(z)=(z- {\b H}_{0} - {\b V}_{3}^{C} )^{-1}
\label{G3C}
\end{equation}
and $[\Gc_{3}^{C\,{\sigma\sigma}}] = [<\varphi_{2}^{\sigma}|{\Gc}^{C}_{3}|\varphi_{2}^{\sigma}>]$,
$\,\,[{V}_{3}^{C\,\sigma\rho}]= [\delta_{\sigma\rho}\,V_{3}^{C}]$.
The effective potentials are diagonal in the upper indices since the internal state of
nucleus 2 cannot be changed by the one-step particle transfer process.
Equations (\ref{AGSMI}) allow one to take into account the
possibility of the excitation of nucleus through the interaction
with the nucleons 1 and 3, respectively, as well as the contribution
of the rescattering of the excited nucleus to the three-body
dynamics. Because our Coulomb $V_{3}^{C}$ interaction potential depends only the 
distance between particle 1 (proton) and the \cm of nucleus $2$, this Coulomb interaction 
cannot excite nucleus and $[{V}_{3}^{C\,\sigma\rho}]$ is diagonal matrix in the upper indices
indicating the target excitation.

Further insight is gained on expanding Eq.\ (\ref{AGSMI}) in a
Neumann series. If in that expansion one neglects the contribution
from all those terms in which the excited nuclear states appear
explicitly, one obtains equations which are of the same form as the
usual AGS equations. The only difference between the latter and the
AGS equations for three point particles is that now the two-particle
amplitudes describing the elastic scattering of each of the nucleons
off nucleus 2 take into account the multistep excitation and
subsequent de-excitation of the nucleus. Therefore, the
corresponding pair potentials must be expected to be
energy-dependent, non-local, and complex. In practical applications
these potentials could be approximated by optical-like potentials
but fitted to the nucleon-nucleus scattering {\em and} bound-state
data.

\section{THREE-BODY EQUATIONS IN THE COULOMB DISTORTED WAVE REPRESENTATION}
\label{Couldistwaverepr}

In this section we obtain the AGS equations in the Coulomb distorted wave representation. To this end we start from the matrix AGS equations (\ref{AGSMI}).
$X^{\sigma\rho}_{\beta \,\zeta_{\beta},\,\alpha \,\zeta_{\alpha}}({\bp}'_\beta,{\bq}_\alpha;z)$ is the reaction amplitude describing the transition from the initial channel $\alpha + (\beta\,\gamma)_{\,\zeta_{\alpha}}$
and nucleus $A$ in the internal state $\rho$, whichever particle it is in the initial channel, to the final channel $\beta + (\alpha\,\gamma)_{\,\zeta_{\beta}}$ with particle $A$ in the internal state $\sigma$. If, for example, $\alpha=2$, the above transition is 
$\alpha^{\rho} + (\beta\,\gamma)_{\,\zeta_{\alpha}} \to \beta + (\alpha^{\sigma}\,\gamma)_{\,\zeta_{\beta}}$.
Because we use separable potentials, the effective potentials $Z^{\rho\rho}_{\beta \zeta_{\beta},\alpha\zeta_{\alpha}}({\bp}'_\beta,{\bq}_\alpha;z)$ are sandwiched by the separable form factors $\chi^\rho_{\alpha \,\zeta_{\alpha}}$ and $\chi^\rho_{\beta \,\zeta_{\beta}}$. In the next section the full AGS equations with explicit indication of  spins and angular momenta will be presented. It is worth mentioning that we assume that all the effective potentials are diagonal over the upper-scripts $\rho$ and $\sigma$, that is the effective potential doesn't change the internal structure of nucleus $A$. This change can occur only in the separable potential matrix $\Delta^{\tau\rho}_{\gamma;\,\zeta'_{\gamma}\,\zeta_{\gamma}}$.
This assumption is justified because the Coulomb $p-A$ interaction depends only on the distance between the proton and the \cm of target $A$. 

Our goal is to obtain the modified Faddeev equations in the AGS form where the Coulomb rescattering of the particles in the initial and final states is explicitly taken into account by sandwiching the transition amplitudes and effective potentials by the corresponding Coulomb scattering wave functions.  That is what we call the Coulomb distorted wave representation. To obtain this representation we follow the strategy outlined in \cite{alt1980}. First, we need to rewrite the matrix AGS equations in a conventional Lippmann-Schwinger form $X = Z + Z\,G_{0}\,X$, where $G_{0}$ is a two-body free Green function. After that we add and subtract the Coulomb potential between the particles in the initial and final channels introducing $Z=Z' + U^{c}$, where $Z'= Z - U^{C}$. After that we will apply the two-potential equation leading to $\,<p^{C}|X|p^{C}> = <p^{C}|Z'|p^{C}> + <p^{C}|Z'\,G^{C}\,X|p^{C}>$, where $\,G^{C}$ is the two-body Coulomb Green function describing the propagation of the particles in the intermediate state and $p^{C}$ is the Coulomb distorted wave. Using the spectral decomposition  of the Coulomb Green function $G^{C}$ we immediately arrive at the desired AGS equations in the Coulomb distorted wave representation, where the reaction amplitudes and the effective potentials in all the channels $\alpha \not=3$ are sandwiched by the Coulomb distorted waves: $\,<p^{C}|X|p^{C}> = <p^{C}|Z'|p^{C}> + <<p^{C}|Z'|p^{C}>\,G_{0}\,<p^{C}|X|p^{C}>>$. We have only quite schematically described our strategy. Its practical implementation has many peculiarities which should be overcome.   
Now we proceed to the practical implementation of the outlined strategy.

First, to rewrite the AGS equation in the Lippmann-Schwinger form, we introduce new effective two-body amplitudes and effective potentials \cite{alt1980}:
\begin{align}
{\tilde X}= X\,\Gcu_{0}\,{\gu}_{0}^{-1}
\label{newtildeX1}
\end{align}
and 
\begin{align}
{\tilde Z}=Z\,\Gcu_{0}\,\gu_{0}^{-1}
\label{newtildeZ1},
\end{align}
with 
\begin{align}
\Gc^{\sigma\rho}_{0;\beta\,\zeta_{\beta},\alpha\,\zeta_{\alpha}}(\bp_{\beta}',\bp_{\alpha};z)= \delta_{\beta\,\alpha}\,\delta(\bp_{\alpha}'-\bp_{\alpha})\,
\Delta^{\sigma\rho}_{\alpha;\zeta'_{\alpha}\,\zeta_{\alpha}}(\hat{z}_{\alpha}),  
\label{Gcrhosigma1}
\end{align}
where $\zeta_{\beta}=\zeta_{\alpha}'$, $\,\,{\hat z}_{\alpha}=z- p_{\alpha}^{2}/(2\,M_{\alpha})$.
For $t_{\alpha} \leq N_{\alpha}$
\begin{align}
\mathit{g}_{0,\,\beta\,\zeta_{\beta}\,\alpha\,\zeta_{\alpha}}^{\sigma\rho}(\bp_{\alpha}',\bp_{\alpha};z)=\delta_{\beta\,\alpha}\,\delta_{\zeta_{\beta}\,\zeta_{\alpha}}\delta_{\sigma\,\rho}\,\delta(\bp_{\alpha}'-\bp_{\alpha})\,\mathit{\tilde g}_{0;\alpha\,\zeta_{\alpha}}(p_{\alpha};z - {\hat E}_{\alpha\, n_{\alpha}}),
\label{g01}
\end{align}
\begin{align}
{\tilde g}_{0;\,\alpha\,\zeta_{\alpha}}(p_{\alpha};z - {\hat E}_{\alpha n_{\alpha}})=\frac{1}{z - {\hat E}_{\alpha\,n_{\alpha}} - p_{\alpha}^{2}/2\,M_{\alpha}}
\label{tildeg01}
\end{align} 
is the two-body free Green function, $z - {\hat E}_{\alpha\,n_{\alpha}} - p_{\alpha}^{2}/2\,M_{\alpha}= q_{\alpha}^{2}/2\,M_{\alpha} - p_{\alpha}^{2}/2\,M_{\alpha} + i\,0$. 

Introducing new reaction amplitudes and effective potentials we are able to rewrite half-off-shell Eqs. (\ref{AGSMI}) in the matrix Lippmann-Schwinger form:
\begin{align}
&{\tilde X}^{\sigma\rho}_{\beta \,\zeta_{\beta},\alpha \,\zeta_{\alpha}}({\bp}'_\beta,{\bq}_\alpha;z)={\tilde Z}^{\sigma\rho}_{\beta \,\zeta_{\beta},\alpha
\,\zeta_{\alpha}}({\bp}'_\beta,{\bq}_\alpha;z)+
\sum_{\gamma}\,\sum\limits_{n_{\gamma}}\,\sum_{\,t_{\gamma}=1}^{N_{\gamma}}\,\sum^{N}_{\tau=1}\int \frac{{\rm d}{\bp}_\gamma}{(2\,\pi)^{3}}
{\tilde Z}^{\sigma\tau}_{\beta \,\zeta_{\beta},\gamma \,\zeta_{\gamma}}({\bp}'_\beta,{\bp}_\gamma;z)
\,\frac{1}{z - {\hat E}_{\gamma \,n_{\gamma}} - {p_{\gamma}}^{2}/(2\,M_{\gamma})} 
\nonumber\\
&\times\,{\tilde X}^{\tau\rho}_{\gamma \,\zeta_{\gamma},\alpha \,\zeta_{\alpha}}({\bp}_\gamma,{\bq}_\alpha;z)  
+  \sum_{\gamma}\sum\limits_{n'_{\gamma},\,n''_{\gamma}}\,\sum_{t'_{\gamma},t''_{\gamma}=N_{\gamma}+1}^{A_{\gamma}}\,\sum^{N}_{\nu=1}\int \frac{{\rm d}{\bp}_\gamma}{(2\,\pi)^{3}}\,
{Z}^{\sigma\sigma}_{\beta \,\zeta_{\beta},\gamma\,\zeta'_{\gamma}}({\bp}'_\beta,{\bp}_\gamma;z)
\,\Delta^{\sigma\nu}_{\gamma,\,\zeta'_{\gamma}\,\zeta''_{\gamma}}(z - p_{\gamma}^{2}/(2\,M_{\gamma})                                \nonumber\\
&\times {X}^{\nu\rho}_{\gamma\,\zeta''_{\gamma},\alpha \,\zeta_{\alpha}}({\bp}_\gamma,{\bq}_\alpha;z).
\label{AGSMII}
\end{align}
Note that the two-body free Green function appears only in the terms describing the physical ${bound\,\,state\,\to\, bound\,\, state}$ transitions. The new effective potentials ${\tilde Z}$ are not diagonal over the upper indices because they contain now $\Gc^{\sigma\rho}$.

The terms with $t'_{\gamma},\,t''_{\gamma} \geq N_{\gamma}+1$ are auxiliary terms included to improve the accuracy. Corresponding auxiliary amplitudes ${\tilde X}^{\tau\rho}_{\gamma\,\zeta''_{\gamma},\alpha \,\zeta_{\alpha}}({\bp}_\gamma,{\bq}_\alpha;z)$ don't describe any physical transition ${bound\,\,state\,\to\, bound\,\, state}$. If we include these auxiliary amplitudes, we need to supplement AGS equations by the equations for these amplitudes.  
 
In \cite{alt2007} the Coulomb $p-A$ interaction was neglected but here we explicitly include it. Allowance for this Coulomb interaction leads to the appearance of the Coulomb-modified separable form factors for the system $(1\,2)$ and the off-shell $1+ 2$ Coulomb scattering amplitude in the four-ray vertex in the triangular diagrams \cite{alt1978,muk2000,muk2001}. The triangular diagrams describing the $d+ A$ and $p +(n\,A)$ elastic scattering with the four-ray vertex, owe to the presence of the off-shell $p+ A$ Coulomb scattering amplitude, contain the Coulomb forward singularity $\Delta^{-2}$ in the transfer momentum plane at $\Delta^{2}= 0$. Coincidence of this singularity with the pole singularity of the two-body Green function leads to a noncompact singularity of the generalized Faddeev equations written in the AGS form \cite{veselova,alt1978,muk2001}. To eliminate it in \cite{alt1978,alt1980} the channel Coulomb potential was added to the diagonal effective potentials and subtracted: 
\begin{align}
&{\tilde Z}^{\sigma\rho}_{\beta \,\zeta_{\beta},\alpha \,\zeta_{\alpha}} = \delta_{\beta\,\alpha}\,{\overline \delta}_{\alpha\,3}\,\delta_{\zeta_{\beta}\,\zeta_{\alpha}}\,\delta_{\sigma\,\rho}\,{\it U}_{\alpha}^{C} 
+ {\tilde Z}^{'\,\sigma\rho}_{\beta\,\zeta_{\beta},\alpha\,\zeta_{\alpha}},     \qquad\qquad t_{\beta}\,\,{\rm or}\,\, t_{\alpha} \leq N_{\alpha},
\label{ZCoulomb1}
\end{align} 
and
\begin{align}
&{\tilde Z}^{\sigma\rho}_{\beta \,\zeta_{\beta},\alpha \,\zeta_{\alpha}} = Z^{\sigma\rho}_{\beta \,\zeta_{\beta},\alpha \,\zeta_{\alpha}},     \qquad\qquad t_{\beta}\,\geq N_{\beta} +1\,\,\, {\rm or}\,\,  t_{\alpha} \geq N_{\alpha} +1.
\label{ZCoulomb2}
\end{align}  
Note that the nondiagonal effective potentials are not affected by the Coulomb channel potential, that is nondiagonal potential ${\tilde Z}^{'\sigma\rho}_{\beta\,\zeta_{\beta},\alpha\,\zeta_{\alpha}}$ coincides with the original effective potential ${\tilde Z}^{\sigma\,\rho}_{\beta\,\zeta_{\beta},\alpha\,\zeta_{\alpha}}$, while its diagonal part in the channel $\alpha \not= 3$  is ${\tilde Z}^{\rho\rho}_{\alpha\,\zeta_{\alpha},\alpha\,\zeta_{\alpha}} - \,{\it U}_{\alpha}^{C}$ for $t_{\alpha}\,\leq N_{\alpha}$; $\,U_{\alpha}^{C}$ is the channel Coulomb potential describing the interaction of particle $\alpha$ and the system $(\beta\,\gamma)$ with its charge concentrated in its {\cm}:
\begin{align}
U_{\alpha}^{C}(\rho_{\alpha})= \frac{Z_{\alpha}\,Z_{\beta\,\gamma}\,e^{2}}{\rho_{\alpha}},
\label{UalphaC1}
\end{align} 
$Z_{\alpha}\,e$ is the charge of particle $\alpha$ and $Z_{\beta\gamma}\,e$ is the charge of the system $(\beta\,\gamma)$, which is equal to $Z_{\beta}\,e$ or $Z_{\gamma}\,e$
depending on which particle of that system has nonzero charge; $\rho_{\alpha}$ is the distance between $\alpha$ and the {\cm} of the system $(\beta\gamma)$. 
The explicit expression for the effective potentials will be given in the next section. 

After adding and subtracting the channel Coulomb potentials, according to the outlined strategy,
we can apply the two-potential theorem, which allows us to rewrite Eqs. (\ref{AGSMII}) in the form, in which the reaction amplitudes and potentials are sandwiched by the Coulomb scattering wave functions (off- and on-shell):   
\begin{align}
&{\tilde X}^{SC\,\sigma\rho}_{\beta\,\zeta_{\beta},\alpha\,\zeta_{\alpha}}({\tilde{\bp}}^{'C(-)}_{\beta},\,{\bq}^{C(+)}_{\alpha};z)=
{\tilde Z}^{'\,SC\,\sigma\rho}_{\beta\,\zeta_{\beta},\alpha\,\zeta_{\alpha}}({\tilde{\bp}}^{'C(-)}_{\beta},{\bq}^{C(+)}_{\alpha};z) +
\sum_{\gamma}\,\sum\limits_{n_{\gamma}}\,\sum\limits_{t_{\gamma}=1}^{N_{\gamma}}\,\sum^{N}_{\tau=1}\int\, \frac{{\rm d}{\bp}_\gamma}{(2\,\pi)^{3}}\,
{\tilde Z}^{'\,SC\,\sigma\tau}_{\beta\, \zeta_{\beta},\gamma\, \zeta_{\gamma}}({\tilde{\bp}}^{'C(-)}_{\beta},{\bp}^{C (-)}_{\gamma};z)\,
    \nonumber\\
&\times \frac{1}{z - {\hat E}_{\gamma\, n_{\gamma}} - {p_{\gamma}}^{2}/(2\,M_{\gamma})} \,{\tilde X}^{SC\,\tau\rho}_{\gamma\, \zeta_{\gamma},\alpha\, \zeta_{\alpha}}({\bp}^{C (-)}_{\gamma},{\bq}^{C(+)}_{\alpha};z)                           \nonumber\\           
&+ \sum_{\gamma}\,\sum\limits_{n_{\gamma},n'_{\gamma}}\,\sum_{t_{\gamma},t'_{\gamma}=N_{\gamma}+1}^{A_{\gamma}}\,\sum^{N}_{\nu=1}\int \frac{ {\rm d}{\bp}_\gamma}{(2\,\pi)^{3}}\,
{Z}^{SC\,\sigma\sigma}_{\beta \,\zeta_{\beta},\gamma \,\zeta_{\gamma}}({\tilde {\bp}}^{'C(-)}_\beta,{\bp}_\gamma;z)\,\Delta^{\sigma\nu}_{\gamma, \,\zeta_{\gamma}\, \zeta'_{\gamma}}(z - p_{\gamma}^{2}/(2\,M_{\gamma}))\,{\tilde X}^{SC\,\nu\rho}_{\gamma \zeta'_{\gamma},\alpha \,\zeta_{\alpha}}({\bp}_\gamma,{\bq}^{C(+)}_{\alpha};z),     \nonumber\\
&\qquad\qquad \,t_{\beta} \leq N_{\beta}, \, \,t_{\alpha} \leq N_{\alpha},
\label{AGSCS1}
\end{align}
\begin{align}
&{\tilde X}^{SC\,\sigma\rho}_{\beta\,\zeta_{\beta},\alpha\,\zeta_{\alpha}}({\bp}'_{\beta},\,{\bq}^{C(+)}_{\alpha};z)=
{\tilde Z}^{'\,SC\,\sigma\rho}_{\beta\,\zeta_{\beta},\alpha\,\zeta_{\alpha}}({\bp}'_{\beta},{\bq}^{C(+)}_{\alpha};z) 
     \nonumber\\
& + \sum_{\gamma}\,\sum\limits_{n_{\gamma}}\,\sum_{t_{\gamma}=1}^{N_{\gamma}}\,\sum^{N}_{\tau=1}\int\, \frac{{\rm d}{\bp}_\gamma}{(2\,\pi)^{3}}\,{\tilde Z}^{SC\,\sigma\tau}_{\beta\, \zeta_{\beta},\gamma\, \zeta_{\gamma}}({\bp}'_{\beta},{\bp}^{C (-)}_{\gamma};z)\,\frac{1}{z - {\hat E}_{\gamma\, n_{\gamma}} - {p_{\gamma}}^{2}/(2\,M_{\gamma})}\,{\tilde X}^{SC\,\tau\rho}_{\gamma\, \zeta_{\gamma},\alpha\, \zeta_{\alpha}}({\bp}^{C (-)}_{\gamma},{\bq}^{C(+)}_{\alpha};z)                                                \nonumber\\
&+ \sum_{\gamma}\,\sum\limits_{n_{\gamma},n'_{\gamma}}\,\sum_{t_{\gamma},t'_{\gamma}=N_{\gamma}+1}^{A_{\gamma}}\,\sum^{N}_{\nu=1}\int \frac{ {\rm d}{\bp}_\gamma}{(2\,\pi)^{3}}\,
{Z}^{\sigma\sigma}_{\beta \,\zeta_{\beta},\gamma \,\zeta_{\gamma}}({\bp}'_\beta,{\bp}_\gamma;z)\,\Delta^{\sigma\nu}_{\gamma, \,\zeta_{\gamma}\, \zeta'_{\gamma}}(z - p_{\gamma}^{2}/(2\,M_{\gamma}))\,{\tilde X}^{SC\,\nu\rho}_{\gamma\,\zeta'_{
\gamma},\alpha \,\zeta_{\alpha}}({\bp}_\gamma,\,{\bq}^{C(+)}_{\alpha};z),                             \nonumber\\
&\qquad\qquad t_{\beta} \geq N_{\beta} + 1, \, t_{\alpha} \leq N_{\alpha}.
\label{AGSCSaux2}
\end{align} 

These equations are the desired Faddeev equations in the AGS form in the Coulomb distorted wave 
representation. They generalize Eqs. (5.35) \cite{alt1980} by taking into account the target excitation. Eqs (\ref{AGSCS1}) determine the components of the reaction amplitude matrix
corresponding to ${bound \,\,state\,\to\,bound\,\,state}$ transitions. From these components one can calculate the observable cross sections. However, to determine these physical matrix elements we need to know also the auxiliary amplitudes ${\tilde X}^{SC\,\tau\rho}_{\gamma\,\zeta'_{\gamma},\alpha\,\zeta_{\alpha}}({\bp}_{\gamma},{\bq}^{C(+)}_{\alpha};z)$  corresponding to transition from the initial bound states ($t_{\alpha} \leq N_{\alpha}$) to the quasiparticle states
($t'_{\gamma} \geq N_{\gamma} +1$). That is why Eqs (\ref{AGSCS1}) are supplemented by Eqs (\ref{AGSCSaux2}), which determine these auxiliary components. We remind that the auxiliary components appear as the result of the separable expansion of the nuclear potential not only over the bound states but also over the quasiparticle states not representing any physical bound states ($t_{\beta} \geq N_{\beta} + 1$).  
 
In these equations the reaction amplitudes are
\begin{align}
&{\tilde X}^{SC\,\sigma\rho}_{\beta\,\zeta_{\beta},\,\alpha\,\zeta_{\alpha}}({\rm {\bf y}}_{\beta},{\bq}^{C(+)}_{\alpha};z)= <{\rm {\bf y}}_{\beta}\big|{\tilde X}^{\sigma\rho}_{\beta\,\zeta_{\beta},\alpha \,\zeta_{\alpha}}(z)\big|\psi^{C(+)}_{{\bq}_\alpha}>,
\label{XSClhs1}
\end{align}
where for $t_{\beta} \leq N_{\beta},\,\,t_{\alpha} \leq N_{\alpha}$ $\,\,{\rm {\bf y}}_{\beta}={\tilde {\bp}}^{'C(-)}_{\beta}= \psi^{C(-)}_{{\bp}'_\beta;q'_{\beta}}$ denotes the off-shell Coulomb scattering wave function  in the exit channel $\beta$ and for $t_{\beta} \geq N_{\beta}+1,\,\,t_{\alpha} \leq N_{\alpha}$ $\,\,{\rm {\bf y}}_{\beta}={\bp}'_{\beta}$; $\,\,{\bq}^{C(+)}_{\alpha}=\psi^{C(+)}_{{\bq}_\alpha}$ stands for the on-shell Coulomb scattering wave functions in the initial channel $\alpha$. 
Also for the amplitudes under the integral
\begin{align}
&{\tilde X}^{SC\,\tau\rho}_{\gamma \,\zeta_{\gamma},\alpha \,\zeta_{\alpha}}({\rm {\bf y}}_{\gamma},{\bq}^{C(+)}_{\alpha};z) \, = \,
<{\rm {\bf y}}_{\gamma} \big|{\tilde X}^{\tau\rho}_{\gamma \,\zeta_{\gamma},\alpha \,\zeta_{\alpha}}(z) \big|\psi^{C(+)}_{{\bq}_\alpha}>,             
\label{XSCint1}
\end{align}
where for $\,t_{\gamma} \leq N_{\gamma},\,\,t_{\alpha} \leq N_{\alpha}$   $\,\,{\rm {\bf y}}_{\gamma}= {\bp}^{C(-)}_{\gamma}= \psi^{C(-)}_{{\bp}_\gamma}$ is the on-shell Coulomb scattering wave function in the intermediate channel $\gamma$ and for $t_{\gamma} \geq N_{\gamma} +1,\,\,t_{\alpha} \leq N_{\alpha}$ $\,\,{\rm {\bf y}}_{\gamma}= {\bp}_{\gamma}$.

The inhomogeneous terms are 
\begin{align}
&{\tilde Z}^{'\,SC\,\sigma\rho}_{\beta\,\zeta_{\beta},\alpha\,\zeta_{\alpha}}({\tilde{\bp}}^{'C(-)}_{\beta},{\bq}^{C(+)}_{\alpha};z)
=<\psi^{C(-)}_{{\bp}'_\beta;\,q_{\beta}} \big|{\tilde Z}^{'\,\sigma\rho}_{\beta \,\zeta_{\beta},\alpha \,\zeta_{\alpha}}(z) \big|\psi^{C(+)}_{{\bq}_\alpha}>,      \qquad\qquad t_{\beta} \leq N_{\beta},\,\,t_{\alpha} \leq N_{\alpha},
\label{ZprSC1}
\end{align}
\begin{align}
&{\tilde Z}^{SC\,\sigma\rho}_{\beta\,\zeta_{\beta},\alpha\,\zeta_{\alpha}}({\bp}'_\beta,{\bq}^{C(+)}_{\alpha};z)
=<{\bp}'_\beta \big|{\tilde Z}^{\sigma\rho}_{\beta \,\zeta_{\beta},\alpha \,\zeta_{\alpha}}(z) \big|\psi^{C(+)}_{{\bq}_\alpha}>,      \qquad\qquad t_{\beta} \geq N_{\beta} +1,\,\,t_{\alpha} \leq N_{\alpha}.
\label{ZprSC2}
\end{align}

The potentials in the integrand are
\begin{align}
&{\tilde Z}^{'\,SC\,\sigma\tau}_{\beta \,\zeta_{\beta},\gamma \,\zeta_{\gamma} }({\tilde{\bp}}^{'C(-)}_{\beta},{\bp}^{ C(-)}_{\gamma};z)
= <\psi^{C(-)}_{{\bp}'_\beta;\,q_{\beta}} \big|{\tilde Z}^{'\,\sigma\tau}_{\beta \,\zeta_{\beta},\gamma \,\zeta_{\gamma}}(z) \big|\psi^{C(-)}_{{\bp}_\gamma}>,  \qquad\qquad t_{\beta} \leq N_{\beta},\,\, t_{\gamma} \leq N_{\gamma},
\label{ZintSC1}
\end{align}
\begin{align}
&{Z}^{SC\,\sigma\sigma}_{\beta \,\zeta_{\beta},\gamma \,\zeta_{\gamma}}({\tilde{\bp}}^{'C(-)}_{\beta},{\bp}_\gamma;z)
= <\psi^{C(-)}_{{\bp}'_\beta;\,q_{\beta}} \big|{Z}^{\sigma\sigma}_{\beta \,\zeta_{\beta},\gamma \,\zeta_{\gamma}}(z) \big|{\bp}_\gamma>,  \qquad\qquad t_{\beta} \leq N_{\beta},\,\, t_{\gamma} \geq N_{\gamma} +1,
\label{ZintSC2}
\end{align}
\begin{align}
&{\tilde Z}^{SC\,\sigma\tau}_{\beta \,\zeta_{\beta},\gamma \,\zeta_{\gamma}}({\bp}'_\beta,{\bp}^{C (-)}_{\gamma};z)
= <{\bp}'_\beta \big|{\tilde Z}^{\sigma\tau}_{\beta \,\zeta_{\beta},\gamma \,\zeta_{\gamma}}(z) \big|\psi^{C(-)}_{{\bp}_\gamma}>,  \qquad\qquad t_{\beta} \geq N_{\beta} +1,\,\, t_{\gamma} \leq N_{\gamma},
\label{ZintSC3}
\end{align}
\begin{align}
&{Z}^{\sigma\sigma}_{\beta \,\zeta_{\beta},\gamma \,\zeta_{\gamma}}({\bp}'_\beta,{\bp}_\gamma;z)
= <{\bp}'_\beta \big|{Z}^{\sigma\sigma}_{\beta \,\zeta_{\beta},\gamma \,\zeta_{\gamma}}(z) \big|{\bp}_\gamma>,  \qquad\qquad t_{\beta} \geq N_{\beta} +1,\,\, t_{\gamma} \geq N_{\gamma} + 1.
\label{ZintSC4}
\end{align}
In each channel, for which two-body state is not a bound state, the Coulomb scattering wave function (off-shell or on-shell) should be replaced by the corresponding plane wave.

The off-shell Coulomb scattering wave function is given by
\begin{align}
|\psi^{C(\pm)}_{{\bp}_\alpha;q_{\alpha}}>\,=\, \big[1+ G_{0}(q_{\alpha}^{2}/(2\,M_{\alpha}) \big]\,T_{\alpha}^{C}(q_{\alpha}^{2}/(2\,M_{\alpha}))|{\bp}_{\alpha}>, \qquad\,\,p_{\alpha} \not= q_{\alpha},
\label{Couloffshscwf1}
\end{align}
\begin{align}
G_{0}(q_{\alpha}^{2}/(2\,M_{\alpha}))= \frac{1}{q_{\alpha}^{2}/(2\,M_{\alpha}) - {\bQ}_{\alpha}^{2}/(2\,M_{\alpha}) + i\,0}
\label{G0alpha1}
\end{align}
is the free Green function describing the propagation of the system of the noninteracting particle 
$\alpha$ and the system $(\beta\,\gamma)$, $\,{\bQ}_{\alpha}^{2}/(2\,M_{\alpha})$ is the kinetic energy operator of their relative motion. Also $T_{\alpha}^{C}({\bp}'_{\alpha},\,{\bp}_{\alpha};\,q_{\alpha}^{2}/(2\,M_{\alpha}))$, $\,\,p'_{\alpha},\,p_{\alpha} \not= q_{\alpha},\,$ is the two-body off-shell Coulomb scattering amplitude of particle $\alpha$ and the {\cm} of the system $(\beta\,\gamma)$ moving with the relative kinetic energy $q_{\alpha}^{2}/(2\,M_{\alpha})$ and interacting via the Coulomb potential $U_{\alpha}^{C}(\rho_{\alpha})$.  

The on-shell Coulomb scattering wave function can be obtained from it by taking the limit 
$p_{\alpha} \to q_{\alpha}$. In this limit we get \cite{vanHaeringen} 
\begin{align}
\lim\limits_{p_{\alpha} \to q_{\alpha}}\,<\psi^{C(-)}_{{\bp}_\alpha;\,q_{\alpha}} 
|\,\Omega(p_{\alpha},\,q_{\alpha})= <\psi^{C(-)}_{{\bq}_\alpha}|,
\label{limpsionsh2}
\end{align}
where 
\begin{align}
\Omega(p_{\alpha},\,q_{\alpha}) = e^{\pi\,\eta_{\alpha}/2}\,[\Gamma(1-i\,\eta_{\alpha})]^{-1}\,\Big(\frac{p_{\alpha} + q_{\alpha}}{p_{\alpha} - q_{\alpha} -i\,0}\Big)^{i\,\eta_{\alpha}},
\label{Omega1}
\end{align}
$\eta_{\alpha}= Z_{\alpha}\,Z_{\beta\gamma}\,e^{2}\,M_{\alpha}/q_{\alpha}$ is the Coulomb parameter in the channel $\alpha$, which characterizes the Coulomb interaction between particle $\alpha$ and the system $\beta + \gamma$ moving with the relative momentum $q_{\alpha}$, $Z_{\beta\gamma}= Z_{\beta} + Z_{\gamma}$. 

We return now to Eqs (\ref{AGSCS1}). These are not integral equations yet and we will address this point.  Let us consider the physical amplitudes (describing the transition 
${\rm bound\,\,state \to bound\,\,state}$).  On the left-hand side of Eq. (\ref{AGSCS1}) 
we have the amplitude ${\tilde X}^{SC\,\sigma\rho}_{\beta\,\zeta_{\beta},\alpha\,\zeta_{\alpha}}({\tilde{\bp}}^{'C(-)}_{\beta},{\bq}^{C(+)}_{\alpha};z)$ given by Eq. (\ref{XSClhs1}), while in the integrand we have different half-off-shell amplitude ${\tilde X}^{SC\,\tau\rho}_{\gamma \,\zeta_{\gamma},\alpha \,\zeta_{\alpha}}({\bp}^{C(-)}_{\gamma},{\bq}^{C(+)}_{\alpha};z)$ given by Eq. (\ref{XSCint1}). Both amplitudes are half-off-shell but the off-shell effects are treated differently in both amplitudes. In ${\tilde X}^{SC\,\sigma\rho}_{\beta\,\zeta_{\beta},\alpha\,\zeta_{\alpha}}({\tilde{\bp}}^{'C(-)}_{\beta},{\bq}^{C(+)}_{\alpha};z)$ in the bra state we have 
the off-shell Coulomb scattering wave function, while in ${\tilde X}^{SC\,\tau\rho}_{\gamma \,\zeta_{\gamma},\alpha \,\zeta_{\alpha}}({\bp}^{C(-)}_{\gamma},{\bq}^{C(+)}_{\alpha};z)$ in the bra state we have the on-shell Coulomb scattering wave function but with momentum ${\bp}_{\gamma}$, which is the integration variable and, hence, $p_{\gamma} \not= q_{\gamma}$, where $q_{\gamma}$ is the on-shell momentum in the channel $\gamma$. Hence the amplitudes in the left-hand side and in the integrand of Eqs. (\ref{AGSCS1}) are not the same functions and these equations cannot be solved as integral equations. 

Let us take the on-shell limit $p'_{\beta} \to q'_{\beta}$ in Eqs (\ref{AGSCS1}). 
Taking into account the on-shell limit of the off-shell scattering wave function, see 
Eq. (\ref{limpsionsh2}), we get for the reaction amplitude and effective potentials:
\begin{align}
\lim\limits_{p'_{\beta} \to q'_{\beta}}\,{\tilde X}^{SC\,\sigma\rho}_{\beta\,\zeta_{\beta},\alpha\,\zeta_{\alpha}}({\tilde{\bp}}^{'C(-)}_{\beta},{\bq}^{C(+)}_{\alpha};z)= [\Omega(p'_{\beta},\,q'_{\beta})]^{-1}\,{\tilde X}^{SC\,\sigma\rho}_{\beta\,\zeta_{\beta},\alpha\,\zeta_{\alpha}}({\bq}^{'C(-)}_{\beta},{\bq}^{C(+)}_{\alpha};z), 
\label{onshXba1}
\end{align}
\begin{align}
\lim\limits_{p'_{\beta} \to q_{\beta}}\,{\tilde Z}^{'SC\,\sigma\rho}_{\beta\,\zeta_{\beta},\alpha\,\zeta_{\alpha}}({\tilde{\bp}}^{'C(-)}_{\beta},{\bq}^{C(+)}_{\alpha};z)= [\Omega(p'_{\beta},\,q'_{\beta})]^{-1}\,{\tilde Z}^{'SC\,\sigma\rho}_{\beta\,\zeta_{\beta},\alpha\,\zeta_{\alpha}}({\bq}^{'C(-)}_{\beta},{\bq}^{C(+)}_{\alpha};z), 
\label{onshZba1}
\end{align}
\begin{align}
\lim\limits_{p'_{\beta} \to q'_{\beta}}\,{\tilde Z}^{'SC\,\sigma\tau}_{\beta\,\zeta_{\beta},\gamma\,\zeta_{\gamma}}({\tilde{\bp}}^{'C(-)}_{\beta},{\bp}^{C(-)}_{\gamma
};z)= [\Omega(p'_{\beta},\,q'_{\beta})]^{-1}\,{\tilde Z}^{'SC\,\sigma\rho}_{\beta\,\zeta_{\beta},\gamma\,\zeta_{\gamma}}({\bq}^{'C(-)}_{\beta},{\bp}^{C(-)}_{\gamma};z)
\label{onshXint1}
\end{align}
and
\begin{align}
\lim\limits_{p'_{\beta} \to q'_{\beta}}\,{\tilde Z}^{SC\,\sigma\tau}_{\beta\,\zeta_{\beta},\gamma\,\zeta_{\gamma}}({\tilde{\bp}}^{'C(-)}_{\beta},{\bp}_{\gamma};z)= [\Omega(p'_{\beta},\,q'_{\beta})]^{-1}\,{\tilde Z}^{SC\,\sigma\rho}_{\beta\,\zeta_{\beta},\gamma\,\zeta_{\gamma}}({\bq}^{'C(-)}_{\beta},{\bp}_{\gamma};z).
\label{onshXint1}
\end{align}

Thus taking limit $p'_{\beta} \to q'_{\beta}$ in Eqs (\ref{AGSCS1}) and multiplying them  by
$[\Omega(p'_{\beta},\,q'_{\beta})]$ we obtain the on-shell limit 
\begin{align}
&{\tilde X}^{SC\,\sigma\rho}_{\beta \,\zeta_{\beta},\alpha \,\zeta_{\alpha}}({\bq}^{'C(-)}_{\beta},\,{\bq}^{C(+)}_{\alpha };z)=
{\tilde Z}^{'SC\,\sigma\rho}_{\beta \,\zeta_{\beta},\alpha \,\zeta_{\alpha}}({\bq}^{'C(-)}_{\beta},\,{\bq}^{C(+)}_{\alpha};z)                          \nonumber\\
&+ \sum_{\gamma}\,\sum\limits_{n_{\gamma}}\,\sum_{t_{\gamma}}^{N_{\gamma}}\,\sum^{N}_{\tau=1}\int\, \frac{{\rm d}{\bp}_\gamma}{(2\,\pi)^{3}}\,
{\tilde Z}^{'\,SC\,\sigma\tau}_{\beta \,\zeta_{\beta},\gamma \,\zeta_{\gamma}}({\bq}^{'C(-)}_{\beta},\,{\bp}^{C(-)}_{\gamma};z)\,
\frac{1}{z - {\hat E}_{\gamma \,n_{\gamma}} - {p_{\gamma}}^{2}/(2\,M_{\gamma})}\,
{\tilde X}^{SC\,\tau\rho}_{\gamma \,\zeta_{\gamma},\alpha \,\zeta_{\alpha}}({\bp}^{C(-)}_{\gamma},\,{\bq}^{C(+)}_{\alpha};z)                                                                              \nonumber\\
&+ \sum_{\gamma}\,\sum\limits_{n_{\gamma},n'_{\gamma}}\,\sum_{t_{\gamma},t'_{\gamma}=N_{\gamma}+1}^{A_{\gamma}}\,\sum^{N}_{\nu=1}\int \frac{ {\rm d}{\bp}_\gamma}{(2\,\pi)^{3}}\,
{Z}^{SC\,\sigma\sigma}_{\beta \,\zeta_{\beta},\gamma \,\zeta_{\gamma}}({\bq}^{'C(-)}_{\beta},\,{\bp}_\gamma;z)\,\Delta^{\sigma\nu}_{\gamma, \,\zeta_{\gamma}\, \zeta'_{\gamma}}(z - p_{\gamma}^{2}/(2\,M_{\gamma}))\,{\tilde X}^{SC\,\nu\rho}_{\gamma \zeta'_{\gamma},\alpha \,\zeta_{\alpha}}({\bp}_\gamma,{\bq}^{C(+)}_{\alpha};z),                           \nonumber\\
&\qquad\qquad t_{\beta} \leq N_{\beta},\,\,t_{\alpha} \leq N_{\alpha}.
\label{AGSCS11}
\end{align}
The on-shell limit of Eqs. (\ref{AGSCSaux2}) is straightforward because the bra state 
${\bp}'_{\beta}$ is the plane wave and we get
\begin{align}
&{\tilde X}^{SC\,\sigma\rho}_{\beta\,\zeta_{\beta},\alpha\,\zeta_{\alpha}}({\bq}'_{\beta},\,{\bq}^{C(+)}_{\alpha};z)=
{\tilde Z}^{SC\,\sigma\rho}_{\beta\,\zeta_{\beta},\alpha\,\zeta_{\alpha}}({\bq}'_{\beta},{\bq}^{C(+)}_{\alpha};z)                                                                    \nonumber\\
&+ \sum_{\gamma}\,\sum\limits_{n_{\gamma}}\,\sum_{t_{\gamma}=1}^{N_{\gamma}}\,\sum^{N}_{\tau=1}\int\, \frac{{\rm d}{\bp}_\gamma}{(2\,\pi)^{3}}\,
{\tilde Z}^{SC\,\sigma\tau}_{\beta n_{\beta},\gamma n_{\gamma}}({\bq}'_{\beta},{\bp}^{C(-)}_{\gamma};z)\,
\frac{1}{z - {\hat E}_{\gamma\, n_{\gamma}} - {p_{\gamma}}^{2}/(2\,M_{\gamma})}\,
{\tilde X}^{SC\,\tau\rho}_{\gamma \,\zeta_{\gamma},\alpha \,\zeta_{\alpha}}({\bp}^{C (-)}_{\gamma},{\bq}^{C(+)}_{\alpha};z)                                                                                        \nonumber\\    
&+ \sum_{\gamma}\,\sum\limits_{n_{\gamma},n'_{\gamma}}\,\sum_{t_{\gamma},t'_{\gamma}=N_{\gamma}+1}^{A_{\gamma}}\,\sum^{N}_{\nu=1}\int \frac{ {\rm d}{\bp}_\gamma}{(2\,\pi)^{3}}\,
{Z}^{\sigma\sigma}_{\beta \,\zeta_{\beta},\gamma \,\zeta_{\gamma}}({\bq}'_\beta,{\bp}_\gamma;z)\,\Delta^{\sigma\nu}_{\gamma, \,\zeta_{\gamma}\, \zeta'_{\gamma}}(z - p_{\gamma}^{2}/(2\,M_{\gamma}))\,{\tilde X}^{SC\,\nu\rho}_{\gamma \zeta'_{\gamma},\alpha \,\zeta_{\alpha}}({\bp}_\gamma,\,{\bq}^{C(+)}_{\alpha};z),     \nonumber\\
&\qquad\qquad t_{\beta} \geq N_{\beta} + 1, \, t_{\alpha} \leq N_{\alpha},
\label{AGSCSaux3}
\end{align}
where $\zeta_{\gamma} = \{n_{\gamma},\,t_{\gamma} \},\,\,\,\zeta'_{\gamma} = \{n'_{\gamma},\,t'_{\gamma} \}$.
Evidently that Eqs (\ref{AGSCS11}) are not integral equations because on the left-hand side we have  the  on-shell transition amplitudes while in the integrand they are half-off-shell. To obtain an integral equations from (\ref{AGSCS11}) we use its off-shell extension, which differs from the one used in (\ref{AGSCS1}):
\begin{align}
&{\tilde X}^{SC\,\sigma\rho}_{\beta\,\zeta_{\beta},\alpha\,\zeta_{\alpha}}({\bp}^{'C(-)}_{\beta},\,{\bq}^{C(+)}_{\alpha};z)=
{\tilde Z}^{'\,SC\,\sigma\rho}_{\beta\,\zeta_{\beta},\alpha\,\zeta_{\alpha}}({\bp}^{'C(-)}_{\beta},{\bq}^{C(+)}_{\alpha};z)                                 \nonumber\\
&+ \sum_{\gamma}\,\sum\limits_{n_{\gamma}}\,\sum_{t_{\gamma}=1}^{N_{\gamma}}\,\sum^{N}_{\tau=1}\int\, \frac{{\rm d}{\bp}_\gamma}{(2\,\pi)^{3}}\,
{\tilde Z}^{'\,SC\,\sigma\tau}_{\beta \,\zeta_{\beta},\gamma \,\zeta_{\gamma}}({\bp}^{'C(-)}_{\beta},{\bp}^{C (-)}_{\gamma};z)\,
\frac{1}{z - {\hat E}_{\gamma n_{\gamma}} - {p_{\gamma}}^{2}/(2\,M_{\gamma})}\,\,
{\tilde X}^{SC\,\tau\rho}_{\gamma \,\zeta_{\gamma},\alpha \,\zeta_{\alpha}}({\bp}^{C (-)}_{\gamma},{\bq}^{C(+)}_{\alpha};z)                                                                   \nonumber\\      &+ \sum_{\gamma}\,\sum\limits_{n_{\gamma},n'_{\gamma}}\,\sum_{t_{\gamma},t'_{\gamma}=N_{\gamma}+1}^{A_{\gamma}}\,\sum^{N}_{\nu=1}\int \frac{ {\rm d}{\bp}_\gamma}{(2\,\pi)^{3}}\,
{Z}^{SC\,\sigma\sigma}_{ \beta \,\zeta_{\beta},\gamma \,\zeta_{\gamma} }({\bp}^{'C(-)}_{\beta},\,{\bp}_\gamma;z)\,\Delta^{\sigma\nu}_{\gamma, \,\zeta_{\gamma}\, \zeta'_{\gamma}}({\hat z}_{\gamma}))\,{\tilde X}^{SC\,\nu\rho}_{\gamma \zeta'_{\gamma},\alpha \,\zeta_{\alpha}}({\bp}_\gamma,{\bq}^{C(+)}_{\alpha};z),     \nonumber\\
&\qquad\qquad t_{\beta} \leq  N_{\beta}, \,t_{\alpha} \leq N_{\alpha},
\label{newAGSCS1}
\end{align}  
\begin{align}
&{\tilde X}^{SC\,\sigma\rho}_{\beta\,\zeta_{\beta},\alpha\,\zeta_{\alpha}}({\bp}'_{\beta},\,{\bq}^{C(+)}_{\alpha};z)=
{\tilde Z}^{'\,SC\,\sigma\rho}_{\beta\,\zeta_{\beta},\alpha\,\zeta_{\alpha}}({\bp}'_{\beta},{\bq}^{C(+)}_{\alpha};z)       \nonumber\\
&+ \sum_{\gamma}\,\sum\limits_{n_{\gamma}}\,\sum_{t_{\gamma}=1}^{N_{\gamma}}\,\sum^{N}_{\tau=1}\int\, \frac{{\rm d}{\bp}_\gamma}{(2\,\pi)^{3}}\,
{\tilde Z}^{'\,SC\,\sigma\tau}_{\beta \,\zeta_{\beta},\gamma \,\zeta_{\gamma}}({\bp}'_{\beta},{\bp}^{C (-)}_{\gamma};z)\,
\frac{1}{z - {\hat E}_{\gamma n_{\gamma}} - {p_{\gamma}}^{2}/(2\,M_{\gamma})}{\tilde X}^{SC\,\tau\rho}_{\gamma \,\zeta_{\gamma},\alpha \,\zeta_{\alpha}}({\bp}^{C (-)}_{\gamma},{\bq}^{C(+)}_{\alpha};z)         \nonumber\\          
&+ \sum_{\gamma}\,\sum\limits_{n_{\gamma},n'_{\gamma}}\,\sum_{t_{\gamma},t'_{\gamma}=N_{\gamma}+1}^{A_{\gamma}}\,\sum^{N}_{\nu=1}\int \frac{ {\rm d}{\bp}_\gamma}{(2\,\pi)^{3}}\,
{Z}^{\sigma\sigma}_{\beta \,\zeta_{\beta},\gamma \,\zeta_{\gamma}}({\bp}'_\beta,{\bp}_\gamma;z)\,\Delta^{\sigma\nu}_{\gamma, \,\zeta_{\gamma}\, \zeta'_{\gamma}}({\hat z}_{\gamma}))\,{\tilde X}^{SC\,\nu\rho}_{\gamma \zeta'_{
\gamma},\alpha \,\zeta_{\alpha}}({\bp}_\gamma,\,{\bq}^{C(+)}_{\alpha};z),     \nonumber\\
&\qquad\qquad t_{\beta} \geq N_{\beta} + 1, \, t_{\alpha} \leq N_{\alpha},
\label{AGSCSaux12}
\end{align}
where ${\hat z}_{\gamma}= z - p_{\gamma}^{2}/(2\,M_{\gamma})$. 
Eqs (\ref{newAGSCS1}) are half-off-shell because in the exit channel $\beta$ the on-shell Coulomb scattering wave function $<{\bp}^{'\,C(-)}_{\beta}| \equiv <\psi_{{\bp}'_{\beta}}^{C(-)}|\,$ is present with momentum $p_{\beta}' \not= q'_{\beta}$. It means that in Eqs (\ref{newAGSCS1}) the off-shell behavior in the exit channel differs from the one in Eqs (\ref{AGSCS1}), where the off-shell Coulomb scattering wave function is used rather than the on-shell one but with the off-shell momentum. But  the on-shell limit of (\ref{newAGSCS1}) coincides with the renormalized on-shell limit of (\ref{AGSCS1}).  
Eqs (\ref{newAGSCS1}) together with (\ref{AGSCSaux12}) are our final equations in the Coulomb distorted wave representation, which will be used to calculate the deuteron stripping amplitudes and cross sections. The advantage of these equations is that one don't need to use Coulomb screening procedure, application of which becomes problematic when the charge of the target increases.
 
\section{Angular momentum decomposition of generalized AGS equations in the Coulomb distorted wave representation}
\label{angmomentdecomp}
Here we present the final expressions for the modified AGS equations (\ref{newAGSCS1}) and (\ref{AGSCSaux12}) after the angular momentum decomposition. We follow the formalism used in \cite{alt2002}. We use the following angular momentum coupling scheme for a given channel $\alpha$: ${\rm {\bf s}}_{\beta} + {\rm {\bf s}}_{\gamma}= {\rm {\bf S}}_{\alpha}$, $\,\,{\rm {\bf L}}_{\alpha} + {\rm {\bf S}}_{\alpha}= {\rm {\bf J}}_{\alpha}$,  $\,\,
{\rm {\bf s}}_{\alpha} + {\rm {\bf J}}_{\alpha}= {\rm {\bf \Sigma}}_{\alpha}$, $\,\,{\rm {\bf l}}_{\alpha} + {\rm {\bf \Sigma}}_{\alpha}= {\rm {\bf J}}$. Here, ${\rm {\bf s}}_{\alpha}$ denotes the spin of particle $\alpha$, $\,{\rm {\bf L}}_{\alpha}$ is the relative orbital angular momentum, ${\rm {\bf S}}_{\alpha}$ is the total spin, and ${\rm {\bf J}}_{\alpha}$ is the total angular momentum of particles $\beta$ and $\gamma$; moreover, ${\rm {\bf l}}_{\alpha}$ denotes the relative orbital angular momentum of particle $\alpha$ and the pair $(\beta\gamma)$, and finally ${\rm {\bf J}}$ is the total angular momentum of the three-body system. 
Also $A_{\alpha} \equiv A^{J_{\alpha}S_{\alpha}}$ is the rank of the separable expansion in the two-body channel $\alpha$ with fixed $J_{\alpha}$ and $S_{\alpha}$; $\,\,N_{\alpha} \equiv N^{J_{\alpha}S_{\alpha}}\,\,(N^{J_{\alpha}S_{\alpha}} \leq A^{J_{\alpha}S_{\alpha}})\,$ is the rank of the separable expansion corresponding to the bound states in the pair $\alpha$ for given quantum numbers $J_{\alpha}$ and $S_{\alpha}$; as before $t_{\alpha}= 1...A^{J_{\alpha}S_{\alpha}}$ enumerates the number of the expansion term of the separable potential in channel $\alpha$ with given $J_{\alpha}$ and $S_{\alpha}$.
As in the previous sections, $n_{\alpha}= \{L_{\alpha},\,S_{\alpha},\,J_{\alpha}\}$ collectively denotes the quantum numbers of the pair $\alpha$, while $\zeta_{\alpha}= \{n_{\alpha},\,t_{\alpha}\} \equiv \{L_{\alpha},\,S_{\alpha},\,J_{\alpha},\,t_{\alpha} \}$ denotes the complete set of quantum numbers characterizing the two-body state $\alpha$, which, in addition to $n_{\alpha}$, includes the number of the separable expansion. A new variable $u_{\alpha}=\{n_{\alpha},\,l_{\alpha},\Sigma_{\alpha} \} \,\,$ collectively denotes all the introduced above spin-angular momentum variables in the channel $\alpha$ except for $t_{\alpha}$. 
As in \cite{alt2007} we don't use the isospin formalism to treat two nucleons.

From Eqs (\ref{newAGSCS1}) we get 
\begin{align}
&{\tilde X}^{SC\,J^{\pi}\,\sigma\rho}_{ u_{\beta}\,t_{\beta},\,u_{\alpha}\,t_{\alpha}}({p}^{'C}_{\beta\,l_{\beta}},\,{q}^{C}_{\alpha\,l_{\alpha}};z)=
{\tilde Z}^{'SC\,J^{\pi}\,\sigma\rho}_{u_{\beta}\,t_{\beta},\,u_{\alpha}\,t_{\alpha}}({p}^{'C}_{\beta\,l_{\beta}},\,{q}^{C}_{\alpha\,l_{\alpha}};z) +
\sum_{\gamma}\,\sum\limits_{u_{\gamma}}\,\sum_{t_{\gamma}=1}^{N_{\gamma}}\,\sum^{N}_{\tau=1}\int\limits_{0}^{\infty}\,\, \frac{{\rm d}{p}_{\gamma}\,p_{\gamma}^{2}}{2\,\pi^{2}}\,
{\tilde Z}^{'\,SC\,J^{\pi}\,\sigma\tau}_{u_{\beta}\,t_{\beta},\,u_{\gamma}\,t_{\gamma}}({p}^{'C}_{\beta\,l_{\beta}},\,{p}^{C}_{\gamma\,l_{\gamma}};z)  \nonumber\\
&\times \frac{1}{z - {\hat E}_{\gamma n_{\gamma}} - {p_{\gamma}}^{2}/(2\,M_{\gamma})}\,{\tilde X}^{SC\,J^{\pi}\,\tau\rho}_{ u_{\gamma}\,t_{\gamma},\,u_{\alpha}\,t_{\alpha}}({p}^{C}_{\gamma\,l_{\gamma}},\,{q}^{C}_{\alpha\,l_{\alpha}};z)       \nonumber\\            
&+ \sum_{\gamma}\,\sum\limits_{u_{\gamma},\,u'_{\gamma}}\,\sum_{t_{\gamma},t'_{\gamma}=N_{\gamma}+1}^{A_{\gamma}}\,\sum^{N}_{\nu=1}\int\limits_{0}^{\infty}\, \frac{{\rm d}{p}_{\gamma}\,p_{\gamma}^{2}}{2\,\pi^{2}}\,
{Z}^{SC\,J^{\pi}\,\sigma\sigma}_{u_{\beta}\,t_{\beta},\,u_{\gamma}\,t_{\gamma}}({p}^{'C}_{\beta\,l_{\beta}},\,{p}_{\gamma\,l_{\gamma}};z),\Delta^{J_{\gamma}S_{\gamma}\,\sigma\nu}_{L_{\gamma}\,t_{\gamma}\,L'_{\gamma}\,t'_{\gamma}}({\hat z}_{\gamma})\,{\tilde X}^{SC\,J^{\pi}\,\nu\rho}_{ u'_{\gamma}\,t'_{\gamma},\,u_{\alpha}\,t_{\alpha}}({p}_{\gamma\,l_{\gamma}},\,{q}^{C}_{\alpha\,l_{\alpha}};z),     \nonumber\\
&\qquad\qquad t_{\beta} \leq N_{\beta}, \, t_{\alpha} \leq N_{\alpha}.
\label{newAGSCS21}
\end{align}  
Here $n_{\gamma}= \{S_{\gamma},L_{\gamma},J_{\gamma} \}$ and $n'_{\gamma}= \{S_{\gamma},L'_{\gamma},J_{\gamma} \}$, $\,\,u_{\gamma}= \{n_{\gamma},l_{\gamma},\Sigma_{\gamma} \}$ and  $u'_{\gamma}= \{n'_{\gamma},l_{\gamma},\Sigma_{\gamma} \}$. Note that in the scattering of particles of the pair $\alpha$  the total momentum $J_{\gamma}$ is conserved. That is why $\Delta^{J_{\gamma}S_{\gamma}\,\tau\nu}_{L_{\gamma}\,t_{\gamma}\,L'_{\gamma}\,t'_{\gamma}}({\hat z}_{\gamma})$ is diagonal over $J_{\gamma}$.  

Also Eq. (\ref{AGSCSaux12}) leads to 
\begin{align}
&{\tilde X}^{SC\,J^{\pi}\,\sigma\rho}_{ u_{\beta}\,t_{\beta},\,u_{\alpha}\,t_{\alpha}}({p}^{'}_{\beta},\,{q}^{C}_{\alpha\,l_{\alpha}};z)=
{\tilde Z}^{SC\,J^{\pi}\,\sigma\rho}_{u_{\beta}\,t_{\beta},\,u_{\alpha}\,t_{\alpha}}({p}^{'}_{\beta},\,{q}^{C}_{\alpha\,l_{\alpha}};z) +
\sum_{\gamma}\,\sum\limits_{u_{\gamma}}\,\sum_{t_{\gamma}=1}^{N_{\gamma}}\,\sum^{N}_{\tau=1}\int\limits_{0}^{\infty}\,\, \frac{{\rm d}{p}_{\gamma}\,p_{\gamma}^{2}}{2\,\pi^{2}}\,
{\tilde Z}^{SC\,J^{\pi}\,\sigma\tau}_{u_{\beta}\,t_{\beta},\,u_{\gamma}\,t_{\gamma}}({p}^{'}_{\beta},\,{p}^{C}_{\gamma\,l_{\gamma}};z)  \nonumber\\
&\times \frac{1}{z - {\hat E}_{\gamma n_{\gamma}} - {p_{\gamma}}^{2}/(2\,M_{\gamma})}\,{\tilde X}^{SC\,J^{\pi}\,\tau\rho}_{ u_{\gamma}\,t_{\gamma},\,u_{\alpha}\,t_{\alpha}}({p}^{C}_{\gamma\,l_{\gamma}},\,{q}^{C}_{\alpha\,l_{\alpha}};z)       \nonumber\\            
&+ \sum_{\gamma}\,\sum\limits_{u_{\gamma},u'_{\gamma}}\,\sum_{t_{\gamma},t'_{\gamma}=N_{\gamma}+1}^{A_{\gamma}}\,\sum^{N}_{\nu=1}\int\limits_{0}^{\infty}\, \frac{{\rm d}{p}_{\gamma}\,p_{\gamma}^{2}}{2\,\pi^{2}}\,
{Z}^{J^{\pi}\,\sigma\sigma}_{u_{\beta}\,t_{\beta},\,u_{\gamma}\,t_{\gamma}}({p}^{'}_{\beta},\,{p}_{\gamma};z),\Delta^{J_{\gamma}S_{\gamma}\,\sigma\nu}_{L_{\gamma}\,t_{\gamma}\,L'_{\gamma}\,t'_{\gamma}}({\hat z}_{\gamma})\,{\tilde X}^{SC\,J^{\pi}\,\nu\rho}_{u'_{\gamma}\,t'_{\gamma},\,u_{\alpha}\,t_{\alpha}}({p}_{\gamma},\,{q}^{C}_{\alpha\,l_{\alpha}};z),     \nonumber\\
&\qquad\qquad t_{\beta} \geq N_{\beta} +1, \, t_{\alpha} \leq N_{\alpha}.
\label{newAGSCSaux21}
\end{align}  
$\Delta^{J_{\gamma}S_{\gamma}\,\sigma\nu}_{L_{\gamma}\,t_{\gamma}\,L'_{\gamma}\,t'_{\gamma}}({\hat z}_{\gamma})$ satisfies the system of equations
\begin{align}
\Delta^{J_{\gamma}S_{\gamma}\,\tau\nu}_{L_{\gamma}\,t_{\gamma}\,L'_{\gamma}\,t'_{\gamma}}({\hat z}_{\gamma})=
\lambda_{L_{\gamma}\,t_{\gamma}\,L'_{\gamma}\,t'_{\gamma}}^{J_{\gamma}S_{\gamma}\,\tau\nu}  +  \sum\limits_{L''_{\gamma}}\,\sum\limits_{t''_{\gamma},t'''_{\gamma}=1}^{ N^{J_{\gamma}S_{\gamma}}}\,\sum\limits_{\omega=1}^{N}\,\lambda_{L_{\gamma}\,t_{\gamma}\,
L''_{\gamma}\,t''_{\gamma}}^{\tau\omega}\,<\chi^{J_{\gamma}S_{\gamma}\,\omega}_{L''_{\gamma}t''_{\gamma}}|\frac{1}{{\hat z}_{\gamma} - \epsilon^{\omega}- K_{\gamma}^{2}/(2\,\mu_{\gamma})}|
g^{J_{\gamma}S_{\gamma}\,\omega}_{L''_{\gamma}\,t'''_{\gamma}}>\,\Delta^{J_{\gamma}S_{\gamma}\,\omega\nu}_{L''_{\gamma}\,t'''_{\gamma}\,L'_{\gamma}\,t'_{\gamma}}({\hat z}_{\gamma}),
\label{Deltapartial1}
\end{align}
where ${\hat z}_{\gamma}= z - p_{\gamma}^{2}/(2\,M_{\gamma})$.

The Coulomb-modified form factor is given by
\begin{align}
g^{J_{\gamma}S_{\gamma}\,\sigma}_{L_{\gamma}\,t_{\gamma}}(k_{\gamma}) = \chi^{J_{\gamma}S_{\gamma}\,\sigma}_{L_{\gamma}\,t_{\gamma}}(k_{\gamma})  +  \delta_{\gamma 3}\,\frac{1}{2\,\pi^{2}}\,\int\limits_{0}^{\infty}\,{\rm d}k'_{{\gamma}}\,{k'_{{\gamma}}}^{2}\,
\frac{T^{C}_{{\gamma}\,L_{{\gamma}}}(k_{{\gamma}},k'_{{\gamma}};{\hat z}_{{\gamma}})\,\chi^{J_{\gamma}S_{\gamma}\,\sigma}_{L_{\gamma}\,t_{\gamma}}(k'_{{\gamma}})}{{\hat z}_{{\gamma}} - {k'_{{\gamma}}}^{2}/(2\,\mu_{{\gamma}})},
\label{gCoulmodif1}
\end{align}
where
\begin{align}
T_{\gamma\,L_{\gamma}}^{C}(k'_{\gamma},k_{\gamma};{\hat z}_{\gamma}) = \frac{1}{2}\,\int\limits_{-1}^{1}\,{\rm d}x\,P_{L_{\gamma}}(x)\,T_{\gamma}^{C}({\rm {\bf k}}'_{\gamma},{\rm {\bf k}}_{\gamma};{\hat z}_{\gamma}),
\label{partprojTCgamma1}
\end{align}
where $x= {\rm {\bf {\hat k}}}'_{\gamma} \cdot {\rm {\bf {\hat k}}}_{\gamma}$, $\,\,{\rm {\bf {\hat k}}}= {\rm {\bf k}}/k$.

The physical reaction amplitude $X_{l_{f} \Sigma_{f} J_{f},l_{i} \Sigma_{i}1}^{SC\,J^{\pi}\sigma\rho}$, which describes the transition from the initial channel $\alpha$, where the particle pair $({\beta \gamma})$ is a deuteron
with $J_{\alpha}=J_{i}=1$, the relative orbital angular momentum of particle $\alpha$ and deuteron is $l_{\alpha}=l_{i}$ and the total channel spin $\Sigma_{\alpha}= \Sigma_{i}$, to the final channel
$\beta$ where the particle pair $(\alpha\gamma)$ is in a bound state $J_{\beta}=J_{f}$ and the channel orbital angular momentum and channel spin $l_{\beta}=l_{f}$ and $\Sigma_{\beta}= \Sigma_{f}$, respectively, can be calculated from the solutions of Eqs (\ref{newAGSCS21}) and (\ref{newAGSCSaux21}) as
\begin{align}
X_{l_{f} \Sigma_{f} J_{f},l_{i} \Sigma_{i}1}^{SC\,J^{\pi}\,\sigma\rho}({p}^{'C}_{\beta\,l_{\beta}},\,{q}^{C}_{\alpha\,l_{\alpha}};E_{+}) = \sum\limits_{u_{\beta}\,t_{\beta}\,u_{\alpha}\,t_{\alpha}}\,\delta_{J_{\beta}J_{f}}\,\delta_{l_{\beta}l_{f}}\,\delta_{\Sigma_{\beta}\Sigma_{f}}\,\delta_{J_{\alpha}1}\delta_{l_{\alpha}l_{i}}\,\delta_{\Sigma_{\alpha}\Sigma_{i}}\,{\tilde X}^{SC\,J^{\pi}\,\sigma\rho}_{ u_{\beta}\,t_{\beta},\,u_{\alpha}\,t_{\alpha}}({p}^{'C}_{\beta\,l_{\beta}},\,{q}^{C}_{\alpha\,l_{\alpha}};E_{+}),
\label{physX1}
\end{align}
$E_{+}= E + i\,0$.
   
Now we proceed to the effective potentials, which are the main input into the AGS equations. Our new defined 
potentials, see Eqs. (\ref{newtildeZ1}), (\ref{Gcrhosigma1}) and (\ref{g01}), are not diagonal over upper indices. We start from generalizing 
Eqs (33) and (34) obtained in \cite{alt2002}, which determine the off-shell effective potentials sandwiched by the plane waves. Here we replace these potentials by the ones in the Coulomb distorted wave representation:
\begin{align}
&{\tilde Z}^{'SC\,J^{\pi}\,\sigma\rho}_{u_{\beta}\,t_{\beta},\,u_{\alpha}\,t_{\alpha}}({p}^{'C}_{\beta\,l_{\beta}},\,{p}^{C}_{\alpha\,l_{\alpha}};z)= \sum\limits_{i=0}^{4}\,{\tilde Z}^{'SC\,J^{\pi}\,\sigma\rho\,(i)}_{u_{\beta}\,t_{\beta},\,u_{\alpha}\,t_{\alpha}}({p}^{'C}_{\beta\,l_{\beta}},\,{p}^{C}_{\alpha\,l_{\alpha}};z), 
\nonumber\\
& t_{\beta} \leq N^{J_{\beta}S_{\beta}},\,\,t_{\alpha} \leq N^{J_{\alpha}S_{\alpha}},
\label{potpartexpZ1}
\end{align}
\begin{align}
&{\tilde Z}^{SC\,J^{\pi}\,\sigma\rho}_{u_{\beta}\,t_{\beta},\,u_{\alpha}\,t_{\alpha}}({p}'_{\beta},\,{p}^{C}_{\alpha\,l_{\alpha}};z)= \sum\limits_{i=0}^{4}\,{\tilde Z}^{SC\,J^{\pi}\,\sigma\rho\,(i)}_{u_{\beta}\,t_{\beta},\,u_{\alpha}\,t_{\alpha}}({p}'_{\beta},\,{p}^{C}_{\alpha\,l_{\alpha}};z), 
\nonumber\\
& t_{\beta} \geq N^{J_{\beta}S_{\beta}} + 1,\,\,t_{\alpha} \leq N^{J_{\alpha}S_{\alpha}},
\label{potpartexpZ11}
\end{align}
and 
\begin{align}
&{Z}^{J^{\pi}\,\sigma\sigma}_{u_{\beta}\,t_{\beta},\,u_{\alpha}\,t_{\alpha}}({p}'_{\beta},\,{p}_{\alpha};z)= \sum\limits_{i=0}^{4}\,{Z}^{J^{\pi}\,\sigma\sigma\,(i)}_{u_{\beta}\,t_{\beta},\,u_{\alpha}\,t_{\alpha}}({p}'_{\beta},\,{p}_{\alpha};z), 
\nonumber\\
& t_{\beta} \geq N^{J_{\beta}S_{\beta}} + 1,\,\,t_{\alpha} \geq N^{J_{\alpha}S_{\alpha}} +1.
\label{potpartexpZ1}
\end{align}
Here
\begin{align}
&{\tilde Z}^{'SC\,J^{\pi}\,\sigma\rho\,(i)}_{u_{\beta}\,t_{\beta},\,u_{\alpha}\,t_{\alpha}}({p}^{'C}_{\beta\,l_{\beta}},\,{p}^{C}_{\alpha\,l_{\alpha}};z) = \int\limits_{0}^{\infty}\frac{{\rm d}p''_{\beta}\,{p''_{\beta}}^{2}}{2\,\pi^{2}}\,\int\limits_{0}^{\infty}\,\frac{{\rm d}p''_{\alpha}\,{p''_{\alpha}}^{2}}{2\,\pi^{2}}\,\psi^{C}_{p'_{\beta}\,l_{\beta}}(p''_{\beta})\,{\tilde Z}^{'J^{\pi}\,\sigma\rho\,(i)}_{u_{\beta}\,t_{\beta},\,u_{\alpha}\,t_{\alpha}}({p}''_{\beta},\,{p}''_{\alpha};z)\,\psi^{C}_{p_{\alpha}\,l_{\alpha}}(p''_{\alpha}),     \nonumber\\ 
& t_{\beta} \leq N^{J_{\beta}S_{\beta}},\,\,t_{\alpha} \leq N^{J_{\alpha}S_{\alpha}},
\label{Zi012}
\end{align}
\begin{align}
&{\tilde Z}^{SC\,J^{\pi}\,\sigma\rho\,(i)}_{u_{\beta}\,t_{\beta},\,u_{\alpha}\,t_{\alpha}}({p}'_{\beta},\,{p}^{C}_{\alpha\,l_{\alpha}};z) = \int\limits_{0}^{\infty}\,\frac{{\rm d}p''_{\alpha}\,{p''_{\alpha}}^{2}}{2\,\pi^{2}}\,{\tilde Z}^{J^{\pi}\,\sigma\rho\,(i)}_{u_{\beta}\,t_{\beta},\,u_{\alpha}\,t_{\alpha}}({p}'_{\beta},\,{p}''_{\alpha};z)\,\psi^{C}_{p_{\alpha}\,l_{\alpha}}(p''_{\alpha}),     \nonumber\\ 
&t_{\beta} \geq N^{J_{\beta}S_{\beta}} +1,\,\,t_{\alpha} \leq N^{J_{\alpha}S_{\alpha}},
\label{Zi0121}
\end{align}
\begin{align}
&{\tilde Z}^{'\,J^{\pi}\,\sigma\rho\,(i)}_{u_{\beta}\,t_{\beta},\,u_{\alpha}\,t_{\alpha}}({p}''_{\beta},\,{p}''_{\alpha};z)= \sum\limits_{\kappa}\,A^{i}_{\kappa}({p}''_{\beta},\,{p}''_{\alpha})\,{\tilde R}^{'\,\sigma\rho\,(i)\,\kappa}_{u_{\beta}\,t_{\beta},\,u_{\alpha}\,t_{\alpha}}({p}''_{\beta},\,{p}''_{\alpha};z),  
\qquad t_{\beta} \leq N^{J_{\beta}S_{\beta}},\,\,\, t_{\alpha}\, \leq N^{J_{\alpha}S_{\alpha}},                                                              
\label{ZdecAR11}
\end{align}
\begin{align}
&{\tilde Z}^{J^{\pi}\,\sigma\rho\,(i)}_{u_{\beta}\,t_{\beta},\,u_{\alpha}\,t_{\alpha}}({p}'_{\beta},\,{p}''_{\alpha};z)= \sum\limits_{\kappa}\,A^{i}_{\kappa}({p}'_{\beta},\,{p}''_{\alpha})\,{\tilde R}^{\sigma\rho\,(i)\,\kappa}_{u_{\beta}\,t_{\beta},\,u_{\alpha}\,t_{\alpha}}({p}'_{\beta},\,{p}''_{\alpha};z),   \qquad t_{\beta} \geq N^{J_{\beta}S_{\beta}} +1,\,\,\, t_{\alpha}\, \leq N^{J_{\alpha}S_{\alpha}},                                                              
\label{Ztilde1}
\end{align}
and
\begin{align}
&Z^{J^{\pi}\,\sigma\sigma\,(i)}_{u_{\beta}\,t_{\beta},\,u_{\alpha}\,t_{\alpha}}({p}'_{\beta},\,{p}_{\alpha};z)= \sum\limits_{\kappa}\,A^{i}_{\kappa}({p}'_{\beta},\,{p}_{\alpha})\,R^{\sigma\sigma\,(i)\,\kappa}_{u_{\beta}\,t_{\beta},\,u_{\alpha}\,t_{\alpha}}({p}'_{\beta},\,{p}_{\alpha};z),       &\qquad t_{\beta} \geq N^{J_{\beta}S_{\beta}} + 1,\,\,\, t_{\alpha}\, \geq N^{J_{\alpha}S_{\alpha}} +1.                                                                
\label{Zreg1}
\end{align}

\begin{figure}
%[tbp] 
\epsfig{file=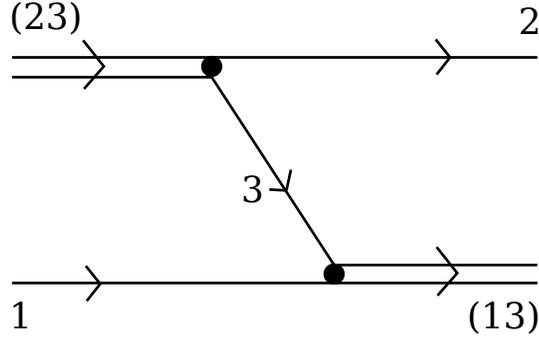,width=7cm}
\caption{
 Pole diagram describing the neutron transfer in the reaction $1 + (23) \to 2 + (13)$.} 
\label{fig_pole01}
\end{figure}

\begin{figure}
%[tbp]
\epsfig{file=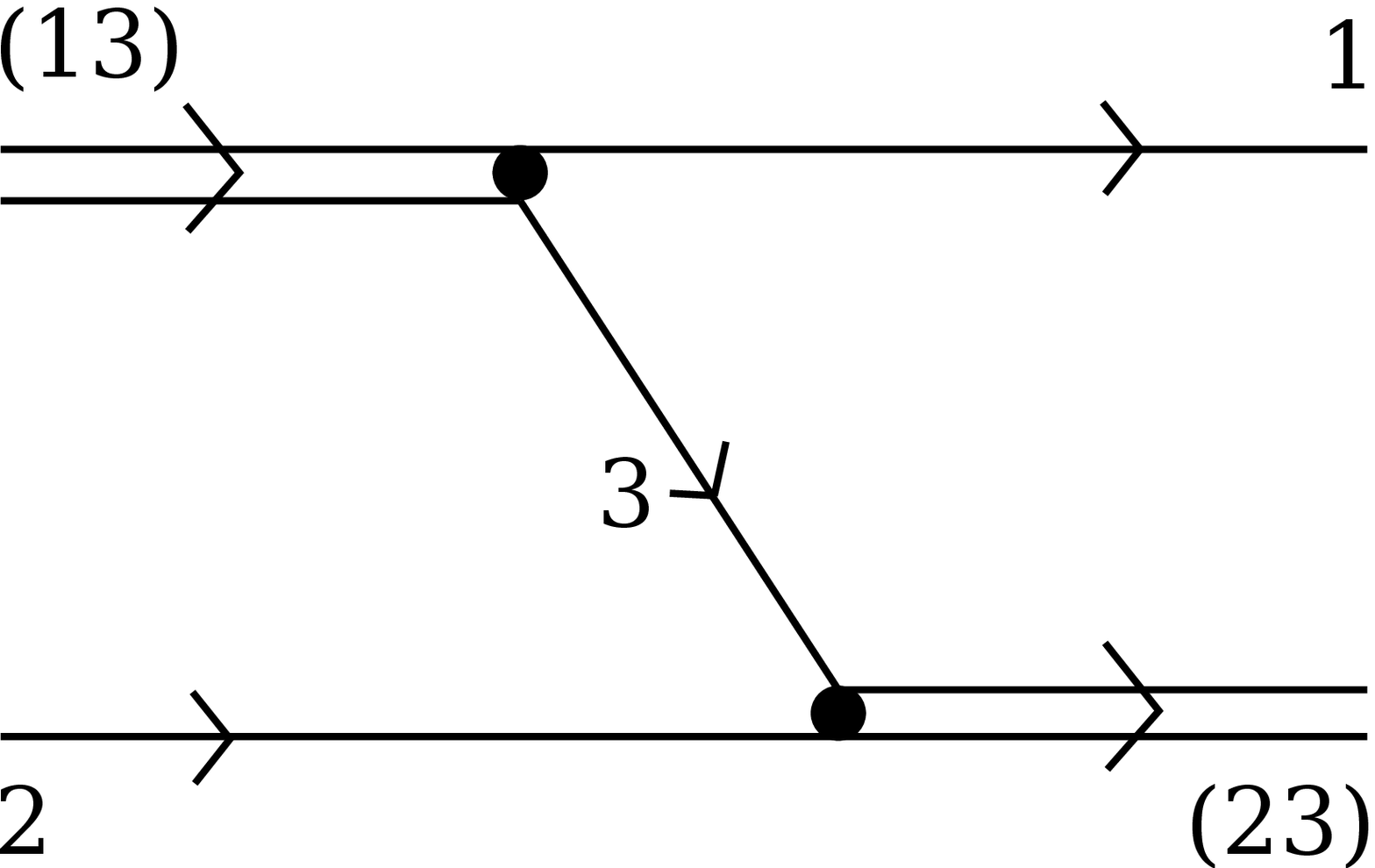,width=7cm}
\caption{
 Pole diagram describing the neutron transfer in the reaction $2 + (13) \to 1 + (23)$.} 
\label{fig_pole02}
\end{figure}

\begin{figure}
%[tbp] 
\epsfig{file=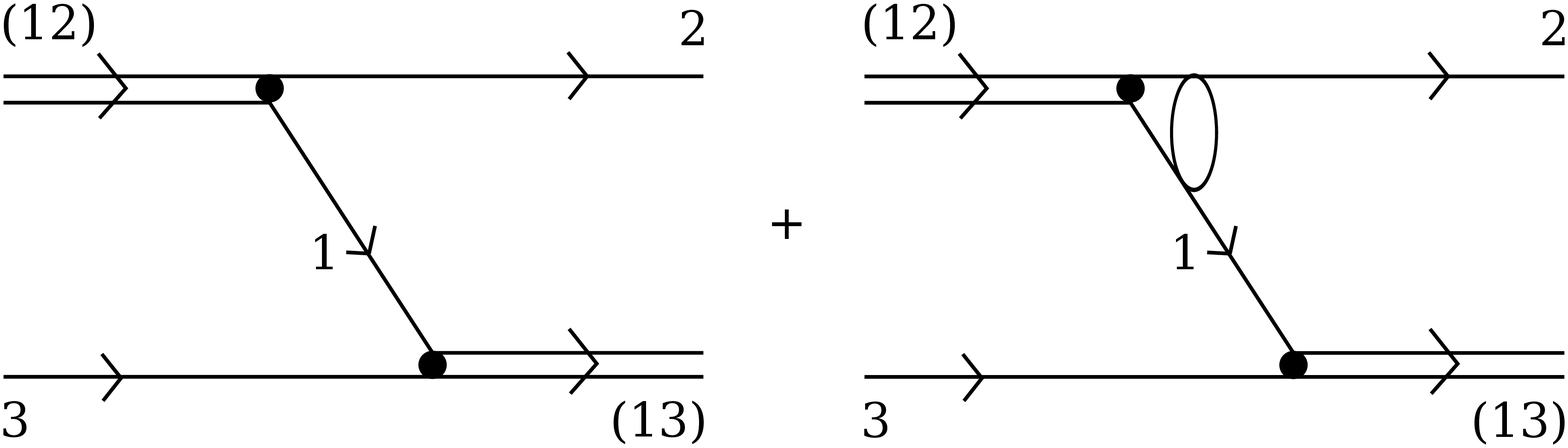,width=12cm}
\caption{
The pole diagram describing the proton transfer in the reaction $3 + (12) \to 2 + (13)$, which contains the Coulomb-modified form factor $(12) \to 1 +2$. Because this form factor consists of two terms, see Eq. (\ref{gCoulmodif1}), the diagram is also represented by the sum of two diagrams. In the first diagram in the vertex form factor $(12) \to 1 +2$ no Coulomb interaction is included, but it is included in the second diagram. Bubble shows the off-shell Coulomb scattering amplitude of proton $1$ and nucleus $2$.} 
\label{fig_Coulmodpole1}
\end{figure}

\begin{figure}
%[tbp] 
\epsfig{file=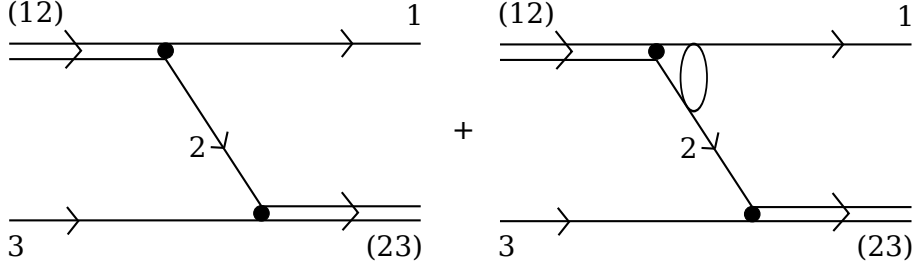,width=12cm}
\caption{
The pole diagram describing the nucleus transfer in the reaction $3 + (12) \to 1 + (23)$, which contains the Coulomb-modified form factor $(12) \to 1 +2$. Because this form factor consists of two terms, the diagram is also represented by the sum of two diagrams. In the first diagram in the vertex form factor $(12) \to 1 +2$ no Coulomb interaction is included, but it is included in the second diagram. Bubble shows the off-shell Coulomb scattering amplitude of proton $1$ and nucleus $2$.} 
\label{fig_Coulmodpole2}
\end{figure}

\begin{figure}
%[tbp] 
\epsfig{file=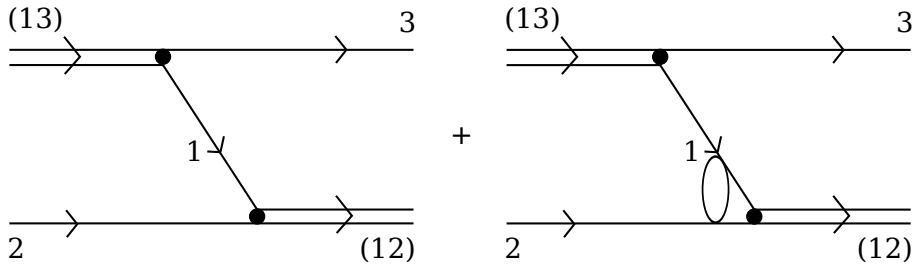,width=12cm}
\caption{
The pole diagram describing the proton transfer in the reaction $2 + (13) \to 3 + (12)$, which contains the Coulomb-modified form factor $1 + 2 \to (12)$. Notations are the same as in Fig. \ref{fig_Coulmodpole1}.} 
\label{fig_poleCoul213}
\end{figure}

\begin{figure}
%[tbp] 
\epsfig{file=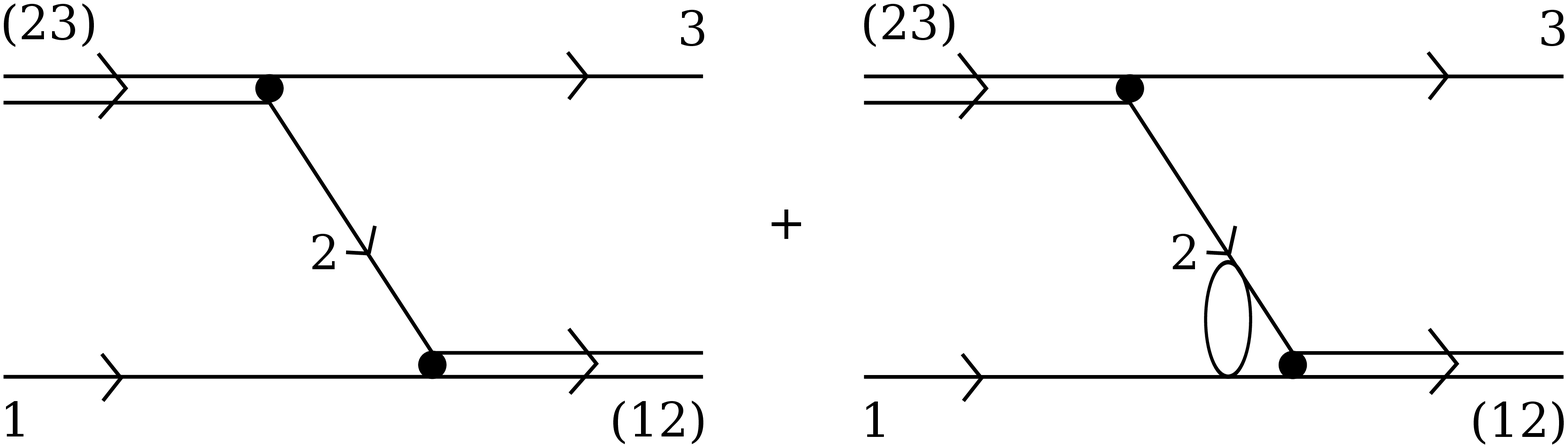,width=12cm}
\caption{
The pole diagram describing the nucleus transfer in the reaction $1 + (23) \to 3 + (12)$, which contains the Coulomb-modified form factor $1 +2 \to (12)$. Notations are the same as in Fig. \ref{fig_Coulmodpole1}.} 
\label{fig_poleCoul123}
\end{figure}

\begin{figure}
%[tbp] 
\epsfig{file=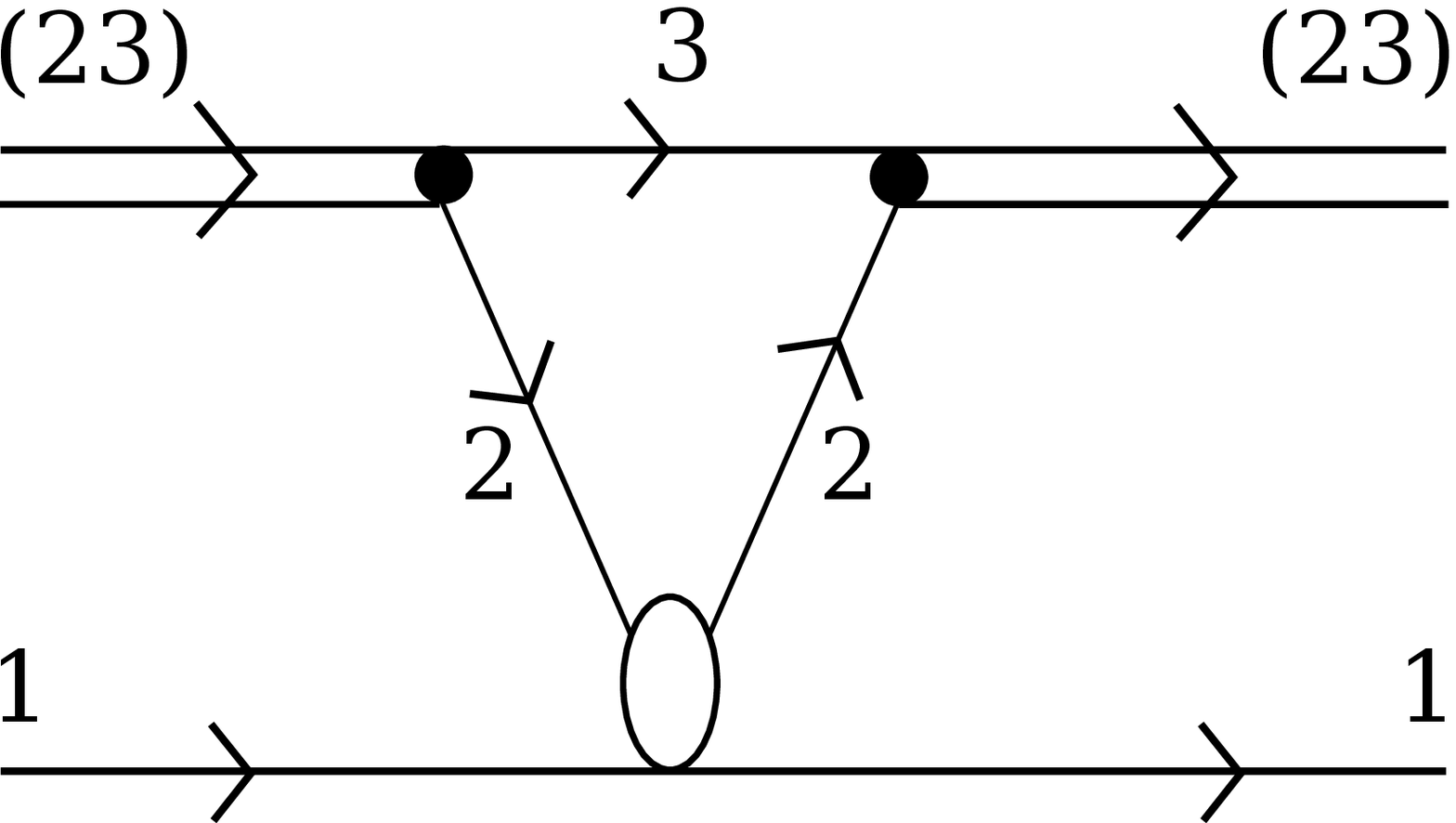,width=7cm}
\caption{
The triangular diagram describing the elastic and inelastic processes in the reaction $1 + (23) \to 1 + (23)$. The four-ray vertex is the off-shell $1+2$ Coulomb scattering amplitude.} 
\label{triang_direct}
\end{figure}

\begin{figure}
%[tbp] 
\epsfig{file=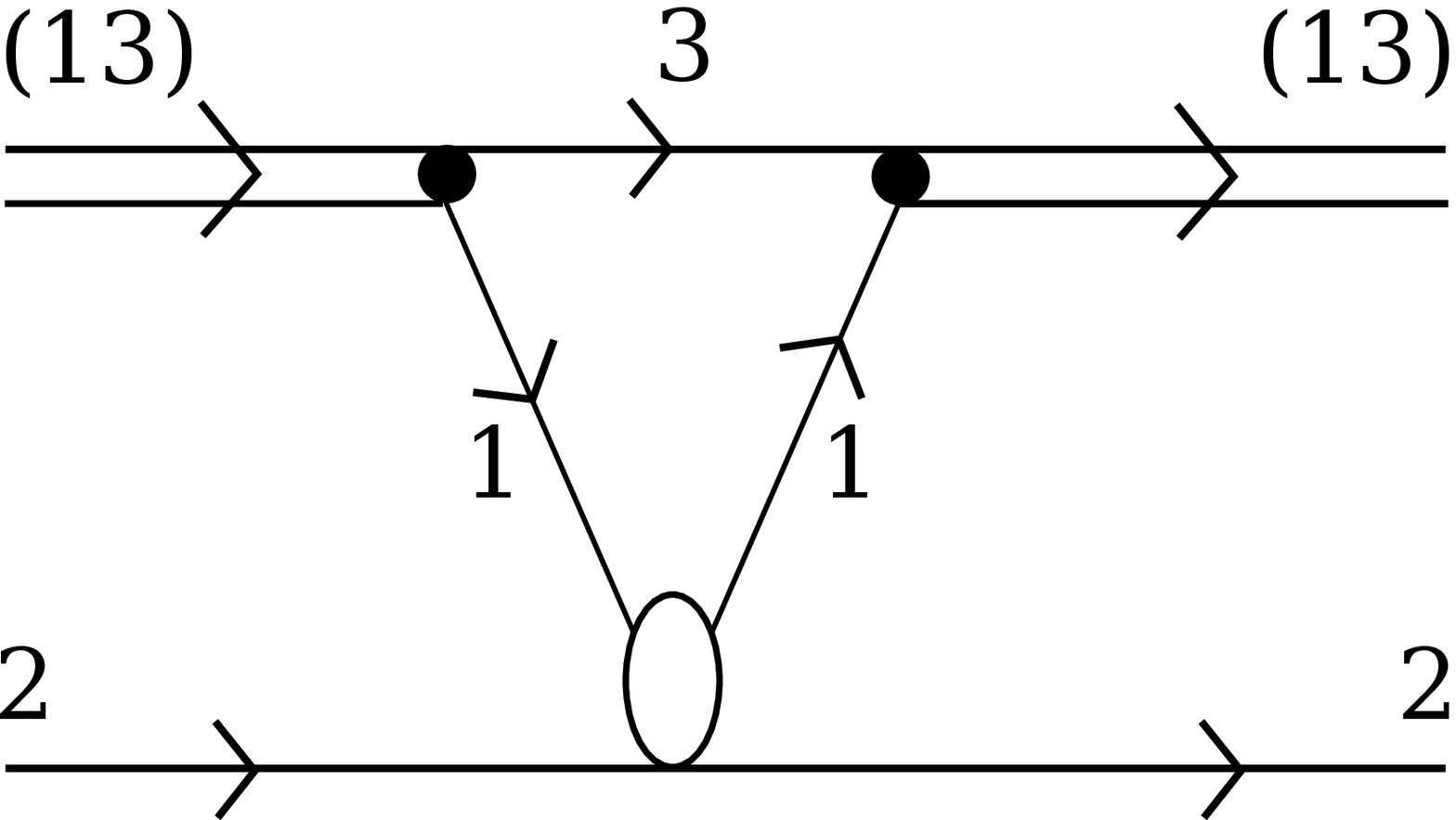,width=7cm}
\caption{
The triangular diagram describing the elastic and inelastic processes in the reaction $2 + (13) \to 2 + (13)$. The four-ray vertex is the off-shell $1+2$ Coulomb scattering amplitude.} 
\label{triang_direct1}
\end{figure}

\begin{figure}
%[tbp] 
\epsfig{file=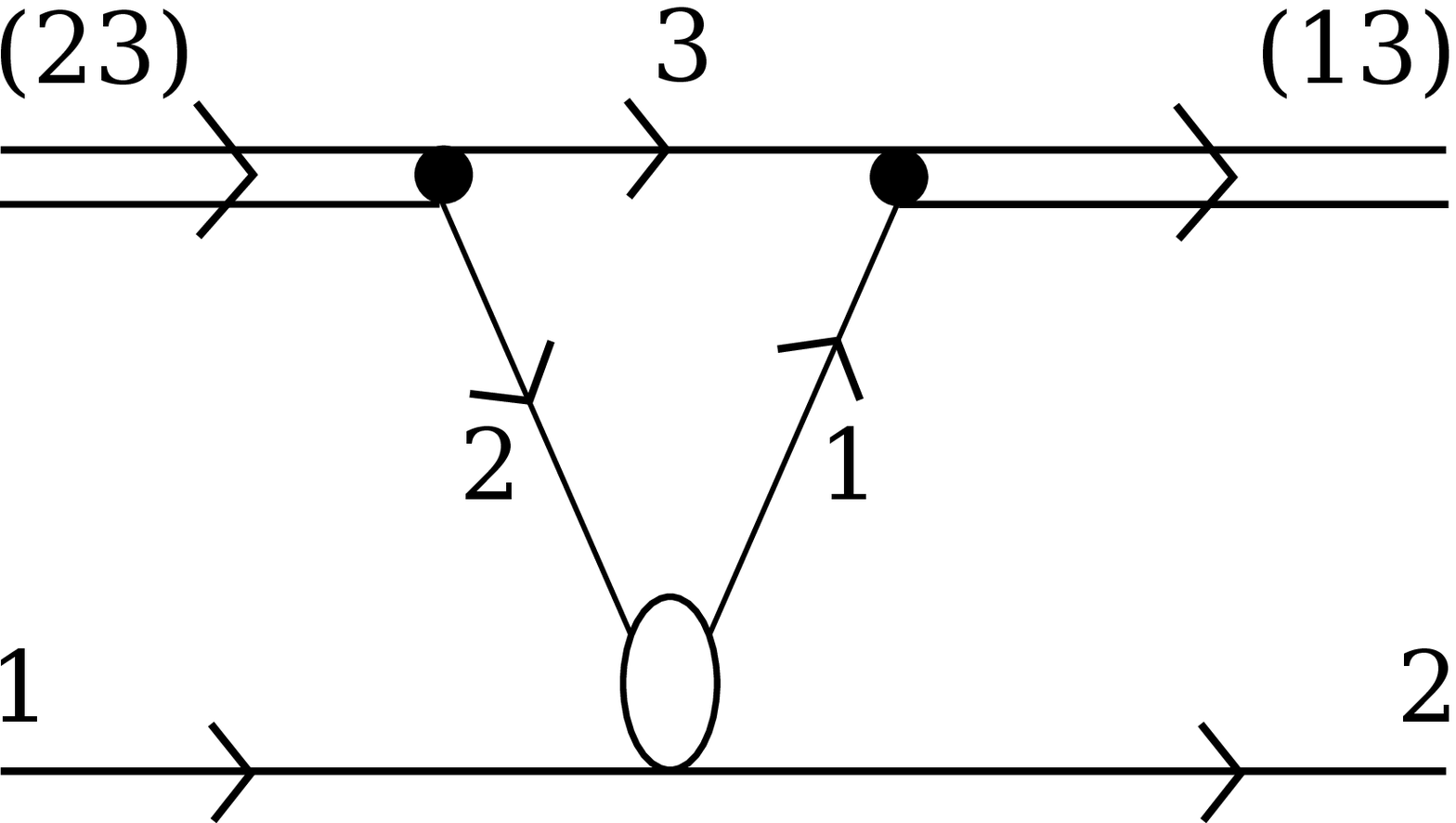,width=7cm}
\caption{
 The triangular diagram describing the exchange processes in the reaction $1 + (23) \to 2 + (13)$. The four-ray vertex is the off-shell $1+2$ Coulomb scattering amplitude.} 
\label{triang_exch}
\end{figure}

\begin{figure}
%[tbp] 
\epsfig{file=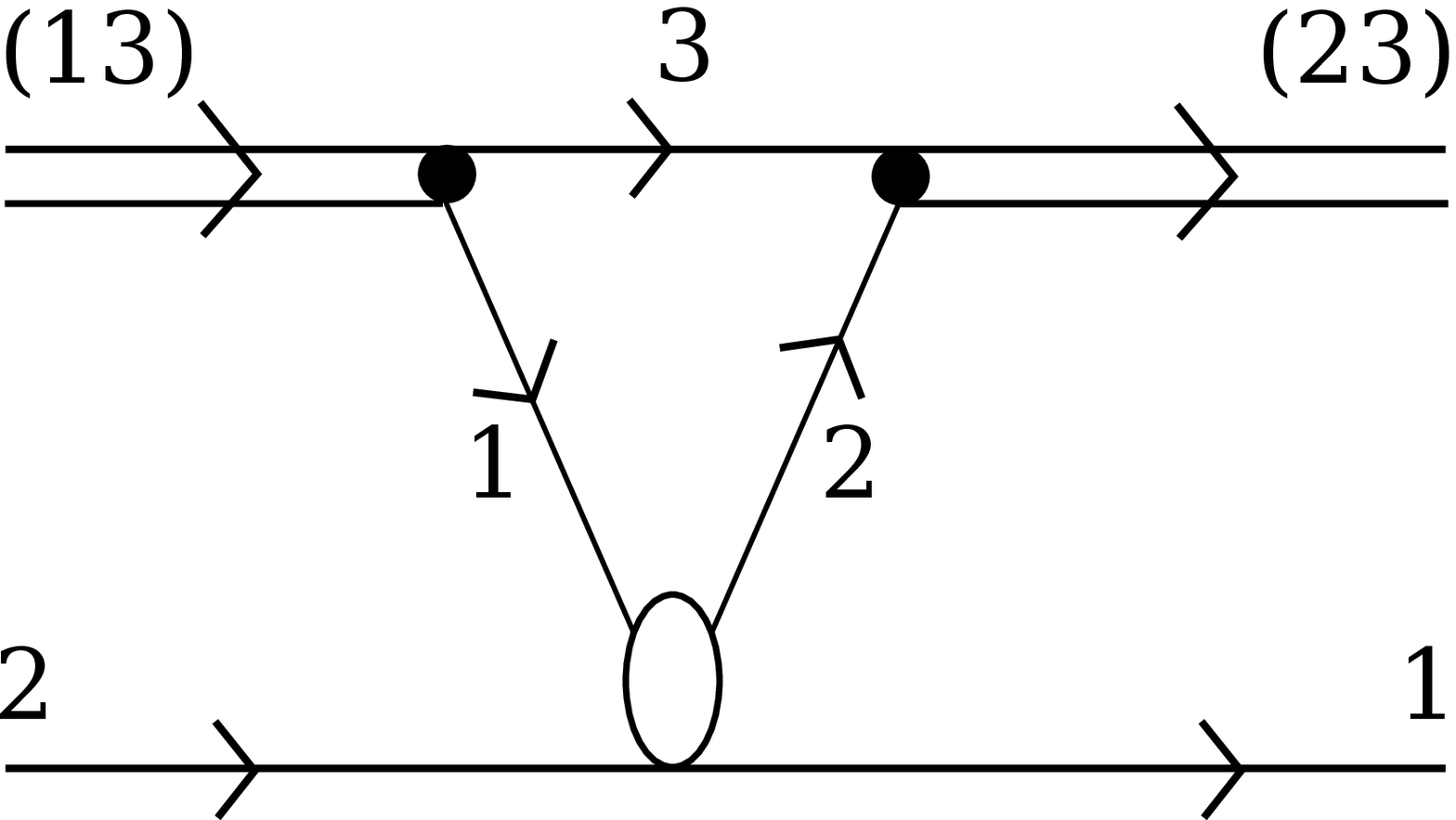,width=7cm}
\caption{
 The triangular diagram describing the exchange processes in the reaction $2 + (13) \to 1 + (23)$. The four-ray vertex is the off-shell $1+2$ Coulomb scattering amplitude.} 
\label{triang_exch1}
\end{figure}

Here 
\begin{align}
&{\tilde R}^{'\,\sigma\rho\,(i)\,\kappa}_{u_{\beta}\,t_{\beta},\,u_{\alpha}\,t_{\alpha}}({p}''_{\beta},\,{p}''_{\alpha};z) = \sum\limits_{L'_{\alpha}}\,\sum\limits_{t'_{\alpha}=1}^{N^{J_{\alpha} S_{\alpha}}}\,R^{'\,\sigma\sigma\,(i)\,\kappa}_{u_{\beta}\,t_{\beta},\,u'_{\alpha}\,t'_{\alpha}}({p}''_{\beta},\,p''_{\alpha};z)\,\Delta^{J_{\alpha}S_{\alpha}\,\sigma\rho}_{L'_{\alpha}\,t'_{\alpha}\,L_{\alpha}\,t_{\alpha}}({\hat z}_{\alpha})\,({\hat z}_{\alpha} -{\hat E}_{\alpha n_{\alpha}}),                                                                         \nonumber\\
&\qquad t_{\beta} \leq N^{J_{\beta}S_{\beta}},\,\,\, t_{\alpha}\, \leq N^{J_{\alpha}S_{\alpha}},
\label{Rprime1}
\end{align}
\begin{align}
&{\tilde R}^{\sigma\rho\,(i)\,\kappa}_{u_{\beta}\,t_{\beta},\,u_{\alpha}\,t_{\alpha}}({p}'_{\beta},\,{p}''_{\alpha};z) = \sum\limits_{L'_{\alpha}}\,\sum\limits_{t'_{\alpha}=1}^{N^{J_{\alpha} S_{\alpha}}}\,R^{\sigma\sigma\,(i)\,\kappa}_{u_{\beta}\,t_{\beta},\,u'_{\alpha}\,t'_{\alpha}}({p}'_{\beta},\,p''_{\alpha};z)\,\Delta^{J_{\alpha}S_{\alpha}\,\sigma\rho}_{L'_{\alpha}\,t'_{\alpha}\,L_{\alpha}\,t_{\alpha}}({\hat z}_{\alpha})\,({\hat z}_{\alpha} - {\hat E}_{\alpha n_{\alpha}}),                                                                         \nonumber\\
&\qquad  t_{\beta} \geq N^{J_{\beta}S_{\beta}}+1,\,\,\,t'_{\alpha}\,\leq N_{\alpha},
\label{R1}
\end{align}
where ${\hat z}_{\alpha} = z - {p''}_{\alpha}^{2}/(2\,M_{\alpha})$                  .                                                                                                                
The nondiagonal potential for $i=0$ describes the pole diagram corresponding to the neutron (particle $3$) transfer, see Figs \ref{fig_pole01} and \ref{fig_pole02}, the potential for $i=1(2)$ describes particle $1(2)$ transfer diagrams which contain one Coulomb-modified form factor, see Figs \ref{fig_Coulmodpole1} and \ref{fig_Coulmodpole2}. 
The pole neutron transfer diagrams describing the inverse processes are shown in Figs \ref{fig_poleCoul213} and  \ref{fig_poleCoul123}.
The effective potential for $i=3$ describes the elastic or inelastic scattering triangle diagram, see Figs \ref{triang_direct} and \ref{triang_direct1}, and the effective potential for $i=4$ describes the exchange triangular diagram leading to the rearrangement in the channels $\alpha \not= 3$, see Figs \ref{triang_exch} and \ref{triang_exch1}. 

Both triangular diagrams contain the $1+2$ off-shell Coulomb scattering amplitude in the four-ray vertex. The amplitude of the elastic scattering triangular diagram contains a strong forward Coulomb singularity generated by the off-shell Coulomb scattering amplitude, which is compensated by subtracting a corresponding Born Coulomb scattering term in the channels $\alpha \not=3$. For the contributions with $i=0,\,1,\,2$  $\,\,\,\kappa \equiv {\Lc}$ is a single index while for $i=3,4\,\,$ it is a multi-index $\,\kappa \equiv ({\Lc}_{1},\,{\Lc}_{2},\,f)$. Explicit equations for $A^{i}_{\kappa}$ and $R^{(i)\,\kappa}_{u_{\beta}\,t_{\beta},\,u_{\alpha}\,t_{\alpha}}$ were given in \cite{alt2002}.  Equations for $A^{i}_{\kappa}$ can be taken from this paper with only two minor modifications: $q'_{\beta}$ should be replaced by $p'_{\beta}$ and $q_{\alpha}$ by $p_{\alpha}$. Equations for $R^{(i)\,\kappa}$ should be modified by including the target excitation indices. In particular, Eq. (A1) \cite{alt2002} for $i=0,\,1,\,2$ takes the form
\begin{align}
R_{\zeta_\beta\,\zeta_\alpha}^{\sigma \sigma\,(i)\,{\Lc}}(p'_{\beta} ,p_\alpha;z) = 
R_{\zeta_\beta\,\zeta_\alpha}^{'\,\sigma \sigma\,(i)\,{\Lc}}(p'_{\beta} ,p_\alpha;z)=
{\overline \delta  _{\beta \alpha }}\,c_{\zeta_{\beta}}^{\sigma\,*}c_{\zeta_{\alpha}}^{\sigma}\,\frac{1}{2}\int\limits_{ - 1}^1 {dx} \,{P_{\Lc}}(x)\frac{{k_\alpha ^{ - {L_\alpha }}k_\beta ^{' - {L_\beta }}{g}_{\zeta_{\beta}}^{\sigma\,* }(k'_{\beta})\,g_{\zeta_{\alpha}}^{\sigma }({k_\alpha })}}{{z - \,\epsilon^{\sigma} - \,\,p_\alpha ^2/(2{M_\alpha }) - k_\alpha ^2/(2{\mu _\alpha })}},    \qquad i=0,\,1,\,2,
\label{Riksigmarho1}
\end{align}
where 
$\,\,\,{\rm {\bf k}}_{\alpha}= \epsilon_{\alpha \beta }(\lambda_{\beta \gamma}\,{\rm {\bf p}}_{\alpha} + {\rm {\bf p}}'_{\beta}),\,\,{\rm {\bf k}}'_{\beta}= \epsilon_{\beta \gamma}(\lambda_{\alpha \gamma}\,{\rm {\bf p}}'_{\beta} + {\rm {\bf p}}_{\alpha}),\,\, x={\rm {\bf {\hat p}}}'_{\beta} \cdot {\rm {\bf {\hat p}}}_{\alpha}$,  $\,\,\lambda_{\alpha\beta}= m_{\alpha}/(m_{\alpha} + m_{\beta})= 1 - \lambda_{\beta \alpha}$, $\,\alpha \not= \beta$. Also $\epsilon_{\alpha \beta}= - \epsilon_{\beta \alpha}$ is the antisymmetric symbol with $\epsilon_{\alpha \beta}= +1$  if $(\alpha, \beta)$ is a cyclic ordering of the indices $(1,2,3)$.  
 Although in Eq. (\ref{Riksigmarho1}) formally two Coulomb-modified form factors are present, utmost only one Coulomb-modified form factor is needed. In the diagram with $i=0$, which describes the neutron transfer (particle $3$), the vertex with $\alpha=3$ doesn't appear. 
Specifically, the amplitude of the pole diagram describing the neutron transfer $1 + (23) \to 2+ (13)$, see Fig. \ref{fig_pole01}, is given by 
\begin{align}
R_{\zeta_\beta\,\zeta_\alpha}^{\sigma \sigma\,(0)\,{\Lc}}(p'_{\beta} ,p_\alpha;z) = 
R_{\zeta_\beta\,\zeta_\alpha}^{'\,\sigma \sigma\,(0)\,{\Lc}}(p'_{\beta} ,p_\alpha;z)=
{\overline \delta  _{\beta \alpha }}\,c_{\zeta_{\alpha}}^{\sigma}\,\frac{1}{2}\int\limits_{ - 1}^1 {dx} \,{P_{\Lc}}(x)\frac{{k_\alpha ^{ - {L_\alpha }}k_\beta ^{' - {L_\beta }}\,\chi_{\zeta_{\beta}}^*(k'_{\beta})\,\chi_{\zeta_{\alpha}}^{\sigma}({k_\alpha })}}{{z - \,\epsilon^{\sigma} - \,\,p_\alpha ^2/(2{M_\alpha }) - k_\alpha ^2/(2{\mu _\alpha })}},    \qquad i=0,
\label{Rpole02}
\end{align}
where $\alpha=1,\,\,\beta=2$ and $\gamma=3$. Only the vertex $\beta + \gamma \to (\beta\gamma)$, which contains nucleus $2$, depends on the nucleus excitation index $\sigma$.
Correspondingly, the amplitude of the inverse process $2+ (13) \to 1 + (23)$, see Fig. \ref{fig_pole02}, is given by
\begin{align}
R_{\zeta_\beta\,\zeta_\alpha}^{\sigma \sigma\,(0)\,{\Lc}}(p'_{\beta} ,p_\alpha;z) = 
R_{\zeta_\beta\,\zeta_\alpha}^{'\,\sigma \sigma\,(0)\,{\Lc}}(p'_{\beta} ,p_\alpha;z)=
{\overline \delta  _{\beta \alpha }}\,c_{\zeta_{\beta}}^{\sigma\,*}\,\frac{1}{2}\int\limits_{ - 1}^1 {dx} \,{P_{\Lc}}(x)\frac{{k_\alpha ^{ - {L_\alpha }}k_\beta ^{' - {L_\beta }}\,\chi_{\zeta_{\beta}}^{\sigma\,* }(k'_{\beta})\,\chi_{\zeta_{\alpha}}({k_\alpha })}}{{z - \epsilon^{\sigma}- \,p_\alpha ^2/(2{M_\alpha }) - k_\alpha ^2/(2{\mu _\alpha })}},    \qquad i=0.
\label{Rpole01}
\end{align}
Here, $\alpha=2,\,\,\beta=1$ and $\gamma=3$. Only the vertex $\alpha + \gamma \to (\alpha\gamma)$, which contains nucleus $2$, depends on the nucleus excitation index $\sigma$.

In the diagrams with $i=1,2$ (proton or nucleus transfer) the vertex with $\alpha=3$ appears only once. 
Also in \cite {alt2002} the Coulomb interaction in the Coulomb-modified factors and in the four-ray vertex of the triangular diagrams is taken into account in the Born approximation. Here the Coulomb Born amplitude is replaced by the full off-shell Coulomb scattering amplitude of particles of the pair $\gamma=3$. Expression for the Coulomb-modified form factor with the off-shell Coulomb scattering amplitude is given above in Eq. (\ref{gCoulmodif1}). 

Now we proceed to the amplitude of the triangular diagram, $i=3$. The elastic scattering triangular diagram has singularity at forward scattering generated by the off-shell Coulomb scattering amplitude. To eliminate this singularity we add and subtract the channel Coulomb scattering potentials in channels $\alpha \not=3$, see Eq. (\ref{ZCoulomb1}).
The added Coulomb potentials can be eliminated by the including Coulomb distorted waves in the initial and final channels $\alpha \not= 3$.   
In Eq. (A3) \cite{alt2002} the Fourier transform of the screened Born Coulomb potential $V^{(R)}_{\gamma}({\rm {\bf \Delta}}'_{\alpha})$ should be replaced by the unscreened $T_{\gamma}^{C}$, $\,\gamma=3$. Then Eq. (A7) \cite{alt2002} for the triangular diagrams in Figs \ref{triang_direct}   and \ref{triang_direct1} takes the form ($\alpha \not=3$):
\begin{align}
&R_{\zeta_{\beta}\,\zeta_{\alpha}}^{(3)\,{\Lc}_{1}\,{\Lc}_{2}\,f\,\sigma\sigma}(p'_{\alpha},\,p_{\alpha};z) 
= \delta_{\beta \alpha}\,{\overline \delta}_{\alpha 3}\delta_{\gamma 3}\,
\Big \{ 4\,\pi\,{\overline F}_{ L_{\alpha}t'_{\alpha}\,L_{\alpha}t_{\alpha} }^{J_{\alpha}S_{\alpha}\,\sigma\sigma}(p_{\alpha};z)\,
V^{C}_{\gamma\,{\Lc}_{2}}({\rm {\bf \Delta}}'_{\alpha})
+ \frac{1}{8\,\pi^{2}}\,\int\limits_{-1}^{1}{\rm d}x_{2}\,P_{{\Lc}_{2}}(x_{2})\,V_{\gamma}^{C}({\rm {\bf \Delta}}'_{\alpha})  \nonumber\\
&\times \Big[F_{L'_{\alpha} t'_{\alpha}\,L_{\alpha}t_{\alpha}\,f}^{J'_{\alpha}S_{\alpha}J_{\alpha}S_{\alpha}\,\sigma\sigma}({\bp}'_{\alpha},\,{\bp}_{\alpha};z) 
- 16\,\pi^{3}\,{\overline F}_{ L_{\alpha}t'_{\alpha}\,L_{\alpha}t_{\alpha} }^{J_{\alpha}S_{\alpha}\,\sigma\sigma}(p_{\alpha};z) \Big] \Big \} + \frac{1}{8\,\pi^{2}}\,\int\limits_{-1}^{1}{\rm d}x_{2}\,P_{{\Lc}_{2}}(x_{2})\,Ftr_{L'_{\alpha}t'_{\alpha}\,L_{\alpha}t_{\alpha}\,f}^{J'_{\alpha}S_{\alpha}J_{\alpha}S_{\alpha}\,\sigma\sigma}({\bp}'_{\alpha},\,{\bp}_{\alpha};z),  \nonumber\\              &\qquad t'_{\alpha} \leq N^{J'_{\alpha}S_{\alpha}},\,\,\,t_{\alpha} \leq N^{J_{\alpha}S_{\alpha}} ,                       \label{R31}
\end{align}
where, owe to $\beta=\alpha$, $\,\,p'_{\beta}=p'_{\alpha}$ and $t'_{\beta}= t'_{\alpha}$,
\begin{align}
&{\overline F}_{L_{\alpha}t'_{\alpha}\,L_{\alpha}t_{\alpha}}^{J_{\alpha}S_{\alpha}\,\sigma\sigma}(p_{\alpha};z)= \frac{1}{16\,\pi^{3}}\,
\,F_{L_{\alpha}t'_{\alpha},\,L_{\alpha}t_{\alpha}\,0}^{J_{\alpha}S_{\alpha}J_{\alpha}S_{\alpha}\,\sigma\sigma}({\bp}_{\alpha},\,{\bp}_{\alpha};z)   \nonumber\\  
&= \frac{1}{{{{(2\pi )}^3}}}\,c_{S_{\alpha},L_{\alpha},J_{\alpha},t'_{\alpha}}^{\sigma\,*}\,c_{S_{\alpha},L_{\alpha},J_{\alpha},t_{\alpha}}^{\sigma}\,\int\limits_0^\infty  {dk\,{k^2}} \frac{{\chi_{{L_\alpha }{t'_\alpha }}^{{J_\alpha }{S_\alpha}\sigma \,*}(k)}{\chi_{{L_\alpha }{t_\alpha }}^{{J_\alpha }{S_\alpha}\sigma }(k)}}{{{{(z - \,\epsilon ^\sigma - \,p_\alpha ^2/(2{M_\alpha }) - {k^2}/(2{\mu _\alpha }))}^2}}},  \qquad \alpha \not= 3,
\label{overlineF11}
\end{align}
\begin{align}
&F_{L'_{\alpha}t'_{\alpha}\,L_{\alpha}t_{\alpha}\,f}^{J'_{\alpha}S_{\alpha}J_{\alpha}S_{\alpha}{\Lc}_1\,\sigma\sigma}({\bp}'_{\alpha},\,{\bp}_{\alpha};z)= c_{\zeta'_{\alpha}}^{\sigma\,*}c_{\zeta_{\alpha}}^{\sigma}\,\int\limits_0^\infty  {\rm d}k\,k^{2 + L'_{\alpha } - f}\,\frac{{\chi_{L_{\alpha} t_{\alpha}}^{J_{\alpha }S_{\alpha}\sigma}(k)}}{{z - \,\epsilon^{\sigma} - \,p_{\alpha}^2/(2\,M_{\alpha}) - k^{2}/(2{\mu_{\alpha })}}}              \nonumber\\
&\times \int\limits_{ - 1}^{1}{\rm d}x \,P_{{\Lc}_1}(x)\frac{{\chi_{{L'_\alpha }{t'_\alpha }}^{{J'_\alpha }{S_\alpha}\sigma\,* }(|{\rm {\bf \Delta}} _{\alpha} + {\rm {\bf k}}|)\,|{\rm {\bf \Delta}}_{\alpha } + {\rm {\bf k}}|^{ - L'_{\alpha}}}}{{z - \,\epsilon^{\sigma} - \,{p'}_{\alpha}^2/(2{M_\alpha }) - {{({\rm {\bf \Delta}} _{\alpha } + {\rm {\bf k}})}^2}/(2{\mu _\alpha })}},
\label{FLbLa1}
\end{align}
\begin{align}
&Ftr_{L'_{\alpha}t'_{\alpha}\,L_{\alpha}t_{\alpha}\,f}^{J'_{\alpha}S_{\alpha}J_{\alpha}S_{\alpha}{\Lc}_1\,\sigma\sigma}({\bp}'_{\alpha},\,{\bp}_{\alpha};z) =  c_{\zeta'_{\alpha}}^{\sigma\,*}c_{\zeta_{\alpha}}^{\sigma}\,\int\limits_0^\infty  {\rm d}k\,k^{2 + L'_{\alpha } - f}\,\frac{{\chi_{L_{\alpha} t_{\alpha}}^{J_{\alpha }S_{\alpha}\sigma}(k)}}{{z - \,\epsilon^{\sigma} - \,p_{\alpha}^2/(2\,M_{\alpha}) - k^{2}/(2{\mu_{\alpha })}}}              \nonumber\\
&\times \int\limits_{ - 1}^{1}{\rm d}x \,P_{{\Lc}_1}(x)\frac{{\chi_{{L'_\alpha }{t'_\alpha }}^{{J'_\alpha }{S_\alpha}\sigma\,* }(|{\rm {\bf \Delta}} _{\alpha} + {\rm {\bf k}}|)\,|{\rm {\bf \Delta}}_{\alpha } + {\rm {\bf k}}|^{ - L'_{\alpha}}}}{{z - \,\epsilon^{\sigma} - \,{p'}_{\alpha}^2/(2{M_\alpha }) - {{({\rm {\bf \Delta}} _{\alpha } + {\rm {\bf k}})}^2}/(2{\mu _\alpha })}}{\tilde T}^{C}_{\gamma}({\rm {\bf k}}'_{\gamma},\,{\rm {\bf k}}_{\gamma}; {\hat z}_{\gamma}).
\label{Ftr1}                         
\end{align}
Here $\zeta_{\alpha}= \{S_{\alpha}
,L_{\alpha},J_{\alpha},t_{\alpha} \}$ and $\zeta'_{\alpha}= \{S_{\alpha},L'_{\alpha},J'_{\alpha},t'_{\alpha} \}$, 
\begin{align}
{\rm {\bf \Delta}}'_{\alpha} = {\rm {\bf p}}_{\alpha} - {\rm {\bf p}}'_{\alpha}, \quad 
{\rm {\bf \Delta}}_{\alpha} =- \lambda_{\gamma \beta}\,{\rm {\bf \Delta}}'_{\alpha}, \quad
x_{2}= {\rm {\bf {\hat p}}}_{\alpha} \cdot {\rm {\bf {\hat p}}}'_{\alpha}, \quad
x_{1}= {\rm {\bf {\hat k}}} \cdot {\rm {\bf {\hat \Delta}}}_{\alpha},
\label{momentvariabl1}
\end{align}
\begin{align}
{\rm {\bf k}}_{\gamma}= \frac{m_{\beta}\,{\rm {\bf p}}_{\alpha} - m_{\alpha}\,{\rm {\bf p}}_{\beta}}{m_{\alpha\beta}}, \qquad {\rm {\bf k}}'_{\gamma}= \frac{m_{\beta}\,{\rm {\bf p}}'_{\alpha} - m_{\alpha}\,{\rm {\bf p}}'_{\beta}}{m_{\alpha\beta}}, \qquad  {\rm {\bf p}}_{\alpha}+ {\rm {\bf p}}_{\beta}= {\rm {\bf p}}'_{\alpha} + {\rm {\bf p}}'_{\beta}, \qquad  {\rm {\bf k}}_{\gamma} - {\rm {\bf k}}'_{\gamma} = {\rm {\bf \Delta}}'_{\alpha}.
\label{kgamma1}
\end{align}
We use the following notations for the particles in the diagram of Fig. \ref{triang_direct} (Fig.  \ref{triang_direct1}): $\alpha=\alpha'=1\,\,$ ($\alpha=\alpha'=2$),   $\,\beta=\beta'=2\,$ ($\beta=\beta'=1$) and $\gamma=3$ in both diagrams. The primed particles on the diagrams are the ones after the Coulomb scattering described by the four-ray vertex.

The amplitude ${\tilde T}^{C}_{\gamma}({\rm {\bf k}}'_{\gamma},\,{\rm {\bf k}}_{\gamma}; {\hat z}_{\gamma})$ is the off-shell Coulomb scattering amplitude of particles $\beta$ and $\gamma$ in the triangular diagram without the Born term, that is
${\tilde T}^{C}_{\gamma}({\rm {\bf k}}'_{\gamma},\,{\rm {\bf k}}_{\gamma}; {\hat z}_{\gamma})
= T^{C}_{\gamma}({\rm {\bf k}}'_{\gamma},\,{\rm {\bf k}}_{\gamma}; {\hat z}_{\gamma}) -
V_{\gamma}^{C}({\rm {\bf k}}_{\gamma} - {\rm {\bf k}}'_{\gamma})$,   $\,\,{\hat z}_{\gamma}= z - p_{\gamma}^{2}/(2\,M_{\gamma})$. 
 
From normalization (\ref{normalizcond1}) on the energy shell $(p_{\alpha}= q_{\alpha})$ we get 
\begin{align}
4\,\pi\,\sum\limits_{\sigma=1}^{N}\,\sum\limits_{t'_{\alpha},t_{\alpha}=1}^{A^{J_{\alpha}S_{\alpha}}}\,{\overline F}_{L_{\alpha}t'_{\alpha}\,L_{\alpha}t_{\alpha}}^{J_{\alpha}S_{\alpha}\,\sigma\sigma}(q_{\alpha};z) =1.
\label{normalcondition2}
\end{align}
Then, according to Eq. (\ref{ZCoulomb1}), after applying the two-potential equation the first term $4\,\pi\,{\overline F}_{ L_{\alpha}t'_{\alpha}\,L_{\alpha}t_{\alpha} }^{J_{\alpha}S_{\alpha}\,\sigma\sigma}(p''_{\alpha};z)\,
V^{C}_{\gamma\,{\Lc}_{2}}({\rm {\bf \Delta}}'_{\alpha})$ in Eq. (\ref{R31}) will be replaced by the corresponding Coulomb distorted wave. 
As a result, we obtain
\begin{align}
&R_{\zeta_{\beta}\,\zeta_{\alpha}}^{'(3)\,{\Lc}_{1}\,{\Lc}_{2}\,f\,\sigma\sigma}(p'_{\beta},\,p_{\alpha};z) 
= \delta_{\beta \alpha}\,{\overline \delta}_{\alpha 3}\delta_{\gamma 3}\,
\frac{1}{8\,\pi^{2}}\,\int\limits_{-1}^{1}{\rm d}x_{2}\,P_{{\Lc}_{2}}(x_{2})\,\Big \{V_{\gamma}^{C}({\rm {\bf \Delta}}'_{\alpha})\,\Big[F_{L'_{\alpha} t'_{\alpha}\,L_{\alpha}t_{\alpha}\,f}^{J'_{\alpha}S_{\alpha}J_{\alpha}S_{\alpha}\,\sigma\sigma}({\bp}'_{\alpha},\,{\bp}_{\alpha};z) 
- 16\,\pi^{3}\,{\overline F}_{ L_{\alpha}t_{\alpha}\,L_{\alpha}t_{\alpha} }^{J_{\alpha}S_{\alpha}\,\sigma\sigma}(p_{\alpha};z)\Big]                                  \nonumber\\
&+ Ftr_{L'_{\alpha}t'_{\alpha}\,L_{\alpha}t_{\alpha}\,f}^{J'_{\alpha}S_{\alpha}J_{\alpha}S_{\alpha}\,\sigma\sigma}({\bp}'_{\alpha},\,{\bp}_{\alpha};z)\Big \},  \qquad t'_{\beta}=t'_{\alpha} \leq N^{J'_{\alpha}S_{\alpha}}, \,\,\, t_{\alpha} \leq N^{J_{\alpha}S_{\alpha}}.                                     
\label{R31}
\end{align}
Also
\begin{align}
&R_{\zeta_{\beta}\,\zeta_{\alpha}}^{(3)\,{\Lc}_{1}\,{\Lc}_{2}\,f\,\sigma\sigma}(p'_{\beta},\,p_{\alpha};z) 
= \delta_{\beta \alpha}\,{\overline \delta}_{\alpha 3}\delta_{\gamma 3}\,
\frac{1}{8\,\pi^{2}}\,\int\limits_{-1}^{1}{\rm d}x_{2}\,P_{{Lc}_{2}}(x_{2})\,F_{L'_{\alpha}t'_{\alpha}\,L_{\alpha}t_{\alpha}\,f}^{J'_{\alpha}S_{\alpha}J_{\alpha}S_{\alpha}\,\sigma\sigma}({\bp}'_{\alpha},\,{\bp}_{\alpha};z),          \qquad t_{\beta}=t'_{\alpha} \geq N^{
J'_{\alpha}S_{\alpha}} +1,\,\,\,t_{\alpha} \leq N^{J_{\alpha}S_{\alpha}}.                       
\label{R311}
\end{align}
The spin-angular part $A_{{{\Lc}_1}{{\Lc}_2}f}^{(3)}$ of the effective potential is given by Eq. (A2) \cite{alt2002}, in which $q_{\alpha}$ and $q'_{\beta}$ should be replaced by $p_{\alpha}$ and $p'_{\alpha}$.
  
The last amplitudes are the exchange triangular diagrams shown in Fig. \ref{triang_exch} and  \ref{triang_exch1}:
\begin{align}
&R_{\zeta_{\beta}\,\zeta_{\alpha}}^{(4)\,{\Lc}_{1}\,{\Lc}_{2}\,f\,\sigma\sigma}(p'_{\beta},\,p_{\alpha};z) 
= {\overline \delta}_{\beta \alpha}\,\delta_{\gamma 3}\,
\frac{1}{8\,\pi^{3}}\,\int\limits_{-1}^{1}{\rm d}x_{2}\,P_{{\Lc}_{2}}(x_{2})\,
\int\limits_{0}^{\infty}\,{\rm d}k_{\alpha}\,k_{\alpha}^{2+ L_{\beta} -f}\,\frac{ \chi_{L_{\alpha}t_{\alpha}}^{J_{\alpha}S_{\alpha}\,\sigma}(k_{\alpha})}{z- \,\epsilon^{\sigma}- p_{\alpha}^{2}/(2\,M_{\alpha}) - k_{\alpha}^{2}/(2\,\mu_{\alpha})}                                 \nonumber\\                                              &\times \int\limits_{-1}^{1}\,{\rm d}x_{1}\,P_{{\Lc}_{1}}(x_{1})\,T_{\gamma}^{C}({\rm {\bf k}}'_{\gamma},{\rm {\bf k}}_{\gamma}; {\hat z}_{\gamma})\,\frac{\chi_{ L_{\beta}t_{\beta}}^{ J_{\beta}S_{\beta}\, * }(|{\rm {\bf k}}_{\alpha} + {\rm {\bf p}}|)\,|{\rm {\bf k}}_{\alpha} + {\rm {\bf p}}|^{-L_{\beta}}}{z- \,\epsilon^{\sigma}- \,p_{\beta}^{'\,2}/(2\,M_{\beta}) - ({\rm {\bf k}}_{\alpha} + {\rm {\bf p}})^{2}/(2\,\mu_{\alpha})}.                                    
\label{R41}
\end{align}
We use the following notations for the particles on the diagrams of Fig.  \ref{triang_exch} (Fig. \ref{triang_exch1}): $\alpha=\alpha'=1\,$ ($\alpha=\alpha'=2$), $\,\beta= \beta'=2\,$ ($\beta=\beta'=1$), $\gamma=3$, where all primed particles are the ones after Coulomb rescattering.
Also
\begin{align}
&{\rm {\bf p}}= \lambda_{\gamma \alpha}\,{\rm {\bf p}}'_{\beta} - \lambda_{\gamma \beta}\,{\rm {\bf p}}_{\alpha}, \qquad 
{\rm {\bf p}}'= {\rm {\bf p}}'_{\beta} + \lambda_{\beta \gamma}\,{\rm {\bf p}}_{\alpha}, \qquad x_{1}= {\rm {\bf {\hat k}}}_{\alpha} \cdot {\rm {\bf {\hat p}}},
\qquad x_{2}= {\rm {\bf {\hat p}}}'_{\beta} \cdot {\rm {\bf {\hat p}}}_{\alpha},  \nonumber\\
&{\rm {\bf k}}'_{\gamma}= \frac{m_{\beta}\,{\rm {\bf p}}'_{\alpha} - m_{\alpha}\,{\rm {\bf p}}'_{\beta}}{m_{\alpha\beta}},   \qquad {\rm {\bf k}}_{\gamma}= \frac{m_{\beta}\,{\rm {\bf p}}_{\alpha} - m_{\alpha}\,{\rm {\bf p}}_{\beta}}{m_{\alpha\beta}},   \qquad   
{\rm {\bf \Delta}}= {\rm {\bf k}}_{\gamma} - {\rm {\bf k}}'_{\gamma} = {\rm {\bf p}}_{\alpha} - {\rm {\bf p}}'_{\alpha} = {\rm {\bf p}}'_{\beta} - {\rm {\bf p}}_{\beta} = {\rm {\bf p}}'_{\beta} + {\rm {\bf p}}_{\alpha} + {\rm {\bf p}}_{\gamma},
\label{kinexchtriang1}
\end{align}
$\,\,{\rm {\bf {\hat p}}}= {\rm {\bf {\hat p}}}/p$. The spin-angular part $\,\,\,A^{(4)}_{{\Lc}_{1}{\Lc}_{2}f}(p'_{\beta},p_{\alpha})\,\,$ of the effective potential is given by Eq. (A8) \cite{alt2002}.  

Using the results of Appendix \ref{Coulrenormexchtr1} we can simplify the calculations of the effective potentials by combining the neutron transfer pole amplitudes with the Coulomb exchange triangular diagram amplitudes taking into account the fact that near the pole the exchange triangular diagram has also a pole singularity. Summing up the pole neutron transfer amplitude and the pole contribution to the triangular exchange diagram and neglecting the regular at the pole part of the triangular exchange diagram, we can replace this two amplitudes by the renormalized neutron transfer pole amplitude, that is  
\begin{align}
&{\tilde Z}^{J^{\pi}\,\sigma\rho\,(0)}_{u_{\beta}\,t_{\beta},\,u_{\alpha}\,t_{\alpha}}(p''_{\beta},\,p''_{\alpha};z)  +  {\tilde Z}^{J^{\pi}\,\sigma\rho\,(4)}_{u_{\beta}\,t_{\beta},\,u_{\alpha}\,t_{\alpha}}(p''_{\beta},\,p_{\alpha};z)  \approx  [1+ D_{\beta \zeta_{\alpha}\,\alpha \zeta_{\alpha}}^{\sigma}]\,{\tilde Z}^{J^{\pi}\,\sigma\rho\,(0)}_{u_{\beta}\,t_{\beta},\,u_{\alpha}\,t_{\alpha}}(p''_{\beta},\,p_{\alpha};z),   \qquad t_{\beta} \leq N^{J_{\beta}S_{\beta}},\,\,\, t_{\alpha}\, \leq N^{J_{\alpha}S_{\alpha}}. 
\label{renormpoleexchtr1}
\end{align}
where $D_{\beta \zeta_{\alpha}\,\alpha \zeta_{\alpha}}^{\sigma}$ is determined in Appendix \ref{Coulrenormexchtr1}. In Eq. (\ref{renormpoleexchtr1}) we took into account that for the pole and triangular exchange diagrams ${\tilde Z}^{'}={\tilde Z}$.  The same renormalization procedure can be applied for any $t_{\beta}$ and $t_{\alpha}$.
Approximation (\ref{renormpoleexchtr1})  will be used to calculate the angular distributions near the main stripping peak, where the pole neutron transfer mechanism gives a dominant contribution, and the results will be compared with the exact approach.

\section{Summary}
We have derived new generalized Faddeev equations in the AGS form taking into account the target excitations and explicitly include the Coulomb interactions. Applying two potential formula we convert the AGS equations to the form, in which the matrix elements are sandwiched by the Coulomb distorted waves in the initial and final states. To obtain the half-off-shell integral equations
we use an off-shell extension. The obtained equations are compact and can be solved. 
We present the final expressions for the modified AGS equations after the angular momentum decomposition. We show how to regularize the matrix elements sandwiched by the Coulomb distorted waves. Besides we investigate the off-shell Coulomb scattering amplitude in different kinematical regions. We also consider the Coulomb-modified form factors and show how to regularize them. After that we investigate the exchange triangular diagram and show that its strongest singularity is the pole of the neutron transfer pole diagram. The strongest singularity of the elastic scattering triangular diagram with Coulomb four-ray vertex is compensated by the subtracted channel Coulomb potential. Thus we have shown that the Coulomb interaction can be taken into account explicitly without Coulomb screening procedure. This will allow us to apply the Faddeev formalism for the analysis of the deuteron stripping on targets with higher charges, at which the Coulomb screening procedure doesn't work. For $NN$ and nucleon-target nuclear interactions we assume the separable potentials what significantly simplifies solution of the AGS equations. 

\appendix
\section{Partial Coulomb scattering wave function in the momentum space and matrix elements in the Coulomb distorted wave representation}
\label{parCoulscwf1}

Here we present the expression for the partial Coulomb scattering wave function in the momentum space. This wave function has singularity and we demonstrate how this singularity can be regularized when calculating the matrix elements, which are given by the sandwiching the diagrams, describing the reaction mechanisms, with the Coulomb scattering wave functions in the initial and/or final states. 

First we start from the definition of the Fourier transform of the Coulomb scattering wave function  \cite{nordsieck}:
\begin{align}
&\psi_{\rm{\bf p}}^{C(+)}({\rm {\bf p}}')= \lim\limits_{\epsilon \to +0}\,\int\,{\rm d}{\rm {\bf r}}\,e^{-\epsilon\,r}\,e^{-i\,{\rm {\bf p}}'\,\cdot {\rm {\bf r}}}\,\psi^{C(+)}_{\rm {\bf p}}({\rm {\bf r}})                                                            \nonumber\\
&= - 4\,\pi\,e^{-\pi\,\eta_{p}/2}\,\Gamma(1 + i\,\eta_{p})\,\lim\limits_{\epsilon \to +0}\,
\frac{{\rm d}}{{\rm d}\,\epsilon}\,\frac{[p'^{2} - (p+i\,\epsilon)^{2}]^{i\,\eta_{p}}}{[({\rm {\bf p}}' - {\rm {\bf p}})^{2} + \epsilon^{2}]^{1+i\,\eta}}                           \nonumber\\
&= 4\,\pi\,\sum\limits_{l\,m_{l}}\,Y^{*}_{lm_{l}}({\rm {\bf {\hat p}}})\,Y_{lm_{l}}({\rm {\bf {\hat p}}}')\,\psi^{C}_{p\,l}(p') = \sum\limits_{l}\,(2\,l+1)\,{\rm P}_{l}({\rm {\bf {\hat p}}} \cdot {\rm {\bf {\hat p}}}')\,\psi^{C}_{p\,l}(p').
\label{FourtransfCoulwf1}
\end{align}
Here $\eta_{p}$ is the Coulomb parameter of the interacting particles moving with the relative momentum $p$, $Y_{lm_{l}}({\rm {\bf {\hat p}}})$ is the spherical harmonic function, ${\rm P}_{l}({\rm {\bf {\hat p}}} \cdot {\rm {\bf {\hat p}}}')$ is the Legendre polynomial. 
We can see from Eq. (\ref{FourtransfCoulwf1}) that the Fourier transform of the Coulomb scattering wave function is a distribution.  
 
The expression for the partial Coulomb scattering wave function in the momentum space was found by one of us (A.M.M.) \cite{dol66}: 
\begin{align}
&\psi^{C}_{p\,l}(p')= -\frac{2\,\pi}{p'}\,e^{-\pi\,\eta_{p}/2
 }\,\Gamma(1+ i\,\eta_{p})\,e^{i\,\phi_{l}^{C}}\,\lim\limits_{\epsilon \to 0 }\,2\,{\rm Im}\Big[e^{-i\,\phi_{l}^{C}}\,\frac{(p'+ p + i\,\epsilon)^{-1+i\,\eta_{p}}}{(p' - p + i\,\epsilon)^{1+i\,\eta_{p}}}\,
{}_2{F}_{1}(-l,l+1;1-i\,\eta_{p}; -\frac{(p'-p)^{2}}{4\,p\,p'}) \Big],
\nonumber\\
\label{psiC1}
\end{align}
where $\phi_{l}^{C}= \sigma_{l}^{C} - \sigma_{0}^{C}\,$, $\,\sigma_{l}^{C}$ is the Coulomb scattering phase shift in the partial wave $l$, $\,\,\eta_{p}= Z_{1}\,Z_{2}\,e^{2}\,\mu_{12}/p\,$ 
is the Coulomb parameter for particles $1$ and $2$ moving with the relative momentum $p = \sqrt{2\,\mu_{12}\,E_{12}}$, 
$\,\,\,{}_2{F}_{1}(-l,l+1,;1-i\,\eta_{p}; -\frac{(p'-p)^{2}}{4\,p\,p'})$ is the hypergeometric function, which reduces to a polynomial of order $l$  in the plane $z= -\frac{(p'-p)^{2}}{4\,p\,p'}$. In particular, for $l=0\,\,$  $\,{}_2{F}_{1}(0,1,;1-i\,\eta_{p}; -\frac{(p'-p)^{2}}{4\,p\,p'})=1$.  

The function $\psi^{C}_{p\,l}(p')$ has singular branching points on the complex plane $p'$ at $p'= p \pm i\,\epsilon$ and $p'=-p \pm i\,\epsilon$. The small imaginary addition $\pm i\,\epsilon$ determines the rules for circuiting around the singularities when integrating. If the integral containing the Coulomb scattering wave functions is calculated in the analytic form, then no difficulties arise, because the presence of the imaginary addend $\pm i\,\epsilon$ shifts the singularities from the integration contour to the complex plane and $\lim\limits_{\epsilon \to 0}$ can be easily taken after carrying out all the integrations. Such a procedure, however, would be highly inconvenient in numerical calculations, for in this case it would be necessary to calculate the integrals for several continuously decreasing values of $\epsilon$ in order to attain a needed accuracy. This procedure, owe to the presence of the singularity of the integrand, may become very unstable and even not converging when $\eta_{p}$ increases.

This difficulty can be readily circumvented if, putting $\epsilon >0$, we regularize the initial integral.
Then the result of the integration will be stable when integration is performed for $\lim\limits_{\epsilon \to 0}$. The regularization method is taken from \cite{gelfand}. To explain the Gel'fand-Shilov method we 
consider the integral
\begin{align}
J(\lambda) = \lim\limits_{\epsilon \to +0}\,\int\limits_{a}^{b}\,{\rm d}x\,f(x)\,
(x+ i\,\epsilon)^{\lambda},
\label{Jint1}
\end{align}
where $b>0$ and $a \leq 0$,  $\,\,{\rm Re} \lambda=-1,\,\,\,{\rm Im}\,\lambda \not=0$ and  $\,\,\,f(0) \not= 0$.
To regularize this integral we assume that $-1 < {\rm Re}\,\lambda$ and $\epsilon >0$. Subtracting $f(0)$ from $f(x)$ and adding it, we can rewrite integral (\ref{Jint1}) as
\begin{align}
&{\tilde J}(\lambda) = \lim\limits_{\epsilon \to +0}\,\big[\int\limits_{0}^{b}\,{\rm d}x\,[f(x) - f(0)]\,(x+ i\,\epsilon)^{\lambda}\big] +  f(0)\,\frac{(b+ i\,0)^{\lambda +1} - (a+i\,0)^{\lambda +1}}{\lambda +1}.
\label{Jint2}
\end{align}
${\tilde J}(\lambda)$ is analytical function of the parameter $\lambda$ in the domain $-2< {\rm Re}\,\lambda$
and $\lambda \not=-1$.  Because $J(\lambda)$ and ${\tilde J}(\lambda)$ coincide in the region $-1 <{\rm Re}\,\lambda$, $\,\,\,{\tilde J}(\lambda)$ is an analytical continuation of $J(\lambda)$ into the domain $-2 < {\rm Re}\,\lambda$. The integrand in (\ref{Jint2}) does not have diverging singularity and the integral can be calculated; the additional term containing $f(0)$ is also not singular because $\lambda = -1+ i\,\eta$.
Eq. (\ref{Jint2}) can be considered as generalization of the famous equation for the integral containing pole singularity: $\int\limits_{a}^{b}\,{\rm d}x\,f(x)/(x- x_{0} - i\,0)= P\,\int\limits_{a}^{b}\,{\rm d}x\,f(x)/(x- x_{0}) + i\,\pi\,f(x_{0})$, where $P$ stands for the Cauchy principal value of the integral. 
By subtracting and adding ${\rm d}f(x)/{\rm d}x|_{x=0}\,x$ in the integral (\ref{Jint2}) we can continue analytically (\ref{Jint1}) into the region $- 3 < {\rm Re}\,\lambda$. Moreover, in this case the integrand vanishes at $x=0$. 

Using the explained regularization we show how it works in practice. As example we consider
\begin{align}
&Z^{SC}_{0}({p}_{\beta},\,{p}^{C}_{{\alpha}\,0}) = \int\limits_{0}^{\infty}\frac{{\rm d}p\,{p}^{2}}{2\,\pi^{2}}\,\,Z_{0}({p}_{\beta},\,p)\,\psi^{C}_{p_{\alpha}\,0}(p).
\label{Zi0ex1}
\end{align}
Here subscript $0$ denotes $l=0$ partial wave. Note that $\psi^{C}_{p_{\alpha}\,l}(p)$ contains $_2{F_1}( - l,l + 1;1 - i\,\eta _{p}; - \frac{{{{(p - {p_{\alpha}})}^2}}}{4\,p\,p_{\alpha}})$, which is expressed in terms the polynomial of $(p-p_{\alpha})^{2n}$ with $0 \leq n \leq l$. Hence the terms with $n \geq 1$ don't require the regularization at the singular point $p = p_{\alpha}$ and the only singular term, which requires regularization, is the one with $n=0$. That is why it is enough to demonstrate how the regularization works for $l=0$. 

To demonstrate the regularization we consider the effective potential $Z({\rm{\bf p}}_{\beta},\,{\rm {\bf p}})$ given by a simple pole propagator:
\begin{align}
Z({\rm{\bf p}}_{\beta},\,{\rm {\bf p}})= -\frac{2\,\mu_{12}\,Z_{1}\,Z_{2}\,e^{2}}{({\rm {\bf p}}_{\beta} - {\rm {\bf p}})^{2} + \kappa^{2}},
\label{polediagram1}
\end{align}
where $\mu_{12}$ is the reduced mass and $Z_{1}\,Z_{2}\,e^{2}$ is the product of charges of the particles $1$ and $2$.
The partial wave pole amplitude at $l=0$ is given by
\begin{align}
Z_{0}(p_{\beta},\,p)= -\,\mu_{12}\,Z_{1}\,Z_{2}\,e^{2}\,\int\limits_{-1}^{1}\,{\rm d}x\,\frac{P_{0}(x)}{({\rm {\bf p}}_{\beta} - {\rm {\bf p}})^{2} + \kappa^{2}}.
= -\frac{\,\mu_{12}\,Z_{1}\,Z_{2}\,e^{2}}{2\,p_{\beta}\,p}\,Q_{0}(\xi),
\label{partwaveexp1}
\end{align}                       
$\xi= (p_{\beta}^{2} + p^{2} +  \kappa^{2})/(2\,p_{\beta}\,p)$.
From equation \cite{nordsieck}
\begin{align} 
Z^{SC}({\rm {\bf p}}_{\beta},{\rm {\bf p}}_{\alpha})= \int\,\frac{{\rm d}{\rm {\bf p}}}{(2\,\pi)^{3}}\,Z({\rm{\bf p}}_{\beta},\,{\rm {\bf p}})\,\psi^{C}_{{\rm {\bf p}}_{\alpha}}({\rm {\bf p}})\, 
= -2\,\mu_{12}\,Z_{1}\,Z_{2}\,e^{2}\,e^{-\pi\,\eta_{\alpha}/2}\,\Gamma(1+ i\,\eta_{\alpha})\,\frac{[p_{\beta}^{2} - (p_{\alpha} + i\,\kappa)^{2}]^{i\,\eta_{\alpha}}}{[{\rm {\bf p}}_{\beta} - {\rm {\bf p}}_{\alpha})^{2} + \kappa^{2}]^{1+ i\,\eta_{\alpha}}},
\label{analexpr1}
\end{align}
where $\eta_{\alpha}= Z_{1}\,Z_{2}\,e^{2}\,\mu_{12}/p_{\alpha}$ is the Coulomb parameter associated with the momentum $p_{\alpha}$, which is the on-shell momentum. 
From Eq. (\ref{analexpr1}) using the partial wave expansion \cite{dol66} we get 
\begin{align}
Z^{SC}({\rm {\bf p}}_{\beta},{\rm {\bf p}}_{\alpha})= \sum\limits_{l=0}^{\infty}\,(2\,l+1)\,
P_{l}({\rm {\bf {\hat p}}}_{\beta} \cdot {\rm {\bf {\hat p}}}_{\alpha})\,Z^{SC}_{l}({p}_{\beta},\,{p}^{C}_{{\alpha}\,l}),
\label{Lpartexp1}
\end{align}
\begin{align}
&Z^{SC}_{l}({p}_{\beta},\,{p}^{C}_{{\alpha}\,l})= \int\limits_{0}^{\infty}\frac{{\rm d}p\,{p}^{2}}{2\,\pi^{2}}\,\,Z_{l}({p}_{\beta},\,p)\,\psi^{C}_{p_{\alpha\,l}}(p)    \nonumber\\
&= -\frac{1}{p_{\beta}}\,e^{-\pi\,\eta_{\alpha}/2}\,\Gamma(1+ i\,\eta_{\alpha})\,e^{i\,\phi_{l}^{C}}\,{\rm Im}\Big[e^{-i\,\phi_{l}^{C}}\Big(\frac{p_{\beta} + p_{\alpha} + i\,\kappa}{p_{\beta} - p_{\alpha} + i\,\kappa}\Big)^{i\,\eta_{\alpha}}\,{}_{2}F_{1}\big(-l,l+1;1-i\,\eta_{\alpha};
-\frac{(p_{\beta} - p_{\alpha})^{2} + \kappa^{2}}{4\,p_{\beta}\,p_{\alpha}}\big)\Big]
\label{Ll2}
\end{align} 
and for $l=0$
\begin{align}
&Z^{SC}_{0}({p}_{\beta},\,{p}^{C}_{{\alpha}\,0})  
= -\frac{1}{p_{\beta}}e^{-\pi\,\eta_{\alpha}/2}\,\Gamma(1+ i\,\eta_{\alpha})\,2\,{\rm Im}\Big[\Big(\frac{p_{\beta} + p_{\alpha} + i\,\kappa}{p_{\beta} - p_{\alpha} + i\,\kappa}\Big)^{i\,\eta_{\alpha}}\,\Big].
\label{Ll10}
\end{align} 
Evidently that analytical expression (\ref{Ll2}) for $Z^{SC}_{l}({p}_{\beta},\,{p}^{C}_{{\alpha}\,l})$, obtained owe to the very simplified approximation for the effective potential $Z$, can be easily calculated.
However, in general case of folding of the partial Coulomb scattering wave functions with the effective potentials there are no analytical expressions for $Z^{SC}_{0}({p}_{\beta},\,{p}^{C}_{{\alpha}\,0})$. In this case the integral must be calculated numerically and regularization of the integrand is required. To show how regularization works we compare the results of the numerical calculation of $Z^{SC}_{0}({p}_{\beta},\,{p}^{C}_{{\alpha}\,0})$ given by the integral representation (\ref{Zi0ex1}) with the analytical expression (\ref{Ll10}).

First we present the regularized integral following the Gel'fand-Shilov method described above:
\begin{align}
&Z^{SC}_{0}({p}_{\beta},\,{p}^{C}_{{\alpha}\,0}) = 
\int\limits_{0}^{p_{1}}\frac{{\rm d}p\,{p}^{2}}{2\,\pi^{2}}\,\,Z_{0}({p}_{\beta},\,p)\,\psi^{C}_{p_{\alpha}\,0}(p) + \frac{\eta_{\alpha}\,p_{\alpha}}{\pi\,p_{\beta}}\,{\rm Im}\,\Bigg\{\int\limits_{p_{\alpha}- \Delta}^{p_{\alpha} + \Delta}\,{\rm d}p\,\frac{FR(p) - FR(p_{\alpha}) - FR'(p_{\alpha})\,(p - p_{\alpha})}{(p - p_{\alpha} + i\,\epsilon)^{1+ i\,\eta_{\alpha}}}            \nonumber\\
& + i\frac{FR(p_{\alpha})}{\eta_{\alpha}}\Big(\frac{1}{(\Delta + i\,\epsilon)^{i\,\eta_{\alpha}}} - \frac{1}{(-\Delta+ i\,\epsilon)^{i\,\eta_{\alpha}}}\Big) + \frac{FR'(p_{\alpha})}{1-i\,\eta_{\alpha}}\,\Big( (\Delta+ i\,\epsilon)^{1-i\,\eta_{\alpha}} - (-\Delta + i\,\epsilon)^{1-i\,\eta_{\alpha}}\Big)\Bigg\} \nonumber\\
&+\,\int\limits_{p_{\alpha}+ \Delta}^{\infty}\,\frac{{\rm d}p\,{p}^{2}}{2\,\pi^{2}}\,\,Z_{0}({p}_{\beta},\,p)\,\psi^{C}_{p_{\alpha}\,0}(p).
\label{Llregularized1}
\end{align} 
Here we use regularization only in the proximity of the singular point $p=p_{\alpha}$. To do it we split the integral into three terms: the integral from $0$ to $p_{\alpha} - \Delta$, from $p_{\alpha} - \Delta$ to 
$p_{\alpha} + \Delta$ and from $p_{\alpha} + \Delta$ to infinity. The regularization is required only in the integral from $p_{\alpha} - \Delta$ to $p_{\alpha} + \Delta$, where
\begin{align}
FR(p)= \frac{Q_{0}(\xi)}{(p+ p_{\alpha})^{1-i\,\eta_{\alpha}}}
\label{FRp1}
\end{align}
and $FR'(p_{\alpha})= {\rm d}\,FR(p)/{\rm d}p\Big|_{p=p_{\alpha}}$. Note that here we use regularization procedure subtracting and adding $FR(p_{\alpha}) + FR'(p_{\alpha})\,(p - p_{\alpha})$. In this case the integrand in the regularized integral just vanishes at the singular point. 

\begin{figure}
[tbp] 
\epsfig{file=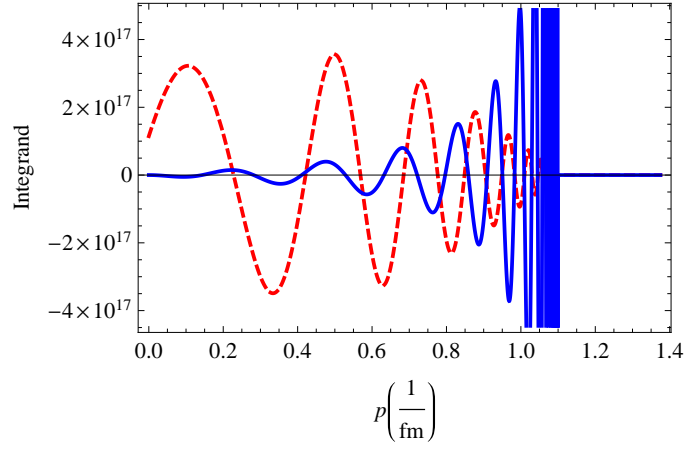,width=9cm}
\caption{(Color online) Comparison of the regularized integrand (dashed red line) in Eq. (\ref{Llregularized1}) with the unregularized one (solid blue line) as function of the integration momentum $p$ for $p \leq p_{\alpha},$  $\,\eta_{\alpha}=12,$ $\,\kappa=0.1$ fm${}^{-1},\,$ $\,\,p_{\alpha}=1,1$ fm${}^{-1},\,$ $p_{\beta}=1.2$ fm${}^{-1}$ and $\,\,\Delta= 0.25\,p_{\alpha}$.}
\label{fig_Integrand1}
\end{figure}

\begin{figure}
[tbp] 
\epsfig{file=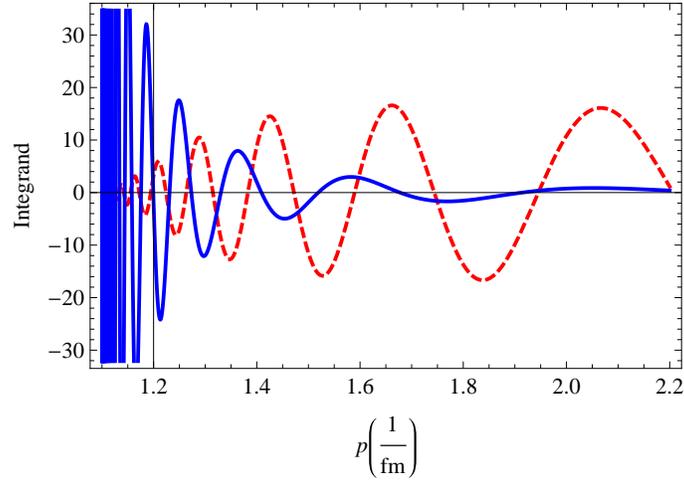,width=9cm}
\caption{(Color online)  The same as in Fig. \ref{fig_Integrand1} but for $p \geq p_{\alpha},\,$.} 
\label{fig_Integrand2}
\end{figure}

To demonstrate how regularization procedure works we performed calculations of analytical Eq. (\ref{Ll10}) 
and regularized Eq. (\ref{Llregularized1}) for parameters given in the captions to Fig. \ref{fig_Integrand1}.
We have chosen the Coulomb parameter $\eta_{\alpha}= 12$, which corresponds to the proton collision with  charge $92$ (uranium) at the relative kinetic energy $\approx 1.5$ MeV. No Coulomb screening procedure would work at such high Coulomb parameter. Results of our calculations are:  $Z^{SC}_{0}({p}_{\beta},\,{p}^{C}_{{\alpha}\,0})$ from Eq. (\ref{Ll10}) gives $-2.009 \times 10^{-12}$ while the regularized Eq. (\ref{Llregularized1}) results in $-2.005 \times 10^{-12}$. In Figs \ref{fig_Integrand1} and \ref{fig_Integrand2}
we demonstrate the behavior of the regularized integrand versus unregularized one. As we can see, the regularization completely change the behavior of the integrand making it possible to perform calculations with singular partial wave Coulomb scattering wave functions in the momentum space.

The same procedure can be used to regularize matrix elements with the Coulomb scattering wave functions in the initial and final states. Succeeding in regularization of the integrals containing the AGS effective potentials sandwiched by the partial Coulomb scattering wave functions, we may conclude that the effective potentials in the AGS equations in the Coulomb distorted wave representation can be calculated without using Coulomb screening procedure. It is the most important result of our regularization procedure.

\section{Off-shell Coulomb scattering amplitude}
\label{OffshTC1}

In this Appendix using the integral representation of the off-shell Coulomb scattering amplitude
we will derive the expressions for this amplitude, which can be used in practical calculations in different kinematical regions. 
We start from the standard expression for the off-shell Coulomb scattering amplitude $T_{\alpha}^{C}({\rm {\bf k}}'_{\alpha},{\rm {\bf k}}_{\alpha},{\hat E}_{\alpha})$ \cite{okubo,schwinger,dol66}:
\begin{align}
T_{\alpha}^{C}({\rm {\bf k}}'_{\alpha},{\rm {\bf k}}_{\alpha},{\hat E}_{\alpha})= 4\,\pi\,
Z_{\beta}\,Z_{\gamma}\,e^{2}\Big[\frac{1}{({\rm {\bf k}}'_{\alpha} - {\rm {\bf k}}_{\alpha})^{2}} - i\,{\hat \eta}_{\alpha}^{0}\,
I({\rm {\bf k}}'_{\alpha},\,{\rm {\bf k}}_{\alpha}; {\hat E}_{\alpha}) \Big],
\label{Talpha1}
\end{align}
\begin{align}
I({\rm {\bf k}}'_{\alpha},\,{\rm {\bf k}}_{\alpha}; {\hat E}_{\alpha}) = \lim\limits_{\epsilon \to 0}\, \int\limits_{0}^{1}\,{\rm d}x\,x^{i\,{\hat \eta}_{\alpha}^{0}}\,\frac{1}{x\,({\rm {\bf k}}'_{\alpha} - {\rm {\bf k}}_{\alpha})^{2} - 
\frac{\mu_{\alpha}}{2\,{\hat E}_{\alpha}}\,\big[{\hat E}_{\alpha} +i\,0 - {k'}^{2}_{\alpha}/(2\,\mu_{\alpha})\big]\,\big[{\hat E}_{\alpha} + i\,0 - k_{\alpha}^{2}/(2\,\mu_{\alpha})\big](1-x)^{2}},
\label{Iint1}
\end{align}
${\rm {\bf k}}_{\alpha}$ and ${\rm {\bf k}}'_{\alpha}$ are the relative off-shell momenta of particles $\beta$ and $\gamma$ before and after scattering moving with the relative kinetic energy ${\hat E}_{\alpha}$, $k_{\alpha}^{0}= \sqrt{2\,\mu_{\alpha}\,{\hat E}_{\alpha}}$. To distinguish the Coulomb parameter of pair $\alpha$ (between particles $\beta$ and $\gamma$) from the Coulomb parameter $\eta_{\alpha}$
in the channel $\alpha$ (between particle $\alpha$ and the bound state of the pair $(\beta \gamma)$) the former is denoted by ${\hat \eta}_{\alpha}$. To underscore that this Coulomb parameter is calculated at $k_{\alpha}^{0}$ we use notation  $\,\,\,{\hat \eta}_{\alpha}^{0}= Z_{\beta}\,Z_{\gamma}\,e^{2}\,\mu_{\alpha}/k_{\alpha}^{0}$. 
In the case under consideration $T_{\alpha}^{C}({\rm {\bf k}}'_{\alpha},{\rm {\bf k}}_{\alpha},{\hat E}_{\alpha}) \not=0$ only for $\alpha=3$.  

As we can see $T_{\alpha}^{C}({\rm {\bf k}}'_{\alpha},{\rm {\bf k}}_{\alpha},{\hat E}_{\alpha})$ 
has the forward singularity at 
\begin{align}
{\rm {\bf {\hat \Delta}}}_{\alpha}= {\rm {\bf k}}'_{\alpha} - {\rm {\bf k}}_{\alpha}=0
\label{Coulforwsing1}
\end{align}
generated by the Coulomb Born term $4\,\pi\,
Z_{\beta}\,Z_{\gamma}\,e^{2}/{{\rm {\bf {\hat \Delta}}}_{\alpha}^{2}}$. This singularity appearing in the triangular elastic scattering amplitude, see Figs \ref{triang_exch} and \ref{triang_exch1}, is dangerous when coinciding with the two-body Green function's singularity. Just to remove this singularity, we added and subtracted the Coulomb channel potentials in the initial channel $\alpha$ and the final channel $\beta$. Applying after that the two-potential equation we obtained the AGS equations in the Coulomb distorted wave representation.  
At ${\rm {\bf {\hat \Delta}}}_{\alpha} \to 0$ $\,\,\,I({\rm {\bf k}}'_{\alpha},\,{\rm {\bf k}}_{\alpha}; {\hat E}_{\alpha})\,\sim 1/|{\rm {\bf {\hat \Delta}}}_{\alpha}|$, that is less singular than the Born term. 

It is worth mentioning that the off-shell Coulomb scattering amplitude doesn't have a definite on-shell limit reflecting the fact that, owe to the infinite range of the Coulomb interaction, charged particles are not free even when the distance between them increases to infinity. 
Assuming that ${k}^{2}_{\alpha}/(2\,\mu_{\alpha}) - {\hat E}_{\alpha} \to 0$ we obtain (the on-shell limit in the entry channel)
\begin{align}
T_{\alpha}^{C}({\rm {\bf k}}'_{\alpha},{\rm {\bf k}}_{\alpha},{\hat E}_{\alpha})\stackrel{k_{\alpha} \to k_{\alpha}^{0}}{=} r_{\alpha}^{C}(k_{\alpha},\,k_{\alpha}^{0})\,
T_{\alpha}^{C(HSH)}({\rm {\bf k}}'_{\alpha},{\rm {\bf k}}_{\alpha}^{0},{\hat E}_{\alpha}), 
\label{halfofshell1}
\end{align}
where the so-called Coulomb renormalization factor
\begin{align}
r_{\alpha}^{C}(k_{\alpha},\,k_{\alpha}^{0})= e^{\pi\,{\hat \eta}_{\alpha}^{0}/2}\,\Gamma(1-i\,{\hat \eta}_{\alpha}^{0})\,\Big(\frac{{k^{0}}^{2}_{\alpha} - k_{\alpha}^{2}}{4\,k_{\alpha}^{2}}\Big)^{i\,{\hat \eta}_{\alpha}^{0}}
\label{Coulrenfctr1}
\end{align}
and the half-off-shell Coulomb scattering amplitude
\begin{align}
T_{\alpha}^{C(HSH)}({\rm {\bf k}}'_{\alpha},{\rm {\bf k}}_{\alpha}^{0},{\hat E}_{\alpha})
= 4\,\pi\,
Z_{\beta}\,Z_{\gamma}\,e^{2}\,e^{-\pi\,{\hat \eta}_{\alpha}^{0}/2}\,\Gamma(1+i\,{\hat \eta}_{\alpha}^{0})\,
\frac{[{k'}^{2}_{\alpha} - (k_{\alpha}^{0} + i\,\epsilon)^{2}]^{i\,{\hat \eta}_{\alpha}^{0}}} {[({\rm {\bf k}}'_{\alpha} - {\rm {\bf k}}_{\alpha}^{0})^{2} + \epsilon^{2}]^{1+ i\,{\hat \eta}_{\alpha}^{0}}}, \qquad \epsilon \to 0.
\label{halfoffshCoulampl1}
\end{align}
Similar equation takes place when $k'_{\alpha} \to k_{\alpha}^{0}$, $\,\,
k_{\alpha}' \not= k_{\alpha}$. 
Taking simultaneous limit $k_{\alpha},\,k'_{\alpha} \to k_{\alpha}^{0}$ we obtain the on-shell Coulomb scattering amplitude. Infinitesimal addend $\epsilon$ is required to correctly bypass 
singularities and for regularization of the Coulomb scattering amplitude. In what follows we assume that energy ${\hat E}_{\alpha}$ has always positive imaginary addend $i\,\epsilon$ with $\epsilon \to 0$, that is we replace ${\hat E}_{\alpha}$ by ${\hat E}_{\alpha\,\epsilon}= {\hat E}_{\alpha} + i\,\epsilon$.  

Let us introduce 
\begin{align}
b_{\epsilon}= \frac{\mu_{\alpha}}{2\,{\hat E}_{\alpha\,\epsilon}}\,\big[{\hat E}_{\alpha\,\epsilon} - {k'}^{2}_{\alpha}/(2\,\mu_{\alpha})\big]\,\big[{\hat E}_{\alpha\,\epsilon} - k_{\alpha}^{2}/(2\,\mu_{\alpha})\big],
\label{b1}
\end{align}
with $b = \lim\limits_{\epsilon \to 0}\,b_{\epsilon}$, and
\begin{align}
{\hat \Delta}_{\alpha\,\epsilon}^{2}= ({\rm {\bf k}}'_{\alpha} - {\rm {\bf k}}_{\alpha})^{2} + \epsilon^{2}.
\label{Deltaalphaeps1}
\end{align}
Then
\begin{align}
T_{\alpha}^{C}({\rm {\bf k}}'_{\alpha},{\rm {\bf k}}_{\alpha},{\hat E}_{\alpha})= \lim\limits_{\epsilon \to 0}\, T_{\alpha\,\epsilon}^{C}({\rm {\bf k}}'_{\alpha},{\rm {\bf k}}_{\alpha},{\hat E}_{\alpha\,\epsilon}),
\label{TTeps1}
\end{align}
where the regularized off-shell Coulomb scattering amplitude is
\begin{align}
T_{\alpha\,\epsilon}({\rm {\bf k}}'_{\alpha},\,{\rm {\bf k}}_{\alpha}; {\hat E}_{\alpha\,\epsilon}) = 
4\,\pi\,Z_{\beta}\,Z_{\gamma}\,e^{2}\Big[\frac{1}{{\hat \Delta}_{\alpha\,\epsilon}^{2}} - i\,{\hat \eta}_{\alpha}\,
I_{\epsilon}({\rm {\bf k}}'_{\alpha},\,{\rm {\bf k}}_{\alpha}; {\hat E}_{\alpha\,\epsilon}) \Big],
\label{Tinteps1}
\end{align}
\begin{align}
I_{\epsilon}=\int\limits_{0}^{1}\,{\rm d}x\,x^{i\,{\hat \eta}_{\alpha\,\epsilon}}\,\frac{1}{x\,{\hat \Delta}_{\alpha\,\epsilon}^{2} - 
b_{\epsilon}\,(1-x)^{2}},
\label{Iinteps1}
\end{align}
${\hat \eta}_{\alpha\,\epsilon}^{0}= Z_{\beta}\,Z_{\gamma}\,e^{2}\,\mu_{\alpha}/\sqrt{2\,\mu_{\alpha}\,{\hat E}_{\alpha\,\epsilon}}$. 

Usage of the regularized Coulomb scattering amplitude allows us to carry out all the calculations
 with different $\epsilon \to 0$ without any problems because the AGS equations are compact at $\epsilon=0$ \cite{muk2000,muk2001}.  

Equation (\ref{Tinteps1}) is not always the most convenient one for practical applications and below we present alternative equations for $T_{\alpha\,\epsilon}^{C}$, which can be used in practical applications depending on the value of ${\hat E}_{\alpha}$.\\
(i) ${\hat E}_{\alpha} < 0$.\\
In this case always $b<0$, that is  $x\,{\hat \Delta}_{\alpha\,\epsilon}^{2} - 
\frac{\mu_{\alpha}}{2\,{\hat E}_{\alpha}}\,({\hat E}_{\alpha} - {k'}^{2}_{\alpha}/(2\,\mu_{\alpha}))({\hat E}_{\alpha} - k_{\alpha}^{2}/(2\,\mu_{\alpha}))(1-x)^{2} > 0$ and $T_{\alpha\,\epsilon}^{C}$ is regular. Note that at ${\hat E}_{\alpha} < 0$, owe to the positive imaginary addend, ${\rm arg} ({\hat E}_{\alpha} + i\,\epsilon) = \pi$ and
$\,\lim\limits_{\epsilon \to 0}\,{\hat \eta}_{\alpha\,\epsilon}^{0}= Z_{\beta}\,Z_{\gamma}\,e^{2}\,\mu_{\alpha}/\sqrt{2\,\mu_{\alpha}\,|{\hat E}_{\alpha}|}\,e^{-i\,\pi/2}$ and $\lim\limits_{\epsilon \to 0}\,x^{i\,{\hat \eta}_{\alpha}^{0}}= x^{Z_{\beta}\,Z_{\gamma}\,e^{2}\,\mu_{\alpha}/\sqrt{2\,\mu_{\alpha}\,|{\hat E}_{\alpha}|}} \geq 0$. \\
(ii)  $b<0$ but ${\hat E}_{\alpha} >0$. In this case also no singularities appear in the integrand of the regularized integral over $x$.  

In general it is more convenient for $b <0$ to use an alternative expression for $T_{\alpha\,\epsilon}^{C}$: 
\begin{align}
&T_{\alpha\,\epsilon}({\rm {\bf k}}'_{\alpha},\,{\rm {\bf k}}_{\alpha}; {\hat E}_{\alpha\,\epsilon}) = 
4\,\pi\,Z_{\beta}\,Z_{\gamma}\,e^{2}\,\frac{\mu_{\alpha}}{2\,{\hat E}_{\alpha\,\epsilon}}\,({\hat E}_{\alpha\,\epsilon} - \frac{ {k'}^{2}_{\alpha} }{ 2\,\mu_{\alpha} })({\hat E}_{\alpha\,\epsilon} -\frac{k_{\alpha}^{2}}{2\,\mu_{\alpha}}) \nonumber\\
&\times \int\limits_{0}^{1}\,{\rm d}x\,x^{i\,{\hat \eta}_{\alpha\,\epsilon}^{0}}\,\frac{(1-x^{2})}{\Big[x\,{\hat \Delta}_{\alpha\,\epsilon}^{2} + \Big|\frac{\mu_{\alpha}}{2\,{\hat E}_{\alpha\,\epsilon}}\,({\hat E}_{\alpha\,\epsilon} - {k'}^{2}_{\alpha}/(2\,\mu_{\alpha}))({\hat E}_{\alpha\,\epsilon} - k_{\alpha}^{2}/(2\,\mu_{\alpha}))\Big|(1-x)^{2}\Big]^{2}}.
\label{Talteps11}
\end{align}
There are no problems in using this explicit equation for $T_{\alpha\,\epsilon}({\rm {\bf k}}'_{\alpha},\,{\rm {\bf k}}_{\alpha}; {\hat E}_{\alpha\,\epsilon})$ in the Coulomb-modified form factors and in the exchange triangular diagram. A strong singularity of $\lim\limits_{\epsilon \to 0}\,T_{\alpha\,\epsilon}({\rm {\bf k}}'_{\alpha},\,{\rm {\bf k}}_{\alpha}; {\hat E}_{\alpha\,\epsilon})$ in the elastic scattering triangular diagram,
as has been explained, is compensated by subtracting the channel Coulomb potential. \\

\begin{figure}
%[tbp] 
\epsfig{file=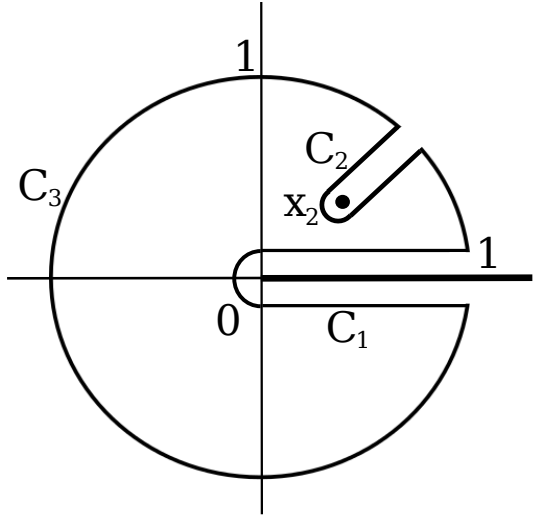,width=7cm}
\caption{
 Integration contour $C= C_{1} + C_{2} + C_{3}$.} 
\label{fig_contour}
\end{figure}

(iii) $b>0$ (it can be only at ${\hat E}_{\alpha} >0$). In this case the integrand in Eq. (\ref{Iinteps1}) has a singularity (zero of the denominator):
\begin{align}
x\,{\hat \Delta}_{\alpha}^{2} - \frac{\mu_{\alpha}}{2\,{\hat E}_{\alpha}}\,({\hat E}_{\alpha} - \frac{ {k'}^{2}_{\alpha} }{ 2\,\mu_{\alpha} })({\hat E}_{\alpha} -\frac{k_{\alpha}^{2}}{2\,\mu_{\alpha}})\,(1-x)^{2}=\,0.
\label{zerodenominator1}
\end{align}
The roots of this equation are:
\begin{align}
x_{1,2}= 1+ \frac{{\hat \Delta}_{\alpha}^{2}}{2\,b} \pm \frac{{\hat \Delta}_{\alpha}^{2}}{2\,b}\,\sqrt{1+ \frac{4\,b}{{\hat \Delta}_{\alpha}^{2}}}.
\label{roots12}
\end{align}
The first root $x_{1}> 1$ lies outside of the integration contour over $x$ in (\ref{Iinteps1}) at ${\hat \Delta}_{\alpha} >0$, while the second one $x_{2} <\,1$ lies on the integration contour  at ${\hat \Delta}_{\alpha} >0$. The integrand in (\ref{Iinteps1}) has a cut connecting the branching point singularities at $x=0$ and $x= \infty$ of the function $x^{i\,{\hat \eta}_{\alpha}^{0}}$. 
Hence we can rewrite $I_{\epsilon}({\rm {\bf k}}'_{\alpha},\,{\rm {\bf k}}_{\alpha}; {\hat E}_{\alpha\,\epsilon})$ as 
\begin{align}
I_{\epsilon}({\rm {\bf k}}'_{\alpha},\,{\rm {\bf k}}_{\alpha}; {\hat E}_{\alpha\,\epsilon}) = 
\frac{1}{1- e^{-2\,\pi{\hat \eta}_{\alpha\,\epsilon}^{0}}}\,
\oint\limits_{{C_1}} {dx} {x^{i{\kern 1pt} {{\hat \eta} _{\alpha\,\epsilon}^{0} }}}{\kern 1pt} \frac{1}{{x{\kern 1pt} {\hat \Delta} _{\alpha\,\epsilon}^2 - \frac{{{\mu _\alpha }}}{{2{\kern 1pt} {{\hat E}_{\alpha\,\epsilon} }}}{\kern 1pt} ({{\hat E}_{\alpha\,\epsilon} } - \frac{{{k'}^2_\alpha}}{{2{\kern 1pt} {\mu _\alpha }}})({{\hat E}_{\alpha\,\epsilon} } - \frac{{k_\alpha ^2}}{{2{\kern 1pt} {\mu _\alpha }}}){{(1 - x)}^2}}},
\label{C11}
\end{align}
where the integral is taken along the contour $C_{1}$, which starts at $x=1$, encircles the branching singularity at $x=0$ and ends at $x=1$. We assume that the root $x_{2}$ has a positive infinitesimal imaginary addend, which shifts the root from the integration contour to the first quadrant. Now we consider the integral along the closed contour $C= C_{1} + C_{2} + C_{3}$ shown in Fig. \ref{fig_contour}. 

Because there are no singularities of the integrand inside and on the integration contour $C$, according to Caushy's theorem the integral taken along $C$ vanishes. Hence, 
\begin{align}
&I_{\epsilon}({\rm {\bf k}}'_{\alpha},\,{\rm {\bf k}}_{\alpha}; {\hat E}_{\alpha\,\epsilon}) 
= -\frac{1}{1- e^{-2\,\pi{\hat \eta}_{\alpha\,\epsilon}^{0}}}\,
\Big[ \oint\limits_{{C_{2}}}\,{\rm d}x\,x^{i\,{\hat \eta}_{\alpha\,\epsilon}^{0}}\,\frac{1}{x\,{\hat \Delta}_{\alpha\,\epsilon}^{2} - 
\frac{\mu_{\alpha}}{2\,{\hat E}_{\alpha\,\epsilon}}\,({{\hat E}_{\alpha\,\epsilon} } - \frac{{{k'}^2_\alpha}}{{2{\kern 1pt} {\mu _\alpha }}})({{\hat E}_{\alpha\,\epsilon} } - \frac{{k_\alpha ^2}}{{2{\kern 1pt} {\mu _\alpha }}})(1-x)^{2}}                                            \nonumber\\
&+ \oint\limits_{{C_{3}}}\,{\rm d}x\,x^{i\,{\hat \eta}_{\alpha\,\epsilon}^{0}}\,\frac{1}{x\,{\hat \Delta}_{\alpha\,\epsilon}^{2} - 
\frac{\mu_{\alpha}}{2\,{\hat E}_{\alpha\,\epsilon}}\,({{\hat E}_{\alpha\,\epsilon} } - \frac{{{k'}^2_\alpha}}{{2{\kern 1pt} {\mu _\alpha }}})({{\hat E}_{\alpha\,\epsilon} } - \frac{{k_\alpha ^2}}{{2{\kern 1pt} {\mu _\alpha }}})(1-x)^{2}}  \Big].
\label{C12}
\end{align}
The integral over $C_{2}$ encircles the pole at $x=x_{2}$ and can be easily taken using Caushy's theorem. The integral over contour $C_{3}$ is taken along the circumference with the radius $|x|=1$.
Using substitution $x= e^{i\,\theta}$ we can simplify the integral over the contour $C_{3}$.
Then the expression for $T_{\alpha\,\epsilon}({\rm {\bf k}}'_{\alpha},\,{\rm {\bf k}}_{\alpha}; {\hat E}_{\alpha\,\epsilon})$ reduces to
\begin{align}
&T_{\alpha\,\epsilon}({\rm {\bf k}}'_{\alpha},\,{\rm {\bf k}}_{\alpha}; {\hat E}_{\alpha\,\epsilon}) =
4\,\pi\,Z_{\beta}\,Z_{\gamma}\,e^{2}\,\Big[\frac{1}{{\hat \Delta}_{\alpha\,\epsilon}^{2}}   
+ \frac{2\,\pi{\hat \eta}_{\alpha\,\epsilon}^{0}}{1- e^{-2\,\pi{\hat \eta}_{\alpha\,\epsilon}^{0}}}\,\Big(\frac{4\,b_{\epsilon}}{\sqrt{{\hat \Delta}_{\alpha\,\epsilon}^{2} + 4\,b_{\epsilon}}    +  {\hat \Delta}_{\alpha\,\epsilon})^{2}}\Big)^{i\,{\hat \eta}_{\alpha\,\epsilon}^{0}}  \nonumber\\
&- \frac{{\hat \eta}_{\alpha\,\epsilon}^{0}}{1- e^{-2\,\pi{\hat \eta}_{\alpha\,\epsilon}^{0}}}\,\int\limits_{0}^{2\,\pi}\,{\rm d}\,\theta\,\frac{e^{-{\hat \eta}_{\alpha\,\epsilon}^{0}\,\theta}}{{\hat \Delta}_{\alpha\,\epsilon}^{2} + 2\,b_{\epsilon}(1- \cos\,\theta)} \Big].
\label{Talphabpos11}
\end{align}
The integrand in the integral is free of singularities and Eq. (\ref{Talphabpos11}) can be used 
to calculate $T_{\alpha\,\epsilon}({\rm {\bf k}}'_{\alpha},\,{\rm {\bf k}}_{\alpha}; {\hat E}_{\alpha\,\epsilon})$ at $b \geq 0\,$. Note that from this equation we can easily find the on-shell limit for 
$T_{\alpha\,\epsilon}({\rm {\bf k}}'_{\alpha},\,{\rm {\bf k}}_{\alpha}; {\hat E}_{\alpha\,\epsilon})$ at $b \to 0\,$
\begin{align}
&T_{\alpha\,\epsilon}({\rm {\bf k}}'_{\alpha},\,{\rm {\bf k}}_{\alpha}; {\hat E}_{\alpha\,\epsilon}) \stackrel{b \to 0}{=}
4\,\pi\,Z_{\beta}\,Z_{\gamma}\,e^{2}\,\frac{2\,\pi\,{\hat \eta}_{\alpha\,\epsilon}^{0}}{1- e^{-2\,\pi{\hat \eta}_{\alpha\,\epsilon}^{0}}}\Big(\frac{b_{\epsilon}}{{\hat \Delta}_{\alpha\,\epsilon}^{2}}\Big)^{i\,{\hat \eta}_{\alpha\,\epsilon}^{0}},
\label{Talphaonsh1}
\end{align}
which coincides with  Eq. (\ref{halfofshell1}).

We have considered different representations of the off-shell Coulomb scattering amplitude and its on-shell limit. In the next Appendices we consider how to calculate the coulomb-modified form factors and the triangular diagrams using the obtained representations of the off-shell Coulomb scattering amplitudes.

\section{Coulomb-modified form factor}
\label{Coulmodformfactor1}

The Coulomb-modified form factor given by Eq. (\ref{gCoulmodif1}) consists of two terms.  
To calculate the integral part we need to regularize it. We see immediately that there is a pole in the integrand at  
\begin{align}
{k'}_{\gamma}= {\hat k}_{\gamma}= \sqrt{2\,\mu_{\gamma}\,{\hat z}_{\gamma}}, \qquad {\hat k}_{\gamma} \geq 0. 
\label{polesingCmf1}
\end{align}
This pole would lead to the logarithmic singularity of the integral when ${\hat k}_{\gamma}=0$ (end-point singularity). However, we recall that, according to Appendix \ref{OffshTC1}, ${k'}_{\gamma} \to {\hat k}_{\gamma}$ corresponds to the on-shell 
limit of the Coulomb scattering amplitude $T^{C}_{{\gamma}\,L_{{\gamma}}}(k_{{\gamma}},k'_{{\gamma}};{\hat z}_{{\gamma}})$, where it has its own singularity- a branching point. According to Eqs. (\ref{halfofshell1}) and (\ref{Coulrenfctr1}), the on-shell limit  of the partial Coulomb scattering amplitude, see Eq. (\ref{partprojTCgamma1}), is 
\begin{align}
T^{C}_{{\gamma}\,L_{{\gamma}}}(k'_{{\gamma}},k_{{\gamma}};{\hat z}_{{\gamma}}) \stackrel{{k'}_{\gamma} \to {\hat k}_{\gamma}}{=} ({\hat k}_{\gamma}^{2} - {k'}_{\gamma}^{2})^{i\,{\hat \eta}_{\gamma}}\,
{\tilde T}^{C}_{{\gamma}\,L_{{\gamma}}}({\hat k}_{{\gamma}},k_{{\gamma}};{\hat z}_{{\gamma}}), 
\label{onshlimitTC2}
\end{align}
where  ${\hat \eta}_{\gamma}= Z_{\alpha}\,Z_{\beta}\,e^{2}\,\mu_{\gamma}/\sqrt{2\,\mu_{\gamma}\,{\hat z}_{\gamma}}$. At ${\hat z}_{\gamma}= {\hat E}_{\gamma \epsilon}= {\hat E}_{\gamma} + i\,\epsilon$, where $\epsilon \to 0$, $\,\,{\hat \eta}_{\gamma}= 
{\hat \eta}^{0}_{\gamma \epsilon}$. Note that in practical calculations 
${\hat z}_{\gamma}= E_{+} - p_{\gamma}^{2}/(2\,M_{\gamma})$, $\,\,E_{+}= E+ i0$.
Then the integrand in the integral part of Eq. (\ref{gCoulmodif1}) has a singular behavior 
\begin{align}
\frac{T^{C}_{{\gamma}\,L_{{\gamma}}}(k'_{{\gamma}},k_{{\gamma}};{\hat z}_{{\gamma}})}{
{\hat z}_{{\gamma}} - {k'_{{\gamma}}}^{2}/(2\,\mu_{{\gamma}})} \stackrel{ {k'}_{\gamma} \to {\hat k}_{\gamma} }{=} \frac{ {\tilde T}^{C}_{\gamma\,L_{\gamma}}({\hat k}_{\gamma},k_{\gamma};{\hat z}_{\gamma})}{
[{\hat z}_{\gamma} - {k'}_{\gamma}^{2}/(2\,\mu_{\gamma})]^{1-i\,{\hat \eta}_{\gamma}}}.
\label{singintg1}
\end{align}
Thus the integrand in the Coulomb-modified form factor (\ref{gCoulmodif1}) has a branching point singularity rather then a simple pole. As it has been demonstrated in Appendix \ref{parCoulscwf1}, this type of singularity can be easily regularized using Gel'fand-Shilov method \cite{gelfand}. We apply here this technique to regularize the integral in the Coulomb-modified form factor. 

First off all such regularization is required only at ${\rm Re}\, {\hat z}_{\gamma} >0$ 
because at ${\rm Re}\, {\hat z}_{\gamma} <0$ and ${\rm Im}\, {\hat z}_{\gamma}=0$ 
$\,\,{\hat z}_{\gamma} - {k'}_{\gamma}^{2} <0 $, that is the integrand is regular function.  To regularize the integral at ${\rm Re}\, {\hat z}_{\gamma} > 0$ and ${\rm  Im}\, {\hat z}_{\gamma} \to +0$ we rewrite the Coulomb-modified form factor (\ref{gCoulmodif1}) in the form 
\begin{align}
&g^{J_{\gamma}S_{\gamma}\,\sigma}_{L_{\gamma}\,t_{\gamma}}(k_{\gamma}) = \chi^{J_{\gamma}S_{\gamma}\,\sigma}_{L_{\gamma}\,t_{\gamma}}(k_{\gamma})  +  {\delta}_{\gamma 3}\,\frac{1}{2\,\pi^{2}}\,\Big[\int\limits_{0}^{{\hat k}_{\gamma}}\,{\rm d}k'_{{\gamma}}\,{k'_{{\gamma}}}^{2}\,
\frac{{\tilde T}^{C}_{{\gamma}\,L_{{\gamma}}}(k_{{\gamma}},k'_{{\gamma}};{\hat z}_{{\gamma}})\,\chi^{J_{\gamma}S_{\gamma}\,\sigma}_{L_{\gamma}\,t_{\gamma}}(k'_{{\gamma}})}{[{\hat z}_{\gamma} - {k'}_{\gamma}^{2}/(2\,\mu_{\gamma})]^{1-i\,{\hat \eta}_{\gamma}}}                       \nonumber\\
&+ \int\limits_{{\hat k}_{\gamma}}^{\infty}\,{\rm d}k'_{{\gamma}}\,{k'_{{\gamma}}}^{2}\,
\frac{{\tilde T}^{C}_{{\gamma}\,L_{{\gamma}}}(k_{{\gamma}},k'_{{\gamma}};{\hat z}_{{\gamma}})\,\chi^{J_{\gamma}S_{\gamma}\,\sigma}_{L_{\gamma}\,t_{\gamma}}(k'_{{\gamma}})}{[{\hat z}_{\gamma} - {k'}_{\gamma}^{2}/(2\,\mu_{\gamma})]^{1-i\,{\hat \eta}_{\gamma}}} \Big].
\label{gCoulmodif2}
\end{align}
In the first integral ${\rm arg} ({\hat z}_{\gamma} - {k'}_{\gamma}^{2}/(2\,\mu_{\gamma})) \to 0\,\,$ if $\,\,{\rm Im}\,{\hat z}_{\gamma} \to 0$. In the second integral ${\hat z}_{\gamma} - {k'}_{\gamma}^{2}/(2\,\mu_{\gamma})= e^{i\,\pi}\,[{k'}_{\gamma}^{2}/(2\,\mu_{\gamma}) - {\hat z}_{\gamma}]$. Then we can rewrite Eq. (\ref{gCoulmodif2}) as
\begin{align}
&g^{J_{\gamma}S_{\gamma}\,\sigma}_{L_{\gamma}\,t_{\gamma}}(k_{\gamma}) = \chi^{J_{\gamma}S_{\gamma}\,\sigma}_{L_{\gamma}\,t_{\gamma}}(k_{\gamma})  +  {\delta}_{\gamma 3}\,\frac{1}{2\,\pi^{2}}\,\Big[\int\limits_{0}^{{\hat k}_{\gamma}}\,{\rm d}k'_{{\gamma}}\,{k'_{{\gamma}}}^{2}\,
\frac{{\tilde T}^{C}_{{\gamma}\,L_{{\gamma}}}(k_{{\gamma}},k'_{{\gamma}};{\hat z}_{{\gamma}})\,\chi^{J_{\gamma}S_{\gamma}\,\sigma}_{L_{\gamma}\,t_{\gamma}}(k'_{{\gamma}})}{[{\hat z}_{\gamma} - {k'}_{\gamma}^{2}/(2\,\mu_{\gamma})]^{1-i\,{\hat \eta}_{\gamma}}}                       \nonumber\\
&- e^{-\pi\,{\hat \eta}_{\gamma}}\,\int\limits_{{\hat k}_{\gamma}}^{2\,{\hat k}_{\gamma}}\,{\rm d}k'_{{\gamma}}\,{k'_{{\gamma}}}^{2}\,
\frac{{\tilde T}^{C}_{{\gamma}\,L_{{\gamma}}}(k_{{\gamma}},k'_{{\gamma}};{\hat z}_{{\gamma}})\,\chi^{J_{\gamma}S_{\gamma}\,\sigma}_{L_{\gamma}\,t_{\gamma}}(k'_{{\gamma}})}{[{k'}_{\gamma}^{2}/(2\,\mu_{\gamma}) -{\hat z}_{\gamma}]^{1-i\,{\hat \eta}_{\gamma}}}          
- e^{-\pi\,{\hat \eta}_{\gamma}}\,\int\limits_{2\,{\hat k}_{\gamma}}^{\infty}\,{\rm d}k'_{{\gamma}}\,{k'_{{\gamma}}}^{2}\,
\frac{{\tilde T}^{C}_{{\gamma}\,L_{{\gamma}}}(k_{{\gamma}},k'_{{\gamma}};{\hat z}_{{\gamma}})\,\chi^{J_{\gamma}S_{\gamma}\,\sigma}_{L_{\gamma}\,t_{\gamma}}(k'_{{\gamma}})}{[{k'}_{\gamma}^{2}/(2\,\mu_{\gamma}) -{\hat z}_{\gamma}]^{1-i\,{\hat \eta}_{\gamma}}} \Big].
\label{gCoulmodif3}
\end{align}

Assuming that  ${\rm Re}\,i\,{\hat \eta}_{\gamma} >0$ we can rewrite Eq. (\ref{gCoulmodif3}) 
in the form, in which the singularity is weakened enough to make it integrable at ${\rm Re} i\,{\hat \eta}_{\gamma} \to 0$:
\begin{align}
&g^{J_{\gamma}S_{\gamma}\,\sigma}_{L_{\gamma}\,t_{\gamma}}(k_{\gamma}) 
 = \chi^{J_{\gamma}S_{\gamma}\,\sigma}_{L_{\gamma}\,t_{\gamma}}(k_{\gamma})  +  {\delta}_{\gamma 3}\,\frac{1}{2\,\pi^{2}}\,\Big[\int\limits_{0}^{{\hat k}_{\gamma}}\,{\rm d}k'_{{\gamma}}\,{k'_{{\gamma}}}\,
\frac{k'_{\gamma}\,{\tilde T}^{C}_{{\gamma}\,L_{{\gamma}}}(k'_{{\gamma}},k_{{\gamma}};{\hat z}_{{\gamma}})\,\chi^{J_{\gamma}S_{\gamma}\,\sigma}_{L_{\gamma}\,t_{\gamma}}(k'_{{\gamma}}) - {\hat k}_{\gamma}\,{\tilde T}^{C}_{{\gamma}\,L_{{\gamma}}}({\hat k}_{{\gamma}},k_{{\gamma}};{\hat z}_{{\gamma}})\,\chi^{J_{\gamma}S_{\gamma}\,\sigma}_{L_{\gamma}\,t_{\gamma}}({\hat k}_{{\gamma}})}{[{\hat z}_{\gamma} - {k'}_{\gamma}^{2}/(2\,\mu_{\gamma})]^{1-i\,{\hat \eta}_{\gamma}}}          \nonumber\\
&+ \frac{\mu_{\gamma}}{i\,{\hat \eta}_{\gamma}}{\hat z}_{\gamma}^{i\,{\hat \eta}_{\gamma}}\,{\hat k}_{\gamma}\,{\tilde T}^{C}_{{\gamma}\,L_{{\gamma}}}({\hat k}_{{\gamma}},k_{{\gamma}};{\hat z}_{{\gamma}})\,\chi^{J_{\gamma}S_{\gamma}\,\sigma}_{L_{\gamma}\,t_{\gamma}}({\hat k}_{{\gamma}})                   \nonumber\\
&- e^{-\pi\,{\hat \eta}_{\gamma}}\, \int\limits_{{\hat k}_{\gamma}}^{2\,{\hat k}_{\gamma}}\,{\rm d}k'_{\gamma}\,{k'_{{\gamma}}}\,
\frac{k'_{\gamma}\,{\tilde T}^{C}_{{\gamma}\,L_{{\gamma}}}(k'_{{\gamma}},k_{{\gamma}};{\hat z}_{{\gamma}})\,\chi^{J_{\gamma}S_{\gamma}\,\sigma}_{L_{\gamma}\,t_{\gamma}}(k'_{{\gamma}})- {\hat k}_{\gamma}\, {\tilde T}^{C}_{{\gamma}\,L_{{\gamma}}}({\hat k}_{{\gamma}},k_{{\gamma}};{\hat z}_{{\gamma}})\,\chi^{J_{\gamma}S_{\gamma}\,\sigma}_{L_{\gamma}\,t_{\gamma}}({\hat k}_{{\gamma}})}{[{k'}_{\gamma}^{2}/(2\,\mu_{\gamma}) - {\hat z}_{\gamma} ]^{1-i\,{\hat \eta}_{\gamma}}}               \nonumber\\
&+  \frac{\mu_{\gamma}}{i\,{\hat \eta}_{\gamma}}\,e^{-\pi\,{\hat \eta}_{\gamma}}\,(3\,{\hat z}_{\gamma})^{i\,{\hat \eta}_{\gamma}}\,{\hat k}_{\gamma}\,{\tilde T}^{C}_{{\gamma}\,L_{{\gamma}}}({\hat k}_{{\gamma}},k_{{\gamma}};{\hat z}_{{\gamma}})\,\chi^{J_{\gamma}S_{\gamma}\,\sigma}_{L_{\gamma}\,t_{\gamma}}({\hat k}_{{\gamma}}) - e^{-\pi\,{\hat \eta}_{\gamma}}\, \int\limits_{2\,{\hat k}_{\gamma}}^{\infty}\,{\rm d}k'_{\gamma}\,{k'_{{\gamma}}}^{2}\,
\frac{{\tilde T}^{C}_{{\gamma}\,L_{{\gamma}}}(k'_{{\gamma}},k_{{\gamma}};{\hat z}_{{\gamma}})\,\chi^{J_{\gamma}S_{\gamma}\,\sigma}_{L_{\gamma}\,t_{\gamma}}(k'_{{\gamma}})}{[{k'}_{\gamma}^{2}/(2\,\mu_{\gamma}) - {\hat z}_{\gamma} ]^{1-i\,{\hat \eta}_{\gamma}}} \Big].
\label{gCoulmodifreg4}
\end{align}
Eq. (\ref{gCoulmodifreg4}) can be used to calculate the Coulomb-modified form factor. 
We have shown how to regularize the Coulomb-modified form factor in the proximity of the singularity 
$k'_{\gamma}= {\hat k}_{\gamma}$.
This regularization allows one to calculate the Coulomb-modified 
form factor.

Now we consider a special case of a simple separable form factor
\begin{align}
\chi^{J_{\gamma}S_{\gamma}\,\sigma}_{L_{\gamma}\,t_{\gamma}}(k_{\gamma}) =
\frac{k_{\gamma}^{L_{\gamma}}}{[k_{\gamma}^{2} + ({\beta_{L_{\gamma}\,t_{\gamma}}^{J_{\gamma}S_{\gamma}\,\sigma}})^{2}]^{L_{\gamma}+1}},
\label{separformfactor1}
\end{align}
for which we will derive analytical expression for the on-shell limit of the Coulomb-modified form factor.
The Coulomb-modified form factor in this case can be written in the integral form \cite{vanhaeringen82}:
\begin{align}
g^{J_{\gamma}S_{\gamma}\,\sigma}_{L_{\gamma}\,t_{\gamma}}(k_{\gamma})=
\chi^{J_{\gamma}S_{\gamma}\,\sigma}_{L_{\gamma}\,t_{\gamma}}(k_{\gamma}) - i\,{\hat \eta}_{\gamma}\,k_{\gamma}^{L_{\gamma}}\Big[\frac{-4\,{\hat k}_{\gamma}^{2}}{([{\beta_{L_{\gamma}\,t_{\gamma}}^{J_{\gamma}S_{\gamma}\,\sigma}}]^{2} + {\hat k}_{\gamma}^{2})(k_{\gamma}^{2} - {\hat k}_{\gamma}^{2})}\Big]^{L_{\gamma}+1}\,
\int\limits_{0}^{1}\,{\rm d}x\,\frac{x^{i\,{\hat \eta}_{\gamma}-1}}{[x\,B+ \frac{1}{x\,B} - a - a^{-1}]^{L_{\gamma}+1}}.
\label{CoulmodformfactrYam1}
\end{align}
Here, 
\begin{align}
a=\frac{k_{\gamma} - {\hat k}_{\gamma}}{k_{\gamma} + {\hat k}_{\gamma}}, \qquad
B= \frac{ \beta_{L_{\gamma}\,t_{\gamma}}^{ J_{\gamma}S_{\gamma}\,\sigma} + i\,k_{\gamma}}{\beta_{L_{\gamma}\,t_{\gamma}}^{J_{\gamma}S_{\gamma}\,\sigma} - i\,k_{\gamma}}.
\label{aB1}
\end{align}

From Eq. (\ref{CoulmodformfactrYam1}) clear that the Coulomb-modified form factor has singularity in the on-shell limit $k_{\gamma} \to {\hat k}_{\gamma}$ corresponding to $a \to 0$. 
For the separable form factor (\ref{separformfactor1}) at $L_{\gamma}=0$ the on-shell limit of the Coulomb-modified form factor was found in \cite{muk2001} (see Appendix C):
\begin{align}
g^{S_{\gamma}S_{\gamma}\,\sigma}_{0\,t_{\gamma}}(k_{\gamma}) \stackrel{k_{\gamma} \to {\hat k}_{\gamma}}{=} \frac{2\,\pi\,{\hat \eta}_{\gamma}}{1- e^{-2\,\pi\,{\hat \eta}_{\gamma}}
\,}\,e^{ 2\,{\hat \eta}_{\gamma}\,\arctan
\,{\hat k}_{\gamma}/\beta_{0\,t_{\gamma}}^{ S_{\gamma}S_{\gamma}\,\sigma}}\,
\Big(\frac{k_{\gamma} - {\hat k}_{\gamma}}{2\,{\hat k}_{\gamma}} \Big)^{i\,{\hat \eta}_{\gamma}}\,
\chi^{S_{\gamma}S_{\gamma}\,\sigma}_{0\,t_{\gamma}}({\hat k}_{\gamma}).
\label{a0limit11}
\end{align}
Here, we extend this expression for arbitrary $L_{\gamma}$. To do it we note that 
Eq. (\ref{CoulmodformfactrYam1})  can be rewritten as
\begin{align}
g^{J_{\gamma}S_{\gamma}\,\sigma}_{L_{\gamma}\,t_{\gamma}}(k_{\gamma})=
\chi^{J_{\gamma}S_{\gamma}\,\sigma}_{L_{\gamma}\,t_{\gamma}}(k_{\gamma}) - i\,{\hat \eta}_{\gamma}\,k_{\gamma}^{L_{\gamma}}\Big[\frac{-4\,{\hat k}_{\gamma}^{2}}{([{\beta_{L_{\gamma}\,t_{\gamma}}^{J_{\gamma}S_{\gamma}\,\sigma}}]^{2} + {\hat k}_{\gamma}^{2})(k_{\gamma}^{2} - {\hat k}_{\gamma}^{2})}\Big]^{L_{\gamma}+1}\,\frac{1}{B^{L_{\gamma}+1}}
\int\limits_{0}^{1}\,{\rm d}x\,\frac{x^{i\,{\hat \eta}_{\gamma} + L_{\gamma}}}{[(x- x_{1})(x - x_{2})]^{L_{\gamma}+1}},
\label{CoulmodformfactrYam2}
\end{align}  
where $\,\,x_{1,2}$ are the roots of equation
$x\,B+ \frac{1}{x\,B} - a - a^{-1}=0$ what is equivalent to $x^{2} - \frac{1}{B}\,( a + a^{-1})\,x + \frac{1}{B^{2}}=0$:
\begin{align}
x_{1,2}= \frac{a + a^{-1}}{2\,B} \pm \frac{\sqrt{(a + a^{-1})^{2} -4}}{2\,B}.
\label{rootseq1}
\end{align}
Evidently that at $k_{\gamma} \to {\hat k}_{\gamma}$, that is $a \to 0$, 
\begin{align}
x_{1} \stackrel{a \to 0}{=} \frac{1}{a\,B} \to \infty, \qquad
x_{2} \stackrel{a \to 0}{=} \frac{a}{B} \to 0.
\label{rootslimita1}
\end{align}
Also $x^{i\,{\hat \eta}_{\gamma}-1}$ in the integrand of Eq. (\ref{CoulmodformfactrYam2}) 
has a branching point singularity at x=0 and $\infty$, which are connected by a cut. Then we can apply the method described in Appendix \ref{OffshTC1}.  First we transform the integral in
(\ref{CoulmodformfactrYam2}) to the integral over contour $C_{1}$, in which the integration contour is taken along the contour starting at $x=1$, encircling the branching singularity at $x=0$ and ending at $x=1$, see Fig.  \ref{fig_contour}.              
 After that we consider the integral along the closed contour $C= C_{1} + C_{2} + C_{3}$. Because there are no singularities of the integrand inside and on the integration contour $C$, according to Caushy's theorem, the integral taken along $C$ vanishes. Hence
\begin{align}
&\int\limits_{0}^{1}\,{\rm d}x\,\frac{x^{i\,{\hat \eta}_{\gamma} + L_{\gamma}}}{[(x- x_{1})(x - x_{2})]^{L_{\gamma}+1}}= -\frac{1}{1- e^{-2\,\pi{\hat \eta}_{\gamma}}}\,
\Big[ \oint\limits_{{C_{2}}}\,{\rm d}x\,\frac{x^{i\,{\hat \eta}_{\gamma} + L_{\gamma}}}{[(x- x_{1})(x - x_{2})]^{L_{\gamma}+1}}+ \oint\limits_{{C_{3}}}\,{\rm d}x\,\frac{x^{i\,{\hat \eta}_{\gamma}  + L_{\gamma}}}{[(x- x_{1})(x - x_{2})]^{L_{\gamma}+1}}    \Big].
\label{C123}
\end{align}
We assume that the root $x_{2}$ has a positive infinitesimal imaginary addend, which shifts the root from the integration contour to the first quadrant. In the integral over contour $C_{2}$ we rewrite
\begin{align} 
\lim\limits_{\epsilon \to 0}\,\frac{1}{[(x - x_{1})(x - x_{2} - \epsilon)]^{L_{\gamma}+1}} = \frac{1}{L_{\gamma}!}\,\frac{1}{(x - x_{1})^{L_{\gamma}+1}}\,\lim\limits_{\epsilon \to 0}\frac{\rm {d}^{L_{\gamma}}}{{\rm d}\,\epsilon^{L_{\gamma}}}\,\frac{1}{x - x_{2} - \epsilon}.
\label{transfpole1}
\end{align} 
We take into account that at $a \to 0$ $\,\,x_{2} \to 0$ while $x_{1} \to \infty$, that is at $a \to 0 $ contour $C_{2}$ encircles the pole at $x=x_{2}+ \epsilon$ and can be reduced to the residue in the pole:
\begin{align}
&-\frac{1}{1- e^{-2\,\pi{\hat \eta}_{\gamma}}}\,\oint\limits_{{C_{2}}}\,{\rm d}x\,\frac{x^{i\,{\hat \eta}_{\gamma} + L_{\gamma}}}{[(x- x_{1})(x - x_{2})]^{L_{\gamma}+1}} = - \frac{1}{1- e^{-2\,\pi{\hat \eta}_{\gamma}}}\,\frac{1}{L_{\gamma}!}\,\lim\limits_{\epsilon \to 0}\,\frac{\rm {d}^{L_{\gamma}}}{{\rm d}\,\epsilon^{L_{\gamma}}}\,\oint\limits_{{C_{2}}}\,{\rm d}x\,\frac{x^{i\,{\hat \eta}_{\gamma} +L_{\gamma}}}{[(x- x_{1})]^{L_{\gamma}+1}\,(x - x_{2} - \epsilon)}                                       \nonumber\\
&= \frac{2\,\pi\,i}{1- e^{-2\,\pi{\hat \eta}_{\gamma}}}\,\frac{1}{L_{\gamma}!}\,\lim\limits_{\epsilon \to 0}\,\frac{\rm {d}^{L_{\gamma}}}{{\rm d}\,\epsilon^{L_{\gamma}}}\,\frac{(x_{2} + \epsilon)^{i\,{\hat \eta}_{\gamma} + L_{\gamma}}}{(x_{2} - x_{1} + \epsilon)^{L_{\gamma} +1}} 
\stackrel{a \to 0}{=}
 \frac{2\,\pi\,i}{1- e^{-2\,\pi{\hat \eta}_{\gamma}}}\,\frac{1}{L_{\gamma}!}\,
(-1)^{L_{\gamma} + 1}\,a^{i\,{\hat \eta}_{\gamma} + L_{\gamma} +1}\,B^{-i\,{\hat \eta}_{\gamma} + L_{\gamma} +1}\frac{1}{{i{{\widehat \eta }_\lambda }}}\,\prod\limits_{n = 0}^{{L_\gamma }} {(i{{\widehat \eta }_\gamma }}  + n),
\label{residpole1}
\end{align}
where $\,\,\prod\limits_{n = 0}^{{L_{\gamma}=0}} {(i{{\hat \eta }_\gamma }}  + n)= i\,{\hat \eta}_{\gamma}$. Also at $a \to 0$ $\,\,a=(k_{\gamma} - {\hat k}_{\gamma})/(2\,{\hat k}_{\gamma})$ and  
$\,\,B= (\beta_{L_{\gamma}\,t_{\gamma}}^{ J_{\gamma}S_{\gamma}\,\sigma} + i\,{\hat k}_{\gamma})/(\beta_{L_{\gamma}\,t_{\gamma}}^{J_{\gamma}S_{\gamma}\,\sigma} - i\,{\hat k}_{\gamma})$.

Taking into account all the factors in front of the integral in Eq. (\ref{CoulmodformfactrYam2}) 
\begin{align}
&- i\,{\hat \eta}_{\gamma}\,k_{\gamma}^{L_{\gamma}}\Big[\frac{-4\,{\hat k}_{\gamma}^{2}}{([{\beta_{L_{\gamma}\,t_{\gamma}}^{J_{\gamma}S_{\gamma}\,\sigma}}]^{2} + {\hat k}_{\gamma}^{2})(k_{\gamma}^{2} - {\hat k}_{\gamma}^{2})}\Big]^{L_{\gamma}+1}\,\frac{1}{B^{L_{\gamma}+1}}\,
\frac{1}{1- e^{-2\,\pi{\hat \eta}_{\gamma}}}\,\oint\limits_{{C_{2}}}\,{\rm d}x\,\frac{x^{i\,{\hat \eta}_{\gamma} + L_{\gamma}}}{[(x- x_{1})(x - x_{2})]^{L_{\gamma}+1}}                     \nonumber\\
&= \frac{2\,\pi\,{\hat \eta}_{\gamma}}{1- e^{-2\,\pi{\hat \eta}_{\gamma}}}\,e^{ 2\,{\hat \eta}_{\gamma}\,\arctan
\,{\hat k}_{\gamma}/\beta_{L_{\gamma}\,t_{\gamma}}^{ J_{\gamma}S_{\gamma}\,\sigma}}\,\Big(\frac{k_{\gamma} - {\hat k}_{\gamma}}{2\,{\hat k}_{\gamma}}\Big)^{i\,{\hat \eta}_{\gamma}}\,\frac{1}{L_{\gamma}!}\,\frac{1}{{i{{\hat \eta }_\lambda }}}\,\prod\limits_{n = 0}^{{L_\gamma }} {(i{{\widehat \eta }_\gamma }}  + n)\,\chi^{J_{\gamma}S_{\gamma}\,\sigma}_{L_{\gamma}\,t_{\gamma}}({\hat k}_{\gamma}).
\label{C22}
\end{align} 

Let us consider the integral over $C_{3}$. In this integral the contour goes along the circumference with $|x|=1$. Hence, along $C_{3}\,$ $\,\,x=e^{i\,\varphi}$ and ${\rm d}x= i\,{\rm d}\,\varphi\,e^{i\,\varphi}$.                      
\begin{align} 
&-\frac{1}{1- e^{-2\,\pi{\hat \eta}_{\gamma}}}\,\oint\limits_{{C_{3}}}\,{\rm d}x\,\frac{x^{i\,{\hat \eta}_{\gamma} + L_{\gamma}}}{[(x- x_{1})(x - x_{2})]^{L_{\gamma}+1}} \stackrel{a \to 0}{=} - i\,\frac{1}{1- e^{-2\,\pi{\hat \eta}_{\gamma}}}\,(-aB)^{L_{\gamma}+1}\,\int\limits_{0}^{2\,\pi}\,{\rm d}\varphi\,e^{-{ \eta}_{\gamma}\,\varphi}                                        \nonumber\\
&= i\,(-1)^{L_{\gamma}}\,(aB)^{L_{\gamma}+1}\,\frac{1}{{\hat \eta}_{\gamma}}.
\label{C31}
\end{align}
Taking into account the factors in front of the integral in Eq. (\ref{CoulmodformfactrYam2})
we get
\begin{align} 
&- i\,{\hat \eta}_{\gamma}\,k_{\gamma}^{L_{\gamma}}\Big[\frac{-4\,{\hat k}_{\gamma}^{2}}{([{\beta_{L_{\gamma}\,t_{\gamma}}^{J_{\gamma}S_{\gamma}\,\sigma}}]^{2} + {\hat k}_{\gamma}^{2})(k_{\gamma}^{2} - {\hat k}_{\gamma}^{2})}\Big]^{L_{\gamma}+1}\,
\frac{1}{1- e^{-2\,\pi{\hat \eta}_{\gamma}}}\,\oint\limits_{{C_{3}}}\,{\rm d}x\,\frac{x^{i\,{\hat \eta}_{\gamma}+ L_{\gamma}}}{[(x- x_{1})(x - x_{2})]^{L_{\gamma}+1}}            \nonumber\\
&\stackrel{a \to 0}{=} - \chi^{J_{\gamma}S_{\gamma}\,\sigma}_{L_{\gamma}\,t_{\gamma}}({\hat k}_{\gamma}).
\label{C33}
\end{align}

Then we arrive at the final equation of this Appendix defining the on-shell limit of the 
Coulomb-modified form factor for the separable form factor (\ref{separformfactor1}):
\begin{align}
&g^{J_{\gamma}S_{\gamma}\,\sigma}_{L_{\gamma}\,t_{\gamma}}(k_{\gamma})=
\chi^{J_{\gamma}S_{\gamma}\,\sigma}_{L_{\gamma}\,t_{\gamma}}(k_{\gamma}) - i\,{\hat \eta}_{\gamma}\,k_{\gamma}^{L_{\gamma}}\Big[\frac{-4\,{\hat k}_{\gamma}^{2}}{([{\beta_{L_{\gamma}\,t_{\gamma}}^{J_{\gamma}S_{\gamma}\,\sigma}}]^{2} + {\hat k}_{\gamma}^{2})(k_{\gamma}^{2} - {\hat k}_{\gamma}^{2})}\Big]^{L_{\gamma}+1}\,\frac{1}{B^{L_{\gamma}+1}}
\int\limits_{0}^{1}\,{\rm d}x\,\frac{x^{i\,{\hat \eta}_{\gamma} + L_{\gamma}}}{[(x- x_{1})(x - x_{2})]^{L_{\gamma}+1}}                                                                \nonumber\\
& \stackrel{k_{\gamma} \to {\hat k}_{\gamma}}{=} \frac{2\,\pi\,{\hat \eta}_{\gamma}}{1- e^{-2\,\pi{\hat \eta}_{\gamma}}}\,e^{ 2\,{\hat \eta}_{\gamma}\,\arctan
\,{\hat k}_{\gamma}/\beta_{L_{\gamma}\,t_{\gamma}}^{ J_{\gamma}S_{\gamma}\,\sigma}}\,\Big(\frac{k_{\gamma} - {\hat k}_{\gamma}}{2\,{\hat k}_{\gamma}}\Big)^{i\,{\hat \eta}_{\gamma}}\,\frac{1}{L_{\gamma}!}\,\frac{1}{{i{{\hat \eta }_\lambda }}}\,\prod\limits_{n = 0}^{{L_\gamma }} {(i{{\widehat \eta }_\gamma }}  + n)\,\chi^{J_{\gamma}S_{\gamma}\,\sigma}_{L_{\gamma}\,t_{\gamma}}({\hat k}_{\gamma}),                 \qquad {\hat k}_{\gamma} >0. 
\label{Coulombfinal1}
\end{align} 
Thus we have shown that the Coulomb-modified form factor has a branching point singularity at $k_{\gamma}={\hat k}_{\gamma}$. This Coulomb-modified form factor is needed to calculate the amplitudes of the diagrams describing the proton and nucleus transfer, see Eq. (\ref{Riksigmarho1}). Assume that the Coulomb-modifed form factor is 
$g_{\zeta_{\alpha}}^{\sigma }({k_\alpha })$, which has the on-shell singularity at $k_{\alpha}= {\hat k}_{\alpha}^{\sigma}$,
where ${\hat k}_{\alpha}^{\sigma}= \sqrt{2\,\mu_{\alpha}\,({\hat z}_{\alpha} - \epsilon^{\rho})}$,  $\,{\hat z}_{\alpha}= z - \,\,p_\alpha ^2/(2{M_\alpha })$.   Then we can rewrite Eq. (\ref{Riksigmarho1}) as
\begin{align}
R_{\zeta_\beta\,\zeta_\alpha}^{\sigma \sigma\,(i)\,{\Lc}}(p'_{\beta} ,p_\alpha;z) = 
R_{\zeta_\beta\,\zeta_\alpha}^{'\,\sigma \sigma\,(i)\,{\Lc}}(p'_{\beta} ,p_\alpha;z)=
{\overline \delta  _{\beta \alpha }}\,c_{\zeta_{\beta}}^{\sigma\,*}c_{\zeta_{\alpha}}^{\sigma}\,\frac{1}{2}\int\limits_{ - 1}^1 {dx} \,{P_{\Lc}}(x)\frac{{k_\alpha ^{ - {L_\alpha }}k_\beta ^{' - {L_\beta }}{\chi}_{\zeta_{\beta}}^{\sigma\,*}(k'_{\beta})\,{\tilde g}_{\zeta_{\alpha}}^{\sigma }({k_\alpha })}}{{\big[{\hat z}_{\alpha} -\epsilon^{\rho} - k_\alpha ^2/(2{\mu _\alpha })\big]^{1-i{{\hat \eta }_\alpha }}}},    \qquad i=0,\,1,\,2,
\label{Riksigmarho11}
\end{align}
where ${\tilde g}_{\zeta_{\alpha}}^{\sigma }({k_\alpha })$  is regular at $k_{\alpha} \to {\hat k}_{\alpha}^{\sigma}$. Thus the integrand, owe to the presence of the Coulomb-modified form factor, has a branching point singularity rather than a pole and the integral can be easily regularized using the Gel'fand-Shilov method \cite{gelfand} applied earlier.   
Note that, although derivation of Eq. (\ref{Coulombfinal1}) has relied on the explicit form 
(\ref{separformfactor1}) of the separable form factor $\chi^{J_{\gamma}S_{\gamma}\,\sigma}_{L_{\gamma}\,t_{\gamma}}(k_{\gamma})$, the result is nevertheless valid for arbitrary nonsingular at the origin form factor since any such form factor can be represented as linear combination of functions of type (\ref{separformfactor1}).  
Thus we have demonstrated that the Coulomb-modified form factor does not create any problems in practical calculations.

Finally, it is worth mentioning that the on-shell limit of the Coulomb-modified form factor at ${\hat k}_{\gamma}=0$ for $L_{\gamma}=0$ was found in \cite{muk2000}: $g^{S_{\gamma}S_{\gamma}}_{0\,t_{\gamma}}(k_{\gamma})
\stackrel{k_{\gamma} \to 0}{\sim} k_{\gamma}^{2}, \,\,$ that is the Coulomb-modified form factor vanishes at the on-shell limit at ${\hat k}_{\gamma}=0$. Similar consideration can be done for $L_{\gamma} >0$.

\section{Pole singularity of the triangular exchange diagram and Coulomb renormalization of its strength} 
\label{Coulrenormexchtr1}

In the previous section we considered the Coulomb-modified form factor and how the presence of the off-shell Coulomb scattering amplitude affects it. The off-shell Coulomb scattering amplitude also is needed to calculate the direct and exchange triangular diagrams. The discussion of the direct triangular diagram has been done   
in Section \ref{angmomentdecomp}. Here we consider the exchange triangular diagrams shown in Figs  \ref{triang_exch} and \ref{triang_exch1}. 

In this section we show that there is a pole singularity of the exchange triangular diagram, which allows us to rewrite the amplitude of this diagram as the renormalized pole neutron transfer diagram plus the nonsingular at the pole term. To show it let us consider, for example, the amplitude of the exchange triangular diagram shown in Fig. \ref{triang_exch}. For simplicity, we neglect the orbital momenta in the three-ray vertices and the spins, although the final result, which we present below, is valid for the general case. The amplitude of the exchange triangular diagram is given by
\begin{align}
Z^{(4)}({\rm {\bf p}}'_{\beta}, {\rm {\bf p}}_{\alpha}) = \int\,\frac{{\rm d}{\rm {\bf p}}_{\gamma}}{(2\,\pi)^{3}}\,\frac{\chi_{\beta}^*({\rm{\bf p}}_{\gamma} + \frac{m_{\gamma}}{m_{\alpha\gamma}}\,{\rm {\bf p}}'_{\beta})}{{\hat k}_{\beta}^{2} - ({\rm{\bf p}}_{\gamma} + \frac{m_{\gamma}}{m_{\alpha\gamma}}\,{\rm {\bf p}}'_{\beta})^{2} }\,T_{\gamma}^{C}({\rm {\bf k}}'_{\gamma},{\rm {\bf k}}_{\gamma};{\hat z}_{\gamma})\,\frac{\chi_{\alpha}({\rm{\bf p}}_{\gamma} + \frac{m_{\gamma}}{m_{\beta\gamma}}\,{\rm {\bf p}}_{\alpha})}{{\hat k}_{\alpha}^{2} - ({\rm{\bf p}}_{\gamma} + \frac{m_{\gamma}}{m_{\beta\gamma}}\,{\rm {\bf p}}_{\alpha})^{2}},
\label{exhtrdiagr11}
\end{align}
where $\alpha=1,\,\beta=2\,$ and $\,\gamma=3$; $\,\,{\hat k}_{\alpha}^{2}= 2\,\mu_{\alpha}\,{\hat z}_{\alpha}\,$ and $\,{\hat z}_{\alpha} =z  - p_{\alpha}^{2}/(2\,M_{\alpha})$;  $\,\,{\hat k}_{\beta}^{2}= 2\,\mu_{\beta}\,{\hat z}_{\beta}\,$ and $\,{\hat z}_{\beta} =z  - {p'}_{\beta}^{2}/(2\,M_{\beta})$; ${\rm{\bf p}}_{\gamma} + \frac{m_{\gamma}}{m_{\beta\gamma}}\,{\rm {\bf p}}_{\alpha}\,$ and $\,\,{\rm{\bf p}}_{\gamma} + \frac{m_{\gamma}}{m_{\alpha\gamma}}\,{\rm {\bf p}}'_{\beta}$ are the relative momenta of particles in the three-ray vertices $(\beta\gamma) \to \beta + \gamma$ and $(\alpha\gamma) \to \alpha + \gamma$ in the triangular diagram. Also ${\rm {\bf k}}_{\gamma}= {\rm {\bf p}}_{\alpha} + (m_{\alpha}/m_{\alpha\beta})\,{\rm {\bf p}}_{\gamma}\,\,$ \big(${\rm {\bf k}}'_{\gamma}= - {\rm {\bf p}}'_{\beta} - (m_{\beta}/m_{\alpha\beta})\,{\rm {\bf p}}_{\gamma}$\big) is the relative momentum of particles $\alpha$ and $\beta$ on the diagram before (after) the Coulomb scattering with the transfer momentum in the four-ray vertex $\Delta'_{\alpha}= {\rm {\bf k}}_{\gamma} - {\rm {\bf k}}'_{\gamma} = {\rm {\bf p}}_{\alpha} + {\rm {\bf p}}'_{\beta} + {\rm {\bf p}}_{\gamma}\,\,$  (see also Eq. (\ref{kinexchtriang1})).

The closest to the physical region and the strongest singularity of the triangular diagram is the one generated by the coincidence of the singularities of the propagators ${\hat k}_{\alpha}^{2} - ({\rm{\bf p}}_{\gamma} + \frac{m_{\gamma}}{m_{\beta\gamma}}\,{\rm {\bf p}}_{\alpha})^{2}=0$  and $\,\,{\hat k}_{\beta}^{2} - ({\rm{\bf p}}_{\gamma} + \frac{m_{\gamma}}{m_{\alpha\gamma}}\,{\rm {\bf p}}'_{\beta})^{2}=0$ and the forward singularity of the off-shell Coulomb singularity $\Delta'_{\alpha}=0$ of the Coulomb scattering amplitude. To show how these singularity of the exchange triangular diagram appears and to simplify consideration we replace the off-shell Coulomb scattering amplitude $T_{\gamma}^{C}({\rm {\bf k}}'_{\gamma},{\rm {\bf k}}_{\gamma};{\hat z}_{\gamma})$ of particles $\alpha$ and $\beta$ by the Born Coulomb amplitude, which is the Fourier transform of the Coulomb potential $4\,\pi/{{\rm {\bf \Delta}}'}_{\alpha}^{2}$. Then the amplitude of triangular exchange diagram simplifies to 
\begin{align}
&Z^{(4)}({\rm {\bf p}}'_{\beta}, {\rm {\bf p}}_{\alpha}) = 4\,\pi\,Z_{\beta}\,Z_{\alpha}\,e^{2}\,\int\,\frac{{\rm d}{\rm {\bf p}}_{\gamma}}{(2\,\pi)^{3}}\,\frac{\chi^*_{\beta}({\rm{\bf p}}_{\gamma} + \frac{m_{\gamma}}{m_{\alpha\gamma}}\,{\rm {\bf p}}'_{\beta})}{{\hat k}_{\beta}^{2} - ({\rm{\bf p}}_{\gamma} + \frac{m_{\gamma}}{m_{\alpha\gamma}}\,{\rm {\bf p}}'_{\beta})^{2} }\,\frac{1}{{{\rm {\bf \Delta}}'}_{\alpha}^{2}}\,\frac{\chi_{\alpha}({\rm{\bf p}}_{\gamma} + \frac{m_{\gamma}}{m_{\beta\gamma}}\,{\rm {\bf p}}_{\alpha})}{{\hat k}_{\alpha}^{2} - ({\rm{\bf p}}_{\gamma} + \frac{m_{\gamma}}{m_{\beta\gamma}}\,{\rm {\bf p}}_{\alpha})^{2}}   \nonumber\\
&= 4\,\pi\,Z_{\beta}\,Z_{\alpha}\,e^{2}\,\int\,\frac{{\rm d}{\rm {\bf \Delta}}'_{\alpha}}{(2\,\pi)^{3}}\,\frac{\chi^*_{\beta}\big({\rm {\bf \Delta}}'_{\alpha} - {\rm {\bf k}}'_{\beta}\big)}{{\hat k}_{\beta}^{2} - ({\rm {\bf \Delta}}'_{\alpha} - {\rm {\bf k}}'_{\beta})^{2} }\,\frac{1}{{{\rm {\bf \Delta}}'}_{\alpha}^{2}}\,\frac{\chi_{\alpha}\big({\rm {\bf \Delta}}'_{\alpha} - {\rm {\bf k}}_{\alpha} \big)}{{\hat k}_{\alpha}^{2} - ({\rm {\bf \Delta}}'_{\alpha} - {\rm {\bf k}}_{\alpha})^{2}},
\label{exhtrdiagr12}
\end{align}
where we used the substitution ${\rm {\bf p}}_{\gamma}=\Delta'_{\alpha}- {\rm {\bf p}}_{\alpha} -{\rm {\bf p}}'_{\beta}$. Also ${\rm {\bf k}}_{\alpha}= {\rm {\bf p}}'_{\beta} + (m_{\beta}/m_{\beta\gamma})\,{\rm {\bf p}}_{\alpha}$ is the relative momentum of particles $\beta$ and $\gamma$ in the three-ray vertex $(\beta \gamma) \to \beta + \gamma$ in the diagram of Fig. \ref{fig_pole01}, with $\alpha=1,\,\beta=2$ and $\gamma=3$. Similarly ${\rm {\bf k}}'_{\beta}= {\rm {\bf p}}_{\alpha} + (m_{\alpha}/m_{\alpha\gamma})\,{\rm {\bf p}}'_{\beta}$ is the relative momentum of particles $\alpha$ and $\gamma$ in the three-ray vertex $(\alpha \gamma) \to \alpha + \gamma$ of the same diagram. Now we rewrite 
\begin{align} 
{\hat k}_{\alpha}^{2} - ({\rm {\bf \Delta}}'_{\alpha} - {\rm {\bf k}}_{\alpha})^{2}=
\sigma_{\alpha} + 2\,{\rm {\bf \Delta}}'_{\alpha} \cdot {\rm {\bf k}}_{\alpha} - {\Delta'}_{\alpha}^{2}= \sigma_{\alpha}\,\big[1 + 2\,{\rm {\bf t}}\cdot {\rm {\bf k}}_{\alpha} - \sigma_{\alpha}\,t^{2} \big],  
\label{sigmaalpha1}
\end{align}
where we introduced $\sigma_{\alpha}= {\hat k}_{\alpha}^{2}- {\rm {\bf k}}_{\alpha}^{2}$
and used the substitution 
\begin{align}
{\rm {\bf \Delta}}'_{\alpha} = \sigma_{\alpha}\,{\rm {\bf t}}.
\label{Deltat1}
\end{align}
Similarly 
\begin{align} 
{\hat k}_{\beta}^{2} - ({\rm {\bf \Delta}}'_{\alpha} - {\rm {\bf k}}'_{\beta})^{2}=
  \sigma_{\alpha}\,\big[\frac{\mu_{\beta}}{\mu_{\alpha}}\, + 2\,{\rm {\bf t}}\cdot {\rm {\bf k}}'_{\beta} - \sigma_{\alpha}\,t^{2} \big] .  
\label{sigmaalpha2}
\end{align}
Here we took into account that from the energy-momentum conservation in both three-ray vertices of the diagram in Fig. \ref{fig_pole01} follows  
\begin{align}
\sigma_{\beta}= \frac{\mu_{\beta}}{\mu_{\alpha}}\,\sigma_{\alpha}.
\label{sigmaalphabeta1}
\end{align}

Because we consider the  singularity of the exchange triangular diagram generated by the coincidence of  zeroes of three denominators (pinch-point singularity) in Eq. (\ref{exhtrdiagr12}), we use the substitution (\ref{Deltat1}) obtaining in the leading order
\begin{align}
Z^{(4)}({\rm {\bf p}}'_{\beta}, {\rm {\bf p}}_{\alpha}) \stackrel{\sigma_{\alpha} \to 0}{=} 
= 4\,\pi\,Z_{\beta}\,Z_{\alpha}\,e^{2}\,\Big[\frac{\chi^*_{\beta}\big( - {\rm {\bf k}}'_{\beta}\big)\,\chi_{\alpha}\big(-{\rm {\bf k}}_{\alpha} \big)}{\sigma_{\alpha}}\Big]\,\int\,\frac{{\rm d}{\rm {\bf t}}}{(2\,\pi)^{3}}\,\frac{1}{\frac{\mu_{\beta}}{\mu_{\alpha}} + 2\,{\rm {\bf t}} \cdot {\rm {\bf k}}'_{\beta}}\,\frac{1}{{\rm {\bf t}}^{2}}\,\frac{1}{1 + 2\,{\rm {\bf t}} \cdot {\rm {\bf k}}_{\alpha}}.
\label{exhtrdiagr13}
\end{align}
Thus we have shown that the  strongest singularity of the amplitude of the exchange triangular diagram is a pole singularity at 
\begin{align}
\sigma_{\alpha}=0. 
\label{sigmaapole1}
\end{align}
The same singularity has the pole diagram in Fig. \ref{fig_pole01}.
Moreover 
\begin{align}
Z^{(0)}({\rm {\bf p}}'_{\beta}, {\rm {\bf p}}_{\alpha}) \sim \Big[\frac{\chi^*_{\beta}\big( - {\rm {\bf k}}'_{\beta}\big)\,\chi_{\alpha}\big(-{\rm {\bf k}}_{\alpha} \big)}{\sigma_{\alpha}}\Big]
\label{poleamplsimple1}
\end{align}
in Eq. (\ref{exhtrdiagr13}) is the amplitude of the pole diagram (we neglect the spins and angular momenta). We can conclude from the simple consideration presented here that near the singularity (\ref{sigmaapole1}) the amplitude of the exchange triangular diagram behaves as renormalized amplitude of the pole diagram in Fig. \ref{fig_pole01}: 
\begin{align}
Z^{(4)}({\rm {\bf p}}'_{\beta}, {\rm {\bf p}}_{\alpha}) \stackrel{\sigma_{\alpha} \to 0}{=} 
D\,Z^{(0)}({\rm {\bf p}}'_{\beta}, {\rm {\bf p}}_{\alpha}) + Z_{reg}^{(4)}({\rm {\bf p}}'_{\beta}, {\rm {\bf p}}_{\alpha}),
\label{exhtrpoleren1}
\end{align}
where $D$ is the renormalization factor determining the strength of the pole singularity. 
The additional term $Z_{reg}^{(4)}({\rm {\bf p}}'_{\beta}, {\rm {\bf p}}_{\alpha})$ is regular at $\sigma_{\alpha}=0$.

A general expression for the renormalization factor for the exchange triangular diagram containing the full $T_{\gamma}^{C}$ Coulomb scattering amplitude rather than the Born Coulomb amplitude was obtained in \cite{mukh1982,blokh84,blokh1986}. Summing up the neutron transfer pole diagram and the corresponding exchange triangular diagram we obtain  the renormalized pole diagram plus the additional term from the exchange triangular diagram, which is regular at $\sigma_{\alpha}=0$. 
For the diagram in Fig. \ref{triang_exch}  the renormalization factor, which determines the strength of the pole singularity of the triangular exchange diagram, is  
\begin{align}
D_{\beta \zeta_{\beta}\,\alpha \zeta_{\alpha}}^{\sigma} 
= -1 + \Bigg[\frac{{\sqrt { - {m_{\beta \gamma }}{m_\alpha }({{\hat z}_\alpha } - {\epsilon^\sigma })}  + \sqrt { - {m_{\alpha \gamma }}{m_\beta }({{\hat z}_\beta } - {\epsilon^\sigma })}  + i\sqrt {{m_\gamma }{m_{\beta \alpha }}\tilde E_{\beta {\zeta _\beta }\alpha {\zeta _\alpha }_\alpha }^\sigma } }}{{\sqrt { - {m_{\beta \gamma }}{m_\alpha }({{\hat z}_\alpha } - {\epsilon^\sigma })}  + \sqrt { - {m_{\alpha \gamma }}{m_\beta }({{\hat z}_\beta } - {\epsilon^\sigma })}  - i\sqrt {{m_\gamma }{m_{\beta \alpha }}\tilde E_{\beta {\zeta _\beta }\alpha {\zeta _\alpha }}^\sigma } }}\Bigg]^{i{\kern 1pt} {\eta _{\beta \alpha }}}, 
\label{Drenfactorgen1}
\end{align}

\begin{align}
{\tilde E}_{\beta \zeta_{\beta}\,\alpha \zeta_{\alpha}}^{\sigma}= E_{\beta \zeta_{\beta}\,\alpha \zeta_{\alpha}}^{\sigma} + \frac{\Mc}{m_{\beta\alpha}}\Big[\frac{{p'}_{\beta}^{2}- {q'}_{\beta}^{2}}{2\,m_{\beta}} + \frac{p_{\alpha}^{2} - q_{\alpha}^{2}}{2\,m_{\alpha}}\Big],	
\label{tildeE1}
\end{align}
\begin{align}
E_{\beta \zeta_{\beta}\,\alpha \zeta_{\alpha}}^{\sigma} = \frac{\Mc}{m_{\beta\alpha}}\,(E+ i\,0) - \frac{m_{\alpha\gamma}}{m_{\beta\alpha}}\,\hat{E}_{\beta n_{\beta}} - \frac{m_{\beta\gamma}}{m_{\beta\alpha}}\,\hat{E}_{\alpha n_{\alpha}}^{\sigma},
\label{Ebetaalpha1}
\end{align}
\begin{align}
\eta_{\beta\alpha} =\frac{ Z_{\beta}\,Z_{\alpha}\,e^{2}\,\mu_{\beta\alpha}}{\sqrt{2\,\mu_{\beta\alpha}\,{\tilde E}_{\beta \zeta_{\beta}\,\alpha \zeta_{\alpha}}^{\sigma}}}.           
\label{etabetaalpha1}
\end{align}
${\hat z}_{\alpha} = z - \frac{ p_{\alpha}^{2}}{2\,M_{\alpha}},
\quad {\hat z}= z-  \frac{{p'}_{\beta}^{2}}{2\,M_{\beta}},$   
and  $\,\,\Mc= m_{\alpha} + m_{\beta} + m_{\gamma}$.
For the diagram in Fig. \ref{triang_exch1} 
\begin{align}
E_{\beta \zeta_{\beta}\,\alpha \zeta_{\alpha}}^{\sigma} = \frac{\Mc}{m_{\beta\alpha}}\,(E+ i\,0) - \frac{m_{\alpha\gamma}}{m_{\beta\alpha}}\,\hat{E}_{\beta n_{\beta}}^{\sigma} - \frac{m_{\beta\gamma}}{m_{\beta\alpha}}\,\hat{E}_{\alpha n_{\alpha}}.
\label{Ebetaalpha2}
\end{align}

Thus we have shown that the off-shell Coulomb scattering amplitude doesn't cause any problem in calculation of the exchange triangular diagram.

\acknowledgments
The work was supported by the US Department of
Energy under Grant Nos. DE-FG02-93ER40773, DE-FG52-
09NA29467, and DE-SC0004958 (topical collaboration
TORUS) and NSF under Grant No. PHY-0852653. The authors appreciate the help from
Dr. N. Upadhyay for checking the manuscript.

\end{document}